\documentclass[prc,twocolumn,showkeys,showpacs,superscriptaddress]{revtex4}
\usepackage{dsfont}
\usepackage{graphicx,amsmath}
\def\rmF{{\rm F}}
\def\rmd{{\rm d}}
\def\om{\omega}

\def\half{{\textstyle \frac12}}

\newcommand{\lsim}{\stackrel{\scriptstyle <}{\phantom{}_{\sim}}}
\newcommand{\gsim}{\stackrel{\scriptstyle >}{\phantom{}_{\sim}}}
\begin{document}
\title{Viscosity of neutron star matter and  $r$-modes in rotating pulsars}
\author{E.\ E.\ Kolomeitsev}\email{Evgeni.Kolomeitsev@umb.sk}
\affiliation{Matej Bel  University, SK-97401 Banska Bystrica, Slovakia}
\author{D.\ N.\ Voskresensky}
\affiliation{National Research Nuclear University (MEPhI), 115409 Moscow, Russia}
\begin{abstract}
We study  viscosity of the neutron star matter and  $r$-mode instability in rotating neutron stars. Contributions to the shear and bulk viscosities from various processes are calculated accounting for in-medium modifications of the nucleon-nucleon
interaction. A softening of the pion mode at densities larger than
the nuclear saturation density $n_0$ and a possibility of the pion
condensation at densities above $3\,n_0$ are included. The
neutron-neutron and proton-proton pairings are incorporated, where
necessary. In the shear viscosity we include the lepton contribution calculated  taking into account the Landau damping in the photon exchange, the nucleon contribution described
by the medium-modified one pion exchange, and some other terms, such as the novel phonon contribution in the 1S$_0$ superfluid neutron phase, and the neutrino term in the neutrino opacity region. The nucleon shear viscosity depends on the density rather
moderately and proves to be much less than the lepton term. On the
contrary, among the terms contributing to the  bulk viscosity,
induced by the delay of the relaxation of lepton concentrations in
the star matter perturbed by the $r$-modes, the term from the
two-nucleon medium modified Urca reactions possesses strongest
density dependence (rising by  several orders of magnitude for
massive stars) because of the pion softening. Also, contributions
to the  bulk viscosity arising from other reactions induced by
charged weak currents, e.g., in the Urca processes on a pion
condensate and in direct Urca processes, are included. The
radiative bulk viscosity induced by charged and neutral weak
currents in the region of the neutrino transparency of the star is
also calculated accounting for in-medium effects. We exploit the
equation of state, which is similar to the
Akmal-Pandharipande-Ravenhall equation of state up to  $4\,n_0$,
but is stiffer at higher densities, producing the maximum neutron
star mass compatible with observations. The direct Urca processes
do not appear up to $5\,n_0$ (corresponding to the star mass
$M\simeq 1.9 M_{\odot}$). Computed with account of in-medium
effects, the frequency boundary of the $r$-mode stability for the
stars with the mass $\gsim 1.8 M_{\odot}$ proves to be above the
frequencies of all rotating young pulsars. However, none of the
conventional contributions to the viscosity are able to explain
the stability of rapid rotation of old recycled pulsars in X-ray
binaries. To solve this problem we propose a novel efficient
mechanism associated with the appearance  of condensates of low-lying
modes of bosonic excitations with finite momentum and/or with an
enhancement of the inhomogeneous pion/kaon condensates in some
parts of the star, if  the angular velocity  exceeds a critical
value.

\end{abstract}
\date{\today}
\pacs{
71.10.Ay,  
21.65.Cd,  
26.60.-c.  
 }
\keywords{nuclear matter, viscosity, r-modes, collective modes}
 \maketitle

\section{Introduction}

The maximum spin frequency of a star, at which the matter can be propelled away from the star surface --- the Kepler frequency, --- is equal to $\nu_{\rm K}\simeq 1.2 (M/M_{\odot})^{1/2}(10{\rm km}/R)^{3/2}$~kHz~\cite{Friedman:1989zzb}, where $M$ is the neutron star mass, $M_{\odot}$ is the solar mass and $R$ is the star radius (typically $M_{\odot}\lsim M\lsim 2M_{\odot}$ and $R\sim 10\mbox{---}13$ km for cold neutron stars). It would be natural to think that many neutron stars are born rotating with frequencies close to $\nu_{\rm K}$ and decelerate then during the further evolution. However, the majority of pulsars of age  $\lsim
10^5$~yr have rotation frequencies below 10~Hz. The fastest observed young pulsar PSR J0537-6910 has the rotation frequency $\nu_{\rm ex}^{\rm max, young}\simeq 62$~Hz~\cite{Marshall}.

The maximal rotation frequency of known pulsars, $\nu_{\rm ex}^{\rm max, old}\simeq $ 716 Hz,  was observed for the old binary radio pulsar PSR J1748-2446ad, see Ref.~\cite{Kaspi} and catalog in Ref.~\cite{Manchester:2004bp}. The old millisecond pulsars situated in the low-mass X-ray binaries (LMXB) are thought to be spun up or ``recycled" through accretion~\cite{Patruno:2010qz}. The majority of rapidly rotating old recycled pulsars have frequencies, $\nu_{\rm ex}^{\rm old}$, in the interval between 200 and 600~Hz. Thus, the Kepler frequency, $\nu_{\rm K}$, occurs much higher than rotation frequencies of young pulsars and even higher than those for the majority of the recycled pulsars.

A possible mechanism, which may be responsible for rapid
diminishing of rotation frequency of young fast pulsars up to
experimentally measured values, is related to the $r$-mode
instability. The $r$-modes, whose restoring force is the Coriolis
force, are closely related to Rossby waves in Earth's atmosphere and
oceans. Their instability was discovered by Andersson, Friedman,
and Morsink~\cite{Andersson:1997xt,Friedman:1997uh}; see also
review~\cite{Anders-Kokkotas00}. If the  instability occurs, the
star rapidly radiates its angular momentum via gravitational waves
until the rotation frequency $\nu$ reaches the critical value
$\nu_{c}$, at which the $r$-modes become stable. This rapid
process is accompanied by the neutron star reheating~\cite{AlfordMS12}.
For frequencies $\nu <\nu_{c}$ $r$-modes become damped and the star cools
down during $\sim 10^5$\,yr by the neutrino radiation and at later times,
by the surface photon radiation, slowly loosing its angular momentum during the
time evolution~\cite{ST83}.

The $r$-modes are damped by a viscosity of warm neutron star matter~\cite{Lindblom:1998wf} or  by other deceleration processes. For angular velocities of our interest, $10^3$ Hz$\lsim \Omega = 2\pi/\mathcal{P} \lsim 10^4$Hz, $\mathcal{P}=1/\nu$ being the rotation period of a pulsar, the
instability is most pronounced at temperatures $10^8\lsim T\lsim 10^{10}$~K.

Different mechanisms, which may eliminate the $r$-mode instability, have been studied, see for example Ref.~\cite{Anders-Kokkotas00} and references therein. Most attempts were made to find appropriate arguments for an increase of the shear and bulk viscosities of the neutron star matter. A common opinion has been formed that traditionally considered dissipation mechanisms are insufficient to provide stability for a large number of old pulsars in LMXBs. This would be at odds with expectations, because then these pulsars would still rapidly decrease their frequencies below the observed values~\cite{Ho:2011tt}. The minimum value of the frequency at the
$r$-mode stability boundary $\nu_c (T)$ is even smaller than the value $\nu_{\rm max}^{\rm young}$ if one uses standard dissipation mechanisms only; e.g., see Refs.~\cite{Anders-Kokkotas00,Owen:1998xg}.

One distinguishes three different contributions to the bulk viscosity of the dense and warm nucleon matter: the kinetic term proportional to the typical nucleon collision time, the soft-mode relaxation term, and the radiative contribution. At the neutron star conditions the kinetic term in the bulk viscosity is much smaller than the shear viscosity term and some other contributions to the bulk viscosity. The processes induced by charged weak currents in matter driven slightly out of the $\beta$-equilibrium occur at very slow rate. As was shown long ago by Mandelstam
and Leontovich~\cite{ML37,LL06}, in a system out of equilibrium the slow processes may essentially contribute to the bulk viscosity. We call this part of the bulk viscosity the``soft-mode" contribution, because it is related to the response function of the nuclear matter at very small frequencies.

References~\cite{Jones-hyp,Lindblom:2001hd,Nayyar:2005th} argued that high values of the soft-mode bulk viscosity might be induced by the reactions $p+\Sigma^- \leftrightarrow n+n$ and $p+\Lambda\leftrightarrow n+p$ in the presence of hyperons in neutron star interiors. However, the realistic equation of state (EoS) should not allow for a large fraction of hyperons in the star; otherwise it would be not able to describe the observable star masses.

When the proton concentration $X_p=n_p/n$ exceeds the critical value $X_p(n_c^{\rm DU}) =0.11\mbox{--}0.14$ the efficient direct Urca (DU) processes $n\to p+l+\bar{\nu}_l$, $p+l\to n+{\nu}_l$, $l=e,\mu^-$ become operative. For densities $n<n_c^{\rm DU}$, the DU processes are forbidden, because it is impossible to fulfill the energy-momentum conservation law. The appearance of the DU processes in the neutron star interiors for $n>n_c^{\rm DU}$ causes significant increase of the soft-mode bulk viscosity that would lead to a strong damping of the
$r$-modes~\cite{Haensel:1992zz,Haensel:2000vz}. The value of the critical density $n_c^{\rm DU}$ essentially depends on the symmetry energy in the EoS. Ordinary relativistic mean-field models and the microscopic Dirac-Brueckner-Hartree-Fock calculations produce low values of $n_c^{\rm DU}$. The work~\cite{Vidana:2012ex} observed that the $r$-mode instability region shrinks for those EoS that have a large symmetry energy slop $L$, because the value of the critical density for the DU
processes decreases with the increase of $L$; here $L=3n_0 (\partial E_{\rm sym}/\partial n)_{n_0}$, the nuclear saturation density  $n_0=0.16$~fm$^{-3}$, and $E_{\rm sym}$ stands
for the symmetry energy. We should notice that low values of $n_c^{\rm DU}$ are hardly compatible with the description of the cooling of neutron stars. According to
Refs.~\cite{Blaschke:2004vq,Grigorian:2005fn,Blaschke:2011gc,Blaschke:2013vma,Klahn:2006ir},
the neutron star cooling data can be hardly explained, if the one-nucleon DU reactions are allowed in stars with masses $M\lsim 1.35\mbox{--}1.5 M_{\odot}$. The microscopic variational
Akmal-Pandharipande-Ravenhall (APR) A18$+\delta v+$UIX$^*$ EoS~\cite{APR} also has a high value of the DU threshold density, $n_c^{\rm DU}\simeq 5 n_0$. Large values of $n_c^{\rm DU}$ are
obtained for some relativistic mean-field models with density-dependent couplings describing well atomic nucleus data~\cite{Typel2005}, and for a model with $\sigma$-field dependent hadron masses and couplings~\cite{Kolomeitsev:2004ff}; see the discussion of astrophysical constraints on the EoS in Ref.~\cite{Klahn:2006ir}.

Reference~\cite{Vidana:2012ex} attempted to set a lower limit on $L$, assuming that the measured high-spin frequency of the LMXB pulsar 4U 1608-52 is safe from the $r$-mode instability, and announced the constraint $L>50$~MeV.  However, in Fig.~6 of Ref.~\cite{Vidana:2012ex} the data lines are set incorrectly low and thereby the announced lower limit of $L$ should be actually set essentially higher. Moreover, simplifying consideration Ref.~\cite{Vidana:2012ex} ignored the nucleon superfluidity, which should be developed in old cold pulsars. The superfluidity effects induce a significant suppression of the DU bulk viscosity~\cite{Haensel:2000vz} and cannot be ignored in a realistic analysis.

Within the ``minimal cooling'' scenario~\cite{Page:2006ly} in the absence of the DU reactions, the most efficient cooling processes in nonsuperfluid matter are modified Urca (MU) processes, $N+n\to N+p+l+\bar{\nu}_l$, $N+p+l\to N+n+{\nu}_l$, for $N=n$ (neutron branch) and $N=p$ (proton branch) with $l=e,\mu^-$. The soft-mode bulk viscosity related to the MU processes was studied in~\cite{Sawyer:1989dp,Haensel:2001mw}. Reference~\cite{Haensel:2001mw}  used the free one-pion exchange (FOPE) model of Ref.~\cite{Friman:1978zq} for the description of the nucleon-nucleon ($NN$) interaction. The efficiency of the so-calculated MU processes ($\sim 10^6$ times less than that for the DU processes) is insufficient to stabilize $r$-modes not only in the recycled old rapidly rotating pulsars but also  in the most rapidly rotating young pulsar. Therefore, in this scenario the problem of the $r$-mode instability becomes even more severe for the stars with $M<M_{c}^{\rm DU}$. In presence of the nucleon pairing the pair-breaking-formation (PBF) neutrino processes incorporated in the minimal cooling scenario may counterbalance the reduction of the MU emissivity and substantially enhance the neutron star cooling~\cite{Voskresensky:1987hm,Senatorov:1987aa}. However, the processes on the neutral currents do not contribute to the soft-mode bulk viscosity term.

In presence of the nucleon superfluidity  there exist low-lying Goldstone modes. These modes only weakly interact with neutrons and with each other. The soft mode bulk viscosity of Goldstone
modes in superfluid matter owing to phonon-phonon interactions has been analyzed in Ref.~\cite{Manuel:2013bwa}. For very low values of the 3P$_2$ neutron gap, which we exploit in the given work following evaluation of Ref. \cite{Schwenk} and for the temperatures of our interest, the phonon contribution to the bulk viscosity in the core can be safely dropped. The phonon-phonon interaction might give a contribution to the bulk viscosity in the region of the 1S$_0$ neutron pairing, but the results are strongly model dependent. Therefore, we ignore this  contribution to the soft-mode bulk viscosity.

However, the neutrino emissivity of the two-nucleon processes  might be essentially enhanced with the density increase because of a modification of the one-pion exchange in nucleon
medium. This so-called pion-softening effect was first incorporated in
Refs.~\cite{Voskresensky:1985qg,Voskresensky:1986af,Senatorov:1987aa}
(also  see \cite{Migdal:1990vm,Voskresensky:2001fd}), and allowed to fit well neutron star cooling data. The ``nuclear medium cooling" scenario based on this effect was developed in
Refs.~\cite{Voskresensky:1986af,Voskresensky:1987hm,Migdal:1990vm,Schaab:1996gd,Voskresensky:2001fd,
Blaschke:2004vq,Blaschke:2011gc,Blaschke:2013vma}. The $NN$ interaction was described within the Landau-Migdal Fermi liquid approach including the long-range p-wave pion-nucleon attraction
and the short-range $NN$ repulsion evaluated with the help of the Landau-Migdal parameters. Being evaluated with relevant values of the Landau-Migdal parameters, the short-range part of the $NN$
interaction amplitude corrected by the loops decreases with a density increase, whereas the more long-range dressed pion exchange term becomes strongly enhanced.  As a consequence of the
softening of the pion mode at growing density,  for $n>n_c^\pi\gsim n_0$  there may appear charged and neutral pion condensates~\cite{Migdal1978,Migdal:1990vm}.

With increase of the squared matrix element of the $NN$ interaction, the nucleon shear viscosity should diminish, whereas the bulk viscosity should increase. The possibility of the pion mode softening,  leading to an increase of the $NN$ interaction amplitude with the growth of the nucleon density, which is found important for the analysis of the neutron star cooling, has not been yet incorporated in the analyses of the viscosity. In this work we reanalyze the stability of $r$-modes in pulsars within the scenario, where the one-nucleon reactions (DU) are suppressed in the majority of neutron stars (with masses $M<1.9M_\odot$), but the two-nucleon reactions on charged current, called  medium modified Urca (MMU) reactions, are enhanced by the pion-mode
softening. Additionally, we include contributions to the bulk viscosity induced by the reactions of the negatively charged pion condensate --- pion Urca (PU) reaction, $n+\pi^{-}_{\rm cond}\to
n+l+\bar{\nu}_l$, $l=e,\mu^-$, --- provided the pion condensate exists for $n>n_c^{\pi}$. We consider two possibilities -- pion condensate occurs for $n>n_c^{\pi}=3n_0$ and it does not occur -- but there is a softening of the pion mode, which degree we vary.

As we have mentioned, the energy of the $r$-mode can be dissipated not only via  the non-equilibrium (soft mode) effects but also via a neutrino radiation. The latter mechanism constitutes another source of the bulk viscosity, the so-called radiative viscosity.
It was studied only recently in Refs.~\cite{Sa'd:2009vx,Yang:2009iw}. Only the DU and MU reactions were incorporated. In-medium effects in these reactions may enhance the resulting radiative contribution to the bulk viscosity. Moreover, the processes with neutral currents, e.g., the
medium $nn$ and $np$ bremsstrahlung (MBn and MBp) reactions, $n+n\to n+n+\nu+\bar{\nu}$ and $n+p\to n+p+\nu+\bar{\nu}$ and the PBF reactions, $n\to n_{\rm pair}+\nu+\bar{\nu}$, $p\to p_{\rm
pair}+\nu+\bar{\nu}$, occurring in the matter with the nucleon pairing, may also contribute to the radiative bulk viscosity. We analyze the mentioned effects in this paper.

The shear viscosity contains two important contributions: one from the lepton scattering and another one from the nucleon scattering. For the nonsuperfluid matter, the nucleon term was first computed in Refs.~\cite{FlowersItoh,Cuttler} and then  recalculated in Ref.~\cite{Shternin:2008es} using vacuum $NN$ cross sections and including Pauli blocking effects. The electron term was first evaluated by Flowers and Itoh in Refs.~\cite{FlowersItoh,Cuttler} and then was recalculated in Ref.~\cite{Shternin:2008es}, taking into account the Landau damping of the intermediate photon and the proton pairing. The importance of the screening effects in the calculation of the shear viscosity at low temperatures was first emphasized in Ref.~\cite{Heiselberg:1993cr}. Note that these loop-screening effects in the photon exchange (dressing of the photon) are similar to the effects we incorporate in the pion exchange in the $NN$ interaction (dressing of the pion). The
lepton shear viscosity computed following Ref.~\cite{Shternin:2008es} proves to be an order of magnitude smaller than that computed in Refs.~\cite{FlowersItoh,Cuttler}. Note that the Flowers-Itoh result has been exploited in many papers studying the $r$-mode stability. These studies should be now redone with new values of the shear viscosities calculated with inclusion of the polarization effects. In the presence of the proton superfluidity the lepton term increases~\cite{Shternin:2008es}, which also should be taken into account.

Although the screening effects reduce the value of the lepton shear viscosity, it still remains larger than the nucleon shear viscosity term, provided the later term is computed with the
vacuum $NN$ cross sections; cf.~\cite{Shternin:2008es}. Below we show that this conclusion does not change after we take into account  the pion-softening effect in the $NN$ contribution to the
shear viscosity. Thus, the lepton shear viscosity proves to be the dominant damping mechanism for low temperatures $T< 10^9$ K, at which the bulk viscosity term is strongly suppressed.

The Goldstone mode contribution to the shear viscosity in superfluid neutron star matter induced by phonon-phonon interactions was studied in Ref.~\cite{Manuel:2011ed}. Even in the ballistic regime, when the phonons move freely within the whole superfluid region, being captured only at its borders, the phonon contribution to the shear viscosity proves to be small for $T\lsim 10^9$~K. For higher temperatures (but still in the superfluid phase) this effect should be included. Phonons in a nucleon superfluid may interact with neutrons. Below we calculate the corresponding novel contribution to the shear viscosity.

Moreover, for $T>T_{\rm opac}$ the neutrinos are trapped and contribute to the shear viscosity. This effect was not included so far. Evaluation of the opacity temperature with the MU processes~\cite{Friman:1978zq} gives $T_{\rm opac}\simeq 22\cdot 10^9$~K. For the MMU processes the quantity $T_{\rm opac}$ decreases~\cite{Voskresensky:1986af,Migdal:1990vm} owing to the pion softening, being strongly density dependent. Below we evaluate the neutrino shear viscosity taking into account the medium effects.

Many alternative mechanisms to get rid of the $r$-mode instability were explored in the literature. It was shown in Ref.~\cite{Rezolla00} that the magnetic field of a particular configuration and strength may preclude the $r$-mode instability.
The mixture of $r$-modes with other more stable resonance superfluid inertial modes in the star has been suggested in Refs.~\cite{Gusakov-ChK-PRL14,Gusakov:2013aza} as a reason for the stability of old recycled pulsars in X-ray binaries. The  authors argue that the mixture can only occur at some fixed ``resonance" stellar temperatures. The fast-spinning stars would cluster, therefore, in the vicinity of this temperature. The spin frequencies of stars are then limited by the instability of octupole ($m=3$) $r$-mode rather than by quadruple ($m=2$) $r$-mode. A neutron star in LMXB may spend a substantial fraction of time in the region of these stellar temperatures and spin frequencies.
The dissipation in the viscous boundary layer between the oscillating fluid in the core and a solid crust and a possible appearance of turbulence were argued in Ref.~\cite{Bildsten:1999zn} to provide an efficient damping of the $r$-modes. In Ref.~\cite{Lindblom:2000gu} it was demonstrated that the $r$-mode instability might be suppressed in neutron stars colder than $\sim 1.5 \cdot 10^{8}$\,K, if the crust were perfectly rigid. Eigenmodes of the crust-core boundary layer with a possible pasta phase were studied in Refs.~\cite{Fattoyev:2012ch}. The main uncertainty in the description of the crust-core transition arises from uncertainties in the parameter $L$ mainly owing to its dependence of the shear modulus. Using the estimated core temperatures of several LMXB pulsars and the EoS that describes the crust-core transition, and assuming that the main dissipation mechanism of the $r$-modes in old recycled pulsars is attributable to the electron-electron scattering at the crust-core boundary, the authors of Ref.~\cite{Fattoyev:2012ch} concluded that these neutron stars can be stabilized against $r$-mode oscillations, only if $L <65$~MeV. However, we should note that stability of $r$-modes at the crust-core boundary does not yet guarantee stability of the star as the whole. A similar constraint on $L$ ($L<70$~MeV) was extracted in Ref.~\cite{Podsiadlowski:2005ig} from the study of the experimental relation for the difference of the gravitational and baryon masses of the pulsar J0737-3039B. This conclusion was reached within a specific scenario of the formation of this pulsar. To explain the data on LMXB rapid pulsars, Ref.~\cite{Alford:2013pma} suggested a tiny $r$-mode scenario, requiring very low $r$-mode amplitudes $a_{\rm sat}\lsim 10^{-8}\div 10^{-7}$ hardly affecting the spin evolution. The currently proposed mechanisms for the $r$-mode
saturation~\cite{Bondarescu:2013xwa} saturate $r$-modes at $a_{\rm sat}\lsim 10^{-6}$ at most.  We note, however, that such a strong suppression of the $r$-modes was questioned in Ref.~\cite{Gusakov:2013aza}. Within the model~\cite{Alford:2012yn} the data on the young pulsar PSR J0537-6910 might be explained with a standard cooling, for $a_{\rm sat}$ being slightly below
$10^{-1}$. Within their scenario the spin-down evolution is generically slower than the thermal evolution, so that a star eventually reaches a steady state where heating equals cooling. Reference~\cite{ZhouWang} included a possibility of a differential rotation of the star. In their analysis the saturation is reached within 100s, for $a_{\rm sat}\sim 0.1-10$, in dependence on the value of the parameter $K$ characterizing an initial amount of the differential rotation, and the thermal evolution is slower than the spindown evolution. These approaches do not include thermal conductivity, nucleon superfluidity, or some other effects. For instance, the heat transport essentially delays the neutrino cooling. References~\cite{Blaschke:2011gc,Blaschke:2013vma} demonstrate that effects of non-zero thermal conductivity are important even for the duration of $330$~yr; their inclusion helps to explain rapid cooling of the pulsar in Cassiopea A. Thus, these interesting investigations  still cannot be considered as unequivocal and further analysis is required.

Another aspect of the problem is that in the rotating neutron star there may appear an inhomogeneous condensate of  bosonic excitations occupying a state with a finite momentum $k\neq 0$. It may occur if there are low-lying soundlike or rotonlike branches  in the  spectrum of  excitations with boson quantum numbers, and the star rotation velocity exceeds the Landau critical value $v_{c,{\rm L}}$ given by the minimum of $\epsilon (k)/k$, where $\epsilon (k)$ is the energy of the excitations~\cite{Voskresensky:1993uw}. A possibility of the condensation of rotons in the $^4$He-filled capillaries was first suggested by Pitaevskii in Ref.~\cite{Pitaev84}. The general
consideration applicable for different non-relativistic and relativistic systems including neutron stars was performed in Ref.~\cite{Voskresensky:1993uw}. A possibility of the condensation of zero-sound-like modes in Fermi liquids was considered in Ref.~\cite{Vexp95}. This reference also pointed out to an extra neutrino radiation owing to condensation of sounds and a possibility of reheating of rapidly rotating neutron stars, provided damped modes are excited. Moreover, the idea of Ref.~\cite{Pitaev84} was recently applied to cold atoms~\cite{BP12}.

In this paper we demonstrate that for sufficiently high angular velocities, $\Omega >\Omega_{c,{\rm L}}$, the condensates of bosonic excitations carrying a nonzero momentum can appear in some regions of the neutron star provided that the momentum is sufficiently large but the corresponding excitation energy is rather low. Then a part of the angular momentum of the star can be transferred to the condensate. As a result a part of the star rotates slower, whereas the remaining angular momentum is contained in the internal motion of the condensate. A similar effect may occur, if in the interior regions there exist nonhomogeneous pion and kaon condensates. Then a momentum might be transferred to these condensates. Below we discuss, whether an account for these possibilities might help explain high rotation frequencies of old recycled pulsars.

Our work is organized as follows. In  Sec.~\ref{configurations} we formulate the model, describe the EoS and the $NN$ interaction with the in-medium modified one-pion exchange contribution, and compute the neutron star configurations. Characteristic relaxation time-scales and the condition of the $r$-mode stability are considered  in Sec.~\ref{instability}. In Sec.~\ref{Shear viscosity} we focus on the calculation of the shear viscosity. First, we use the results of Refs.~\cite{Shternin:2008es} to calculate the lepton term with and without the proton pairing. This calculation incorporates polarization effects in the lepton-lepton interaction. We provide a simple numerical parametrization for a full lengthy analytical expression. Then we focus on the nucleon term. Here we consider the $NN$ interaction taking into account nuclear polarization effects including the pion softening.  Our result is compared with result of Ref.~\cite{Shternin:2008es} based on the use of the free cross-sections. Then we calculate the phonon-neutron interaction contribution to the shear viscosity in the 1S$_0$ neutron pairing phase. Finally, in the star regions, where neutrinos are trapped, for sufficiently high temperatures, we evaluate the neutrino contribution to the shear viscosity. In Sec.~\ref{Bulk viscosity} we focus on the soft-mode bulkviscosity. First, within our model we recalculate the DU and MU contributions to the soft-mode bulk viscosity in the absence and in the presence of the pairing and then incorporate the in-medium effects. It is demonstrated that the MMU and the PU processes strongly increase the resulting soft-mode bulk viscosity. Then we focus on the analysis of the radiative bulk viscosity.  The DU and MU  radiative bulk viscosity terms and then the MMU, PU MBn, MBp, and the PBF ones are calculated. The critical spin frequency with all processes included is computed in Sec.~\ref{spin}. Limits on the star rotation frequencies owing to inhomogeneous condensates of the bosonic excitations are studied in Sec.~\ref{Bose condensates}. An effect on the critical spin frequency is demonstrated. A mechanism of an acceleration of the rotation of old pulsars in LMXB because of the formation of inhomogeneous pion and/or kaon condensates in the course of the accretion is discussed. Our conclusions are formulated in
Sec.~\ref{Conclusion}. Some calculation details are deferred to the Appendix ~\ref{app}.

Throughout the paper we use the system of units in which the Planck constant, $\hbar$, and the speed of light, $c$, are equal to unity.

\section{Equation of state, neutron star configurations, and  $NN$ interaction}\label{configurations}

\subsection{Equation of state}\label{subsec:eos}

As the nuclear matter EoS, for nucleon densities $n>0.6\, n_0$  we use the phenomenological HDD EoS proposed in Ref.~\cite{Blaschke:2013vma}. We include contributions from neutrons, protons, electrons and muons. Our EoS is based on the Heiselberg-Hjorth-Jensen (HHJ) parametrization \cite{HHJ} (with the parameter $\delta =0.2$) that fits the microscopic APR A18+$\delta v+$UIX$^*$ EoS \cite{APR} for symmetric nuclear matter at baryon densities up to $4n_0$. This yields an acceptable (although not perfect) fit of the APR EoS and at the same time it makes it possible to avoid problems with causality at high densities. As a result, at high densities the HHJ EoS is softer than the APR EoS and does not support a neutron star with a mass $2~M_\odot$. To cure this drawback, it was suggested in Ref.~\cite{Blaschke:2013vma} to use a rescaled baryon density in
the HHJ EoS as an effective account for a missing repulsion at short distances,
$E^{\rm(HDD)}(n,X_p)=E^{\rm(HHJ)}(n\,\Phi(n/n_0),X_p)$, where the scaling function is chosen as
$\Phi(u)=1/[1-\alpha\,u\,\exp(-(\beta/u)^\sigma)]$ with the parameters $\alpha=0.02$, $\beta=6$, and $\sigma=4$, and $X_p=n_p/n$ is the proton concentration. This rescaling makes the EoS stiffer for densities larger than $5\,n_0$. The dependence of the effective nucleon mass on the nucleon density we parametrize as $m_N^*=m_n^*=m_p^* =(1-0.15\sqrt{n/n_0})\, m_N$, where $m_N
=938$~MeV is the nucleon mass in vacuum, as suggested in Ref.~\cite{V89}.

For the density $n\simeq 0.6\, n_0$ we match the HDD EoS with the Friedman-Pandharipande-Skyrme (FPS) EoS from Ref.~\cite{HP04}, which we use for lower densities. With our parameter choice, the HDD EoS produces a maximum mass $M_{\rm max}\simeq 2.05M_{\odot}$, being in agreement with observations~\cite{Demorest:2010bx,Antoniadis:2013pzd}. The causality condition is also satisfied.

\subsection{Neutron star configurations}

\begin{figure}
\begin{center}
\parbox{6cm}{\includegraphics[width=6cm]{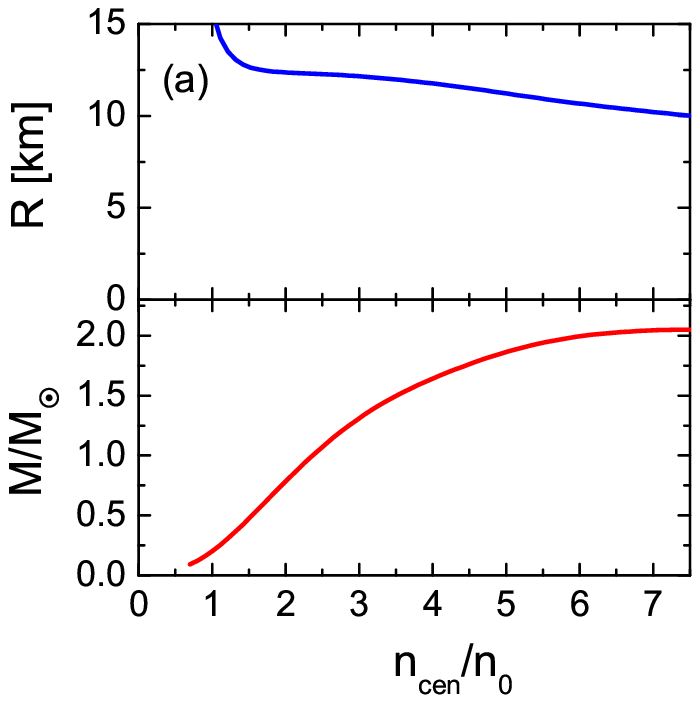}}\\
\parbox{6cm}{\includegraphics[width=6cm]{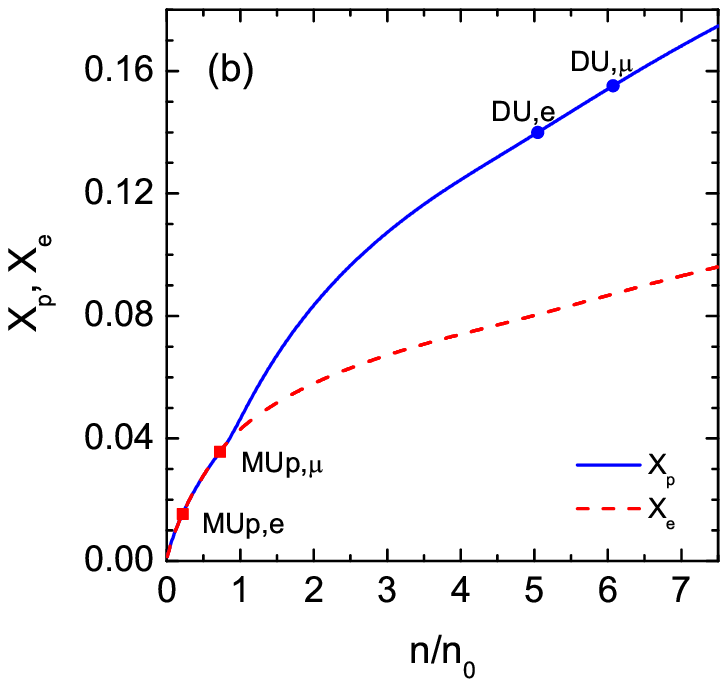}}
\end{center}
\caption{\label{fig:RM-HDD}
(Color online)  (a) Dependence of the neutron star mass and radius on the central nucleon  density for the HDD EoS~\cite{Blaschke:2013vma} with the FPS crust EoS~\cite{HP04}. (b) Concentrations of protons and electrons as functions of nucleon  density for the HDD EoS. Circles and squares indicate thresholds for  the $e$- and $\mu$- DU processes and for the MU processes (proton branch), respectively.}
\end{figure}

The masses and radii of neutron stars resulting from the solution of the Tolman-Oppenheimer-Volkoff equation are shown in Fig.~\ref{fig:RM-HDD} \,(a) as functions of the central nucleon density. For $1.1 M_{\odot}<M<M_{\rm max}\simeq 2.05 M_{\odot}$ the radius changes within the interval 10~km $<R<$13~km. Figure~\ref{fig:RM-HDD}\,(b) shows concentrations of protons and electrons as functions of nucleon density for the HDD EoS. The thresholds for the DU processes with participation of electrons and muons are indicated by circles and the thresholds for the MU processes (proton branch) by squares. As for the APR EoS, the one-nucleon DU processes with electrons, $n\to p+e+\bar{\nu}$, start to contribute only for densities $n>5n_0$, i.e.,  for stars with masses $M>M_c^{\rm DU}\simeq 1.9M_{\odot}$.

\begin{figure}
\begin{center}
\parbox{6cm}{\includegraphics[width=6cm]{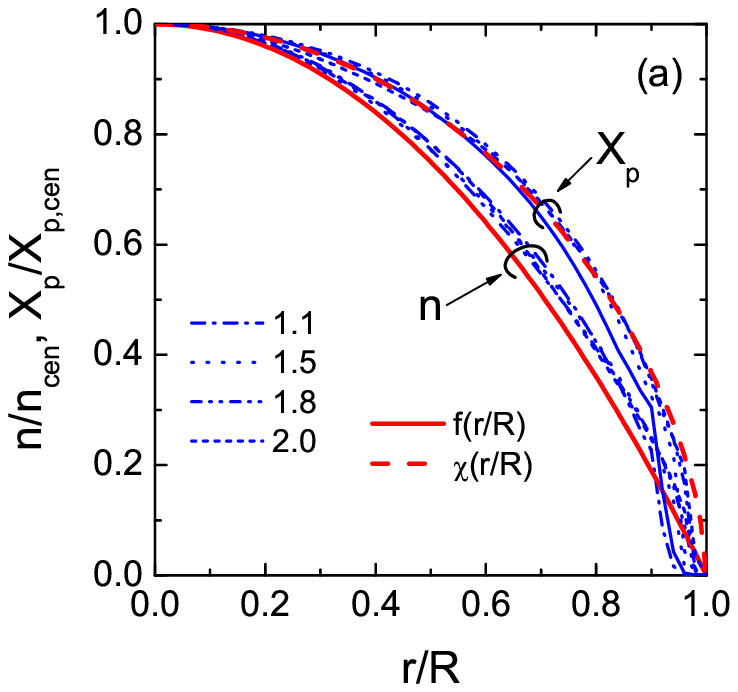}}\\
\parbox{6cm}{\includegraphics[width=6cm]{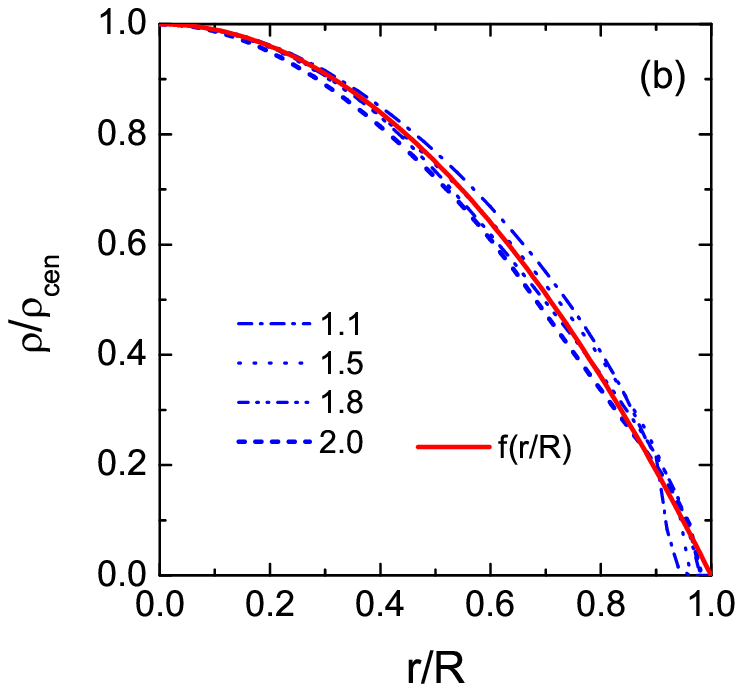}}
\end{center}
\caption{\label{fig:profiles} (Color online) Profiles of the nucleon density, proton concentration [panel (a)] and the mass density [panel (b)] for neutron stars of different masses, calculated with the HDD+FPS EoS in comparison with the simple parameterizations given by Eqs.
(\ref{f-profile}) and (\ref{fx-profile}). }
\end{figure}

In Fig.~\ref{fig:profiles} we show the profiles of the normalized nucleon number density, $n(r)/n_{\rm cen}$, and the normalized proton concentration, $X_p (r)/X_{p,{\rm cen}}$ [panel (a)], and the normalized mass density, $\rho(r)/\rho_{\rm cen}$ [panel (b)] for neutron stars of different masses calculated with the HDD+FPS EoS. Here $n_{\rm cen}$ and $\rho_{\rm cen}$ are the central number and mass densities and $n_{p,{\rm cen}}$ is the central proton number density, $X_{p,{\rm cen}}=n_{p,{\rm cen}}/n_{\rm cen}$. We see that the profiles of the number and mass densities depend weakly on the neutron star mass and  with appropriate accuracy can be parameterized by the universal function
\begin{eqnarray}
f(x)= 1-x^2\,,\quad x=r/R. \label{f-profile}
\end{eqnarray}
Such a mass distribution in the star corresponds to one of the analytical solutions found by Tolman~\cite{Tolman39}; see also discussion in Ref.~\cite{LattPrak01}. The profile of the proton fraction can be parametrized by the function $\chi (x)\simeq (1-x^2)^{3/5}\,,$ and the profile of the proton number density is then given by
\begin{eqnarray}
n_p (x)/n_{p,{\rm cen}}=f(x)\chi (x)=(1-x^2)^{8/5}\,.
\label{fx-profile}
\end{eqnarray}
The density profile (\ref{f-profile}) sets the relation between the central density of the star and its average density,
\begin{eqnarray}
\rho_{\rm cen}=\frac{5}{2}\bar{\rho}\,, \quad
\bar{\rho}=\frac{3M}{4\pi R^3}\,. \label{rho-aver}
\end{eqnarray}
Note that in the literature are often used profiles taken {\em ad hoc} without reference to a specific EoS. Our simple analytical fit is helpful for calculations with the HDD EoS. Below we exploit the above parametrizations.

As in all previous papers discussing $r$-mode instability, we use a simplifying assumption of a homogeneous distribution of the temperature in the star core. More general consideration would require a simultaneous solution of the heat transport problem (calculation of the heat conductivity) and an analysis of the $r$-mode instability (calculation of the viscosity). This is a much more ambitious problem lying beyond the scope of our research in this work.

\subsection{$NN$ interaction}\label{sec:NN-int}

\begin{figure}
\includegraphics[width=0.48\textwidth]{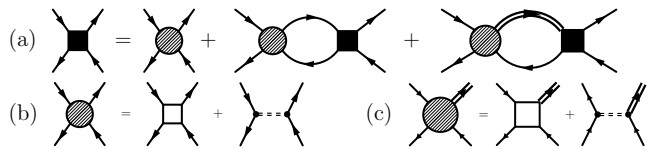}
\caption{Diagrams describing the $NN$ interaction in nuclear medium  (a), the particle-hole--irreducible nucleon-nucleon-hole interaction (b) and the particle-hole--irreducible coupling between nucleon--nucleon-hole and $\Delta$--nucleon-hole (c)}
\label{fig:NN-int}
\end{figure}

A key role in the development of the Fermi-liquid theory for atomic nuclei belongs to A.B. Migdal~\cite{M67}. Originally, nucleon particle-hole excitations were taken into account
explicitly, whereas other processes were hidden in a phenomenological parametrization.  The  pion mode is soft ($m_{\pi}\ll m_N$, where $m_\pi$ and $m_N$ are the pion and nucleon masses) and, therefore, should be considered explicitly on equal footing with the nucleon particle-hole terms. With an increase of the density the pion mode becomes softer owing to the momentum-dependent pion-nucleon attraction. The Fermi-liquid approach with explicit incorporation of the in-medium pion exchange was constructed by A.B. Migdal for zero temperature; see Ref.~\cite{Migdal1978}. Then, the approach was generalized for equilibrium systems at finite temperature and for nonequilibrium systems, see works~\cite{VM78,Migdal:1990vm,Voskresensky:1993ud,Kolomeitsev:2010pm}. For excitation energies of our interest ($\epsilon^{*}\ll \epsilon_{{\rm F},N}$) nucleons are only slightly excited above
their Fermi surfaces and all processes occur in a narrow vicinity of the nucleon Fermi energy $\epsilon_{{\rm F},N}$. Within this approach all the most important long-range processes are treated explicitly. At low excitation energies the $NN$ interaction amplitude is presented by diagrams (a)--(c) in Fig.~\ref{fig:NN-int} defined either on the Schwinger--Keldysh contour (or in matrix notation) in non-equilibrium or as retarded quantities in equilibrium. Below we deal only with equilibrium (retarded) quantities. The solid line stands for a nucleon; the double-line stands for a $\Delta$ isobar. Although the mass difference between the $\Delta$ and $N$, $m_{\Delta}-m_N \simeq 2.1 m_{\pi}> \epsilon_{{\rm F},N}(n)$, the $\Delta$-nucleon-hole term is
numerically rather large, because the $\pi N\Delta$ coupling constant is twice as larger as the $\pi NN$ one, and the $\Delta$ spin-isospin degeneracy factor is 4 times larger than for nucleons. The double-dashed line corresponds to the exchange of the free pion with inclusion of the contributions of the residual s-wave $\pi NN$ interaction and $\pi\pi$ scattering, i.e., the
residual irreducible interaction to the nucleon particle-hole and $\Delta$--nucleon--hole insertions. The block in Fig.~\ref{fig:NN-int}(b), depicted by the empty square, is irreducible with respect to particle--hole, $\Delta$--nucleon-hole, and pion states and is, by construction, essentially more local than the contributions given by explicitly presented graphs. One
reduces it to the set of functions dependent on the direction of the momenta of incoming and outgoing nucleon and hole at the Fermi surface,
\begin{align}
\label{localint}
\Gamma^{\omega}_{{\vec n}\alpha\beta,{\vec n}'\gamma\delta}=F_{{\vec n},{\vec
n}'}\delta_{\alpha\beta}\delta_{\gamma\delta}+G_{{\vec n},{\vec
n}'}\vec{\sigma}_{\alpha\beta}\vec{\sigma}_{\gamma\delta}\,,
\end{align}
where $\vec{\sigma}$ is the Pauli matrix. One introduces also dimensionless amplitudes $f_{{\vec n},{\vec n}'}=C_0^{-1} F_{{\vec n},{\vec n}'}$ and $g_{{\vec n},{\vec n}'}=C_0^{-1} G_{{\vec
n},{\vec n}'}$, where  $C_0^{-1}= m_N^* (n_0)p_{{\rm F},N}(n_0)/\pi^2$ is the density of states at the Fermi surface for $n=n_0$ and $p_{{\rm F},N}$ is the nucleon Fermi momentum. Dimensionless amplitudes $f_{{\vec p},{\vec p}'}$ and $g_{{\vec p},{\vec p}'}$ are expanded in the Legendre polynomials with the coefficients  known as Landau-Migdal parameters. It is sufficient to deal only with zero and first harmonics. These harmonics can be extracted from comparison with the data or can be computed within some models.

\begin{figure}
\includegraphics[width=0.48\textwidth]{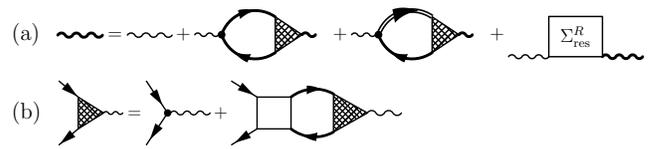}
\caption{(a)~Diagram representation of the Dyson equation for the pion propagator  in nuclear medium. (b)~Diagram equation for the $\pi N N$ vertex. }
\label{fig:pion-dress}
\end{figure}

The irreducible part of the interaction involving $\Delta$ isobars is given by diagram (c) in Fig.~\ref{fig:NN-int}. The main part of the $N\Delta$ interaction is attributable to the pion exchange. Therefore, and also for  simplicity one may neglect the first graph on diagram (c) of Fig.~\ref{fig:NN-int} readjusting, where necessary, other parameters.

Resummation of diagrams shown in Fig.~\ref{fig:NN-int} yields the following Dyson equation for the retarded full pion Green's function, diagram (a) in Fig.~\ref{fig:pion-dress}. Here $\Sigma_{\rm res}^{R}$ is the residual retarded pion self-energy that includes the contribution of all diagrams, which are not presented explicitly in Fig.~\ref{fig:pion-dress}(a): For instance, the
s-wave $\pi N$ scattering and the $\pi\pi$ scattering [shown by double-dashed line in diagram (c) in Fig.~\ref{fig:NN-int}]. Mainly, $\Sigma_{\rm res}^{R}$ is determined by the Weinberg-Tomazawa
term. The full vertex $\Gamma (n,\omega,k)$,  as shown by diagram (b) in Fig.~\ref{fig:pion-dress}, takes into account the $NN$ correlations.

The key effect that we incorporate  is the softening of the pion mode with increasing density \cite{Migdal1978,Migdal:1990vm}. Only with the inclusion of this softening effect the phase transition to a pion condensation state in dense nucleon matter appear. The approach, being often used in the literature, that exploits the FOPE model for description of the $NN$-interaction
and simultaneously incorporates processes, occurring in the presence of a pion condensate, violates unitarity. Indeed, when calculating the MU emissivity perturbatively one may use both the Born $NN$ interaction amplitude given by the FOPE and the optical theorem, considering the imaginary part of the pion self-energy \cite{Voskresensky:1986af,Voskresensky:1987hm,Voskresensky:2001fd}. In the latter case, at low densities  one may expand the exact pion Green's function,
\begin{align}
D_{\pi}^R(\omega,k)=[\omega^2 -m_{\pi}^2 -k^2 -\Sigma^R(\omega,k,n)]^{-1}
\end{align}
up to second order, i.e. one may use for the pion self-energy function $\Sigma^R (\omega,k,n)$ the perturbative one-loop diagram, $\Sigma_0^R(\omega,k,n)$. For  $k=k_0$, which is the pion momentum at the minimum of the effective pion gap $\widetilde{\omega}$ defined as
\begin{align}
\widetilde{\omega}^{\,2}=-D_{\pi}^{R,-1}(\omega =0,k =k_0)\,,
\end{align}
the self-energy function $\Sigma_0^R (\omega;k=k_0\simeq p_{{\rm F},N};n)$ yields a strong  p-wave attraction already for $n\ll n_0$. The attraction proves to be so strong that $\widetilde{\omega}^{\,2}$ becomes negative already at $n \sim 0.3\, n_0$ for  isospin-symmetric matter, which would trigger the pion condensation, in disagreement with experimental data indicating that there is no pion condensation in atomic nuclei. Note that the perturbative calculation (with $\Sigma_0^R$ instead of full $\Sigma^R $) contains  no free parameters. The paradox is resolved by observing that one needs to include a short-range repulsion arising from the dressed $\pi NN$ vertices. Then $\widetilde{\omega}^{\,2}$ becomes larger and the critical point of the pion condensation is moved to $n>n_0$.

A consistent description of the $NN$ interaction in matter should, thus, use a medium-modified one-pion exchange (MOPE), characterized by the fully dressed pion Green's function (depending on $\Sigma$ rather than on $\Sigma_0$) and dressed vertices $\Gamma(n)$ and a residual residual $NN$ interaction. Nevertheless, exploiting this approach, we should mention that the $NN$ interaction at high densities contains many quantities, which are poorly known. Therefore, quantitatively, the value of the vertex $\Gamma (n)$ and the value of the pion gap $\widetilde{\omega} (n)$ may vary from model to model  and such a variation may significantly affect the resulting values of the
$NN$ interaction amplitude. Here we continue to realize the program started in Refs.~\cite{Migdal1978,Migdal:1990vm} and continued in many subsequent works: (i) Include the important medium effects mentioned above; (ii) fit parameters to satisfy available experimental data for $n\leq n_0$, (iii) using $\widetilde{\omega} (n)$ and $\Gamma (n)$ as phenomenological quantities make predictions for higher $n$.

\begin{figure}
\includegraphics[width=6cm]{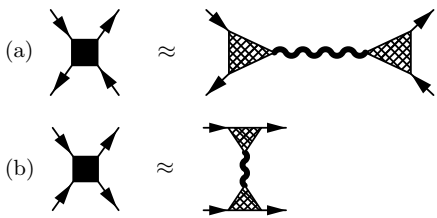}
\caption{Effective $NN$ interaction in the particle-hole (a) and particle-particle (b) channels at densities $n > n_0$ dominated by MOPE. The wavy lines and hatched vertices are determined as in Fig.~\ref{fig:pion-dress}.}
\label{fig:NN-int-mope}
\end{figure}

Microscopic calculations of the residual interaction are very cumbersome. However, according to  evaluations in Refs.~\cite{Migdal1978, Voskresensky:1986af,Migdal:1990vm}, the main contribution for $n>n_0$ is given by MOPE, whereas the relative contribution of the residual interaction diminishes with increasing density owing to polarization effects. Thus, one can evaluate the $NN$ interaction for $n> n_0$ with the help of the MOPE, see diagram (a) in Fig.~\ref{fig:NN-int-mope}. Here the bold wavy line relates to the in-medium pion. In the soft-pion approximation the same MOPE determines also interaction in the particle-particle channel, diagram (b) in Fig.~\ref{fig:NN-int-mope}. Namely, this quantity determines the $NN$ interaction entering neutrino emissivities of the MMU and MBn and MBp processes.

From Fig.~\ref{fig:pion-dress}(b) using the explicit expression for the nucleon-nucleon-hole loop for $\omega \ll kv_{{\rm F},N}$, one may evaluate
\begin{align}
\label{Gameval} \Gamma (n,\omega\ll k_0v_{{\rm F},N})\simeq
 [1+C (n/n_0)^{1/3}]^{-1}\,.
\end{align}
Here $ v_{{\rm F},N}=p_{{\rm F},N}/m_N^*$ is the nucleon Fermi velocity. Factor $C$ depends on values of the Fermi-liquid spin-spin Landau-Migdal parameters $g_{nn}$, $g_{np}$. As in
Refs.~\cite{Blaschke:2004vq,Grigorian:2005fn,Blaschke:2011gc,Blaschke:2013vma}
we use $C\simeq 1.6$.

\begin{figure}[!tbh]
\parbox{6cm}{
\includegraphics[width=6cm]{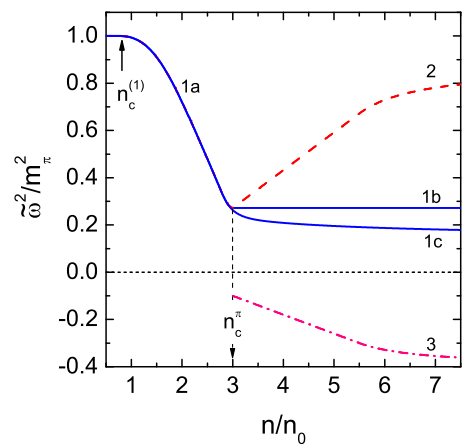}
} \caption{(Color online) Square of the effective pion gap $\widetilde{\omega}$
with  account for pion condensation at $n>n^\pi_{c}=3\,n_0$
(branches 1a+2 and 3) and without (1a+ 1b) or (1a+1c).
\label{omegatil} }
\end{figure}

The pion-softening effect starts for $n>n_{c}^{(1)}$  (taken here to be $0.8~n_0$) because for $n<n_{c}^{(1)}$ the value $-D^{-1}(\omega =0,k)$ has no minimum at $k\neq 0$. The density
dependence of the effective pion gap $\widetilde{\omega}$, which we use for $n>n_{c}^{(1)}$, is demonstrated in Fig.~\ref{omegatil} adapted from Fig.~1 in Ref.~\cite{Blaschke:2004vq} (also see
Ref.~\cite{Migdal:1990vm}). Curve 1a in Fig.~\ref{omegatil} shows behavior of the pion gap for $n_{c}^{(1)}<n<n_{c}^{\pi}$, where $n_{c}^{\pi}$, taken  to be $3 n_0$,  is the critical
density for the pion condensation. For simplicity we do not distinguish between different possibilities of $\pi^0$, $\pi^{\pm}$ condensations; see Ref.~\cite{Migdal:1990vm} for a more
general description. Curve 1b  demonstrates the possibility of a saturation of the pion softening in the absence of the pion condensation for $n>n_{c}^{\pi}$ (this possibility could be
realized, e.g., if the Landau-Migdal parameters increase with the density).  The curve~1c corresponding to another choice of parameters demonstrates continuing softening of the pion mode in absence of the pion condensation. Thus, both choices 1a$+$1b and 1a$+$1c determine behavior of the Green's function for the pion excitations in the absence of the condensation. Curves~2 and~3
demonstrate the possibility of the pion condensation for $n>n_{c}^{\pi}$. The continuation of branch 1a for $n>n_{c}^{\pi}$ along  branch 2, shows the reconstruction of the pion dispersion relation in the presence of the condensate state.  Then, for $n>n_{c}^{\pi}$ the values of
$\widetilde{\omega}$ from curve 2 determine the values of the MMU emissivity. The jump from branch 1a to branch 3 at $n=n_{c}^{\pi}$ is attributable to the first-order phase transition to the
$\pi$ condensation; see Ref.~\cite{Migdal:1990vm}. The $|\widetilde{\omega}|$ value on branch 3 (for the effective pion-pion coupling $\lambda \sim 1$) determines the amplitude of the pion condensate field. Simplifying as in Ref.~\cite{Voskresensky:2001fd}, we take
\begin{align}
|a_\pi|=f_\pi\,\sin\theta_{\pi} \simeq \sqrt{2}\,f_\pi\, |\tilde{\om}|/m_\pi\,,
\label{pi-cond}
\end{align}
where $\theta_{\pi}$ is the chiral angle and $f_\pi=92$~MeV is the pion decay constant.

We stress that dressing of the pion mode, which we exploit, is similar to the ordinary dressing of the photon mode in  plasma. Computing similar diagrams results in the dielectric constant $\varepsilon (\omega, q)$ in plasma, which may essentially deviate from unity. Moreover, only  including dressed vertices one is able to describe zero-sound modes in Fermi liquids.

\begin{figure}
\includegraphics[width=0.49\textwidth]{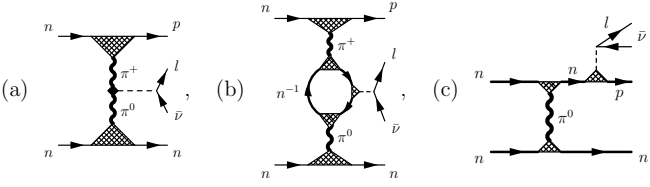}
\caption{Set of diagrams determining the MMU reactions.}
\label{fig:MMU-react}
\end{figure}

In the case when the $NN$ interaction amplitude is mainly controlled by the soft  pion exchange, the MU matrix element should be replaced with the MMU one. For $n\gsim n_0$ the main contributions to the MMU matrix element follow from the diagrams presented in Fig.~\ref{fig:MMU-react}.
Evaluations~\cite{Voskresensky:1986af,Voskresensky:2001fd} have shown that the dominating contribution to MMU rate comes from the first two diagrams in Fig.~\ref{fig:MMU-react}, whereas the third diagram, which naturally generalizes the corresponding MU (FOPE) contribution, gives only a small correction for $n \gsim n_0$. The strongest density dependence comes from the first diagram in Fig.~\ref{fig:MMU-react}. Note that  this term and the second term are absent, if one approximates the $NN$ interaction by a two-body potential.

Estimated for $n\gsim n_0 >n_c^{(1)}$ with the help of the first diagram in Fig.~\ref{fig:MMU-react}, the ratio of the MMU to MU neutrino production rates and the emissivities is equal to
\begin{align}
F_{\rm MMU}(n)=
3\Big(\frac{n}{n_0}\Big)^{10/3}\frac{[\Gamma(n)/\Gamma(n_0)]^6}{(\widetilde{\om}/m_{\pi})^8}\,.
\label{Ffun}
\end{align}
Note that in this estimation we use the vacuum weak coupling vertices, because  for $q_0 \gsim q$ in neutrino vertices the $NN$ correlation corrections are rather suppressed; see
Refs.~\cite{Migdal:1990vm,Voskresensky:2001fd} for details, but we correct the strong interaction vertices.

Analogously, the neutrino emissivity in the bremsstrahlung reactions $n+n\to n+n +\nu +\bar{\nu}$, being calculated with the MOPE instead of FOPE interaction, is enhanced by the factor, see
Ref.~\cite{Blaschke:2011gc},
\begin{align}
F_{{\rm MB}}(n) = 3\Big(\frac{n}{n_0}\Big)^{4/3}\frac{[\Gamma(n)/\Gamma(n_0)]^6}
{(\widetilde{\om}/m_{\pi})^3}\,.
\label{F-BS}
\end{align}
In this case only diagram similar to the diagram (c) in Fig.~\ref{fig:MMU-react} contributes to the reaction rate. For $n>n_0$,  $F_{{\rm MB}}(n)<F_{\rm MMU}(n)$. Contribution of the
$n+p\to n+p +\nu +\bar{\nu}$ reaction is still smaller.

\begin{figure}[!tbh]
\parbox{6cm}{
\includegraphics[width=5.5cm]{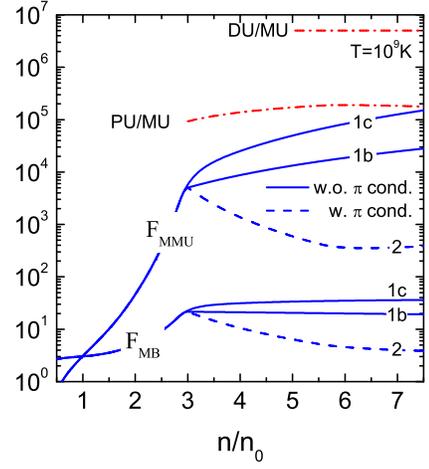}
}
\caption{(Color online) Enhancement factors for MMU and MB reactions given by Eqs.~(\ref{Ffun}) and (\ref{F-BS}), respectively, and those for PU and DU, cf. Eq.~(\ref{FDUPU}),  for $T =10^9$K as functions of the density for various types of the density dependence of the pion gap $\widetilde{\om}$ with (w. $\pi$ cond.) and without (w.o. $\pi$ cond.) pion condensation for $n>n_c^{\pi}$, labeled as in Fig.~\ref{omegatil}.
}
\label{fig:F-fact}
\end{figure}

The enhancement factors for the MMU and MB reactions given by Eqs.~(\ref{Ffun}) and (\ref{F-BS}) as functions of the density for various types of the density dependence of the pion gap $\widetilde{\om}$ are demonstrated in Fig.~\ref{fig:F-fact} for the typical temperature $T_9 = T/10^9$K $=1$. With values $\Gamma(n)$ and $\widetilde{\om}$ that we use, which allow to fit
measured surface temperatures of pulsars~\cite{Blaschke:2004vq,Blaschke:2011gc,Blaschke:2013vma}, we get $F_{\rm MMU}\approx 3$ for $n=n_0$ and $F_{\rm MMU}\approx 5\cdot 10^3$ for $n=3n_0$. We see that $F_{\rm MMU}\gg F_{\rm MB}$ for $n\gg n_0$. We plot also the ratios of the PU and DU
emissivities to the emissivity of the MU process, which can be estimated as
\begin{align}\label{FDUPU}
&F_{\rm DU}=\frac{\epsilon_{\rm DU}}{\epsilon_{\rm MU}}\simeq
5\cdot 10^{6}\,, \nonumber\\ &F_{\rm PU}=\frac{\epsilon_{\rm
PU}}{\epsilon_{\rm MU}}\simeq 7\cdot 10^5
\,\frac{|a_\pi|^2}{m_\pi^2}\frac{\Gamma^2(n)}{\Gamma^2(n_0)}
\Big(\frac{n_n}{n_e}\Big)^{\frac13}\,,
\end{align}
where $n_n$ is the neutron density and $n_e$ is the electron density. Note that in spite of a strong enhancement of the MMU emissivity with a density increase compared to the MU emissivity, the ratio $F_{\rm MMU}$ is less than both ratios $F_{\rm PU}$ and $F_{\rm DU}$. $F_{\rm MMU}(3n_0)/F_{\rm PU}(n>3n_0)\sim 0.05$, and only for the strongest softening among the choices we consider (following curve 1c of Fig.~\ref{omegatil}), the enhancement factor $F_{\rm MMU}$ tends to that for  the PU reactions at the maximal available density ($n_{\rm cen}\simeq 7.5 n_0$). In other cases (for curves 1b and 2 in Fig.~\ref{omegatil}), we have $F_{\rm MMU}\ll F_{\rm PU}$.

The following remarks are in order. First, we note that the ratio of the $NN$ cross sections calculated with the MOPE and FOPE interactions proves to be $\sigma [\rm{MOPE}]/\sigma[\rm{FOPE}]
\sim \Gamma^4(n) p_{{\rm F},N}^4/\widetilde{\omega}^4(n)$, being $\sim 0.3\mbox{--}0.5$ for isospin-symmetric matter at the density $n=n_0$ and for values of $\widetilde{\omega}$ and $\Gamma$ used above. The subsequent increase of the cross section with an increasing density is attributable to the win of the pion-mode softening (i.e., a decrease of $\widetilde{\om}$) induced by  $\pi NN$ and $\pi N\Delta(1236)$ $p$-wave attractions over the suppression of vertices by repulsive $NN$ correlations~\cite{Migdal:1990vm,Voskresensky:2001fd}. Thus, the known suppression~\cite{Blaschke:1995va,Bacca:2008yr} of the in-medium $NN$ cross section at $n\lsim n_0$ compared to that computed with the pure FOPE interaction does not conflict with the strong enhancement of the MMU emissivity at the increasing density; cf. Eq.~(\ref{Ffun}) and Fig.~\ref{fig:F-fact}. The estimated strong density dependence of the in-medium
neutrino-processes motivated authors of Ref.~\cite{Voskresensky:1986af} to suggest that difference in surface temperatures of neutron stars is explained by different masses of the objects. (At that time  only upper limits on surface temperatures were known).

Now let us discuss the influence of the pion-mode softening on the EoS. Although the long-range part of the $NN$ interaction might suffer a strong modification, the  EoS can remain rather insensitive to this. A similar situation often occurs in various problems in condensed matter physics. For example, in the case of a second-order phase transition, only the second derivative of a Ginzburg-Landau free energy undergoes a jump, whereas  the free energy and its first derivative are continuous in the critical point. Thereby, the EoS proves to be insensitive to the soft modes, whereas the particle scattering might be strongly enhanced in the presence of the soft mode. In the theory of finite Fermi systems by Migdal~\cite{M67} the scalar interaction $f_0$ is
responsible for the stiffness of the EoS rather than the long-range pion term. In the so-called self-consistent version of the theory of finite Fermi systems~\cite{Khodel-Saper} the EoS is constructed with the help of the density-dependent $f_0$ parameter. In the relativistic mean-field models for the EoS  all quantum effects are usually ignored, including effects of the pion mode.

The in-medium pion propagator is strongly enhanced only in the narrow range of exchanged frequencies and momenta. Therefore if this propagator is a part of some complicated diagram with many internal lines, the pion enhancement is averaged out. Such diagrams are assumed to be hidden in the Landau-Migdal parameters. So, in Ref.~\cite{Voskresensky:2001fd} it is shown that the number of virtual pions in nuclear matter at small temperatures is not singular even for $\widetilde{\om}\to 0$ and rather insensitive to the pion softening. The soft pion contribution to the energy per nucleon ${\cal{E}}_{\rm MOPE}$ (the leading Fock contribution) counted from the FOPE term, $\mathcal{E}_{\rm FOPE}$, can be estimated following Ref.~\cite{Senatorov-Yaf} as
\begin{align}
&\delta \mathcal{E}_{\rm MOPE}=\mathcal{E}_{\rm
MOPE}-\mathcal{E}_{\rm FOPE}\\ &\sim -(3/8)n f^2_{\pi NN}m^2_\pi
\left[\Gamma^2(n)/\widetilde{\om}^2(n)-1/(m_{\pi}^2 +p_{{\rm
F},N}^2) \right]\,,\nonumber
\end{align}
where $f_{\pi NN}\simeq1/m_\pi$ is the $\pi N N$ coupling constant. Thus, for $n\sim n_0$ and for isospin-symmetric matter we have $|\delta{\cal{E}}_{\rm MOPE}|\sim $~MeV. For $n\sim 3n_0$ we
evaluate $\delta\mathcal{E}_{\rm MOPE}\sim -10$~MeV, which is much less in comparison with  both the kinetic energy and the potential energy terms at this density.  For $n>3n_0$, the value
$|\delta\mathcal{E}_{\rm MOPE}|$ decreases with an increase of the density provided there appears pion condensate (see curve 2 in Fig.~\ref{omegatil}) or remains of the same order as for $n\sim 3n_0$ if the pion softening is saturated, see curves 1b and 1c.

For $n>n_{c}^{\pi}$ there appears an additional contribution of the pion condensation to the EoS. The value of the energy density-depends substantially on the poorly known values of the density
dependent Landau-Migdal parameters. Thereby it is difficult to judge if the effect is strong or weak. Some works speculated about possibility of the strong first-order phase transition, at which the neutron star may come to the dense pion condensate state; others assumed a tiny energy release, being sufficient only to explain strong star quakes in Vela pulsar; see Refs.~\cite{Migdal1978,Migdal:1990vm} and references therein. Within microscopic APR A18+$\delta v+$UIX$^*$ EoS, Ref.~\cite{APR} estimated the effect of the pion condensation on EoS  as tiny.
With the effective pion gap that we exploit (see curve 3 in Fig.~\ref{omegatil}) and for a typical value for the  effective pion-pion coupling $\Lambda\sim 1$ the energy gain per nucleon can be evaluated as $\delta{\cal{E}}_{\rm cond}^{\pi}\simeq -\widetilde{\omega}^4/(2\Lambda n)\sim -3$\,MeV for $n\simeq 7n_0$.

With these arguments, in the present work we exploit a phenomenological EoS described in Sec.~\ref{subsec:eos}, which ignores effects of the pion softening and the pion condensation on
the EoS.

\subsection{Nucleon pairing}

A possibility of the nucleon pairing in neutron stars was suggested by A.B. Migdal in Ref.~\cite{M59}. In spite of many calculations performed so far, the values of nucleon gaps in dense neutron star matter remain poorly known. This is the consequence of the exponential dependence of the gaps on the potential of the in-medium $NN$ interaction. The latter potential is not known sufficiently well. There are many calculations of the 1S$_0$ neutron ($nn$) and proton ($pp$) gaps and there are some evaluations of 3P$_2$ neutron gaps, for references see, e.g.,
Refs.~\cite{Shen:2002pm,Zuo:2004mc,Zhou:2004fz,Baldo:2007jx,Zuo:2006gn,Zuo:2008zza,Dong:2013sqa}.
Here, to be specific for the description of a superfluid phase in a neutron star we use parametrizations of critical temperatures, $T_{\rm c}$, from Ref.~\cite{Kaminker}: models 1ns and 1p for the neutron and proton 1S$_0$ pairing, respectively. These singlet pairing parametrizations have been exploited in Refs.~\cite{Blaschke:2004vq,Grigorian:2005fn,Blaschke:2011gc,Blaschke:2013vma}, as one of two  choices used there for fitting the neutron star cooling data within the nuclear medium cooling scenario developed in these works. In our numerical simulations below we exploit the above
mentioned set of gaps (named set I in~\cite{Blaschke:2004vq,Grigorian:2005fn}). As follows from the analysis of Ref.~\cite{Grigorian:2005fn} the cooling proved to be not very sensitive to the value and density dependence of the 1S$_0$ neutron gap mainly because the neutron pairing is restricted to the region of rather low densities, but it is sensitive to the values and density dependence of the proton gap. Below, when necessary, we demonstrate sensitivity of our results to the choice of gaps.

For the  temperature dependence of the pairing gap $\Delta (T)$ for the $^1$S$_0$ pairing we use the parametrization from Ref.~\cite{LYa94}
\begin{align}
\Delta(T)/T=\sqrt{1-\tau}\Big(1.456-\frac{0.157}{\sqrt{\tau}}+\frac{1.764}{\tau}\Big)\,.
\label{gap-T}
\end{align}
with  $\tau=T/T_{\rm c}$.

The value of $T_{\rm c}$ for the $^3$P$_2$  neutron pairing calculated according to the model 1nt of Ref.~\cite{Kaminker} (with the vacuum $NN$ interaction) is rather small, $\sim 0.25\times 10^9$~K. However, Ref.~\cite{Schwenk}  argued that the critical temperature for the triplet neutron pairing can be still strongly suppressed down to values $\sim 10^8$~K as an effect of the medium-induced spin-orbit interaction. This choice has been exploited in the nuclear medium cooling scenario of~\cite{Blaschke:2004vq,Grigorian:2005fn,Blaschke:2011gc,Blaschke:2013vma} for the description of the neutron star cooling. Because we focus here mainly on the description of the temperature interval $10^8\lsim T\lsim 10^{10}$K  and to avoid extra uncertainties, we
ignore the 3P$_2$ $nn$ pairing.

\begin{figure}
\centerline{
\includegraphics[width=7.5cm]{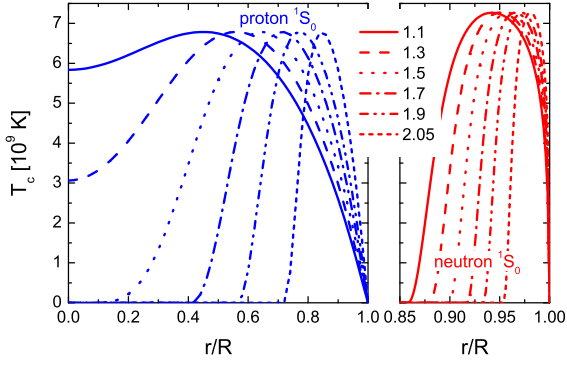}
} \caption{\label{fig:gap-profiles} (Color online) Profiles of critical temperatures, $T_c$, for 1S$_0$ proton and neutron pairings in neutron stars with various masses indicated by labels in the solar mass units. The density dependence of  $T_c$ is taken according to models 1ns and 1p of Ref.~\cite{Kaminker}. }
\end{figure}

Using the functions $f(x)$ and $\chi(x)$ from Eqs.~(\ref{f-profile}) and (\ref{fx-profile}) for the number density profile and the proton concentration profile, we calculate profiles of $T_{\rm c}(r/R)$ for different star masses and show them in Fig.~\ref{fig:gap-profiles}. The proton $^1$S$_0$ paring reaches deep inside the star core. Owing to this circumstance, within our model with strongly suppressed 3P$_2$ $nn$ pairing gaps the emissivities of the processes on charged weak currents are exponentially suppressed in a broader region of $r$ than emissivities of the processes on neutral currents.

Note that the angular velocities of the known most rapidly rotating old pulsars and even the angular velocity of the most rapidly rotating young pulsar PSR J0537-6910 exceed the values of the critical Landau angular velocities $\Omega_{c,{\rm L}}\sim \Delta_{NN}/(p_{{\rm F},N}R)$. Indeed, for the 1S$_0$ $nn$ and $pp$ pairings a rough estimate produces $\Omega_{c,{\rm L}}\sim 10^2$\,Hz. For the 3P$_2$ neutron pairing with the gap $\Delta \sim 10^8$\,K we get $\Omega_{c,{\rm L}}\sim$\,Hz. This, however, does not mean a complete destruction of the superfluidity, because for angular velocities in a broad interval $\Omega_{c1}\ll \Omega_{c,{\rm L}}\ll \Omega_{c2}$ there exists a mixed state consisting of the vortices of the normal matter surrounded by a superfluid. In the presence of vortices superfluid and normal components may co-rotate. Rough estimates yield~\cite{Tilli} $\Omega_{c1}\sim (\pi m_N R^2)^{-1}\ln (R\Delta)\sim 10^{-15}$\,Hz,
whereas $ \Omega_{c2}\sim \Delta^2/m_N$ and for the 1S$_0$ pairing $\Omega_{c2}(1{\rm S}_0) \sim 10^{18}$ Hz, and for 3P$_2$ neutron pairing with the gap $\Delta \sim 10^{8}$\,K, $ \Omega_{c2}(3{\rm P}_2) \sim 10^{14}$\,Hz.

\section{Time relaxation scales and $r$-mode instability}\label{instability}

The $r$-modes of rotating stars are associated with solutions of the perturbed fluid equations in the presence of (Eulerian) velocity perturbations in the form
\begin{eqnarray}
\delta v =a\, R\Omega\left(\frac{r}{R}\right)^l \vec{r}\times
\vec{\nabla}Y_{ll}e^{i\om t-t/\tau}+O(\Omega^3)\,,
\end{eqnarray}
where $R$ and $\Omega$ are the radius and angular velocity of the unperturbed star, $a$ is a dimensionless amplitude of the mode, $a\ll 1$, $Y_{ll}$ is the $ll$- spherical harmonic, and $\om
=-\Omega (l-1)(l+2)/(l+1)+ O(\Omega^3)$ is the circular frequency of the $r$-mode in the inertial frame; see Ref.~\cite{Anders-Kokkotas00}.

In the linear approximation the $r$-mode amplitude $a$, for $a\lsim 1$, follows equation ${\rmd a}/{\rmd t}=-a/\tau$; see Ref.~\cite{Owen:1998xg}. The characteristic time scale of the evolution
of the mode amplitude is given by
\begin{eqnarray}
\tau^{-1}=-\tau^{-1}_G +\tau^{-1}_{\eta}+\tau^{-1}_{\zeta}\,.
\label{t-mode}
\end{eqnarray}
Here $\tau_G$ is the typical time of the gravitational radiation, $\tau_{\eta}$ is the relaxation time owing to the shear viscosity, and $\tau_{\zeta}$ is the relaxation time owing to the bulk viscosity. The $r$-modes become stable if $\tau^{-1}>0$.

The time scale of the $r$-mode growth owing to the emission of gravitational waves is given by \cite{Lindblom:1998wf}
\begin{align}
\frac{1}{\tau_G} &=\frac{32\pi(l-1)^{2l}(l+2)^{2l+2}}{[(2l+1)!!]^2
(l+1)^{2l+2}} \, G \Omega^{2l+2}R^{2l+3} \langle\rho \rangle_{2l+2}
\,, \label{tG}\\
&\qquad \big\langle\dots\big \rangle_n =
\frac{1}{R^{n+1}} \int_0^R (\dots)\,r^{n} \,\rmd r \,,
\label{aver}
\end{align}
where $G=6.674\times 10^{-8}~{\rm cm^3\cdot  g^{-1}\cdot s^{-2}}$ is the gravitation constant.

The time scale of the $r$-mode damping owing to the shear viscosity is~\cite{Lindblom:1998wf}
\begin{eqnarray}
\tau_\eta^{-1}
=(l-1)(2l+1)\, R^{-2}\langle\eta\rangle_{2l}\big/\langle\rho\rangle_{2l+2}\,,
\label{tau-eta}
\end{eqnarray}
and the time scale of the $r$-mode damping owing to the bulk viscosity is
\begin{eqnarray}
\tau_\zeta^{-1}
=\frac{4\pi}{690}\left(\frac{\Omega^2}{\pi G\bar{\rho}}\right)^2
\frac{ \langle \zeta\, (1+0.86
(r/R)^2)\rangle_{8}}{R^2\langle\rho\rangle_{2l+2}}\,,
\label{tau-zeta}
\end{eqnarray}
where $\bar{\rho}$ is given by Eq.~(\ref{rho-aver}); see Refs.~\cite{Nayyar:2005th,Vidana:2012ex}.

The integral over the mass density profile with the function (\ref{f-profile}) can be easily calculated using the approximated profile (\ref{f-profile}),
\begin{eqnarray}
\langle\rho\rangle_{n} \simeq \rho_{\rm cen} \langle
f(r/R)\rangle_n =\frac{2\, \rho_{\rm cen}}{(n+1)\,(n+3)}\,.
\end{eqnarray}
In the general case the profile integral can be expressed through the Euler beta function
$\langle f^m(r/R)\rangle_n=\frac12 B(1+m,(1+n)/2)$\,.

According to Ref.~\cite{Lindblom:1998wf} the $r$-modes with $l=m=2$ are dominant ones. Then the gravitational time  can be estimated as
\begin{eqnarray}
\tau_G^{-1}=\frac{6.43\cdot 10^{-2}}{{\rm s}} R_{6}^7\, \Omega_4^6
\, \frac{\rho_{\rm cen}}{\rho_0}\,,
\label{t-G-1}
\end{eqnarray}
where $R_6=R/({\rm 10^6\,cm})$, $\Omega_4=\Omega/({\rm 10^4 Hz})$ and $\rho_0=2.63\cdot 10^{14}\,{\rm g/cm^3}$ stands for the mass density of the nuclear matter at saturation. The central density of the star depends on the neutron  star mass, cf. Fig.~\ref{fig:RM-HDD}(a).

The damping times of the r-modes (\ref{tau-eta}) and (\ref{tau-zeta}) can be rewritten as
\begin{align}
\tau_\eta^{-1} &= \frac{5.98\cdot 10^{-5}}{{\rm s}}\,
\frac{\langle \eta_{20} \rangle_4}{R_6^2 \, \rho_{\rm
cen}/\rho_0}\,, \label{t-eta-1}\\
\tau_\zeta^{-1} &= \frac{2.20\cdot 10^{-7}}{{\rm s}} R_6^4\, \Omega^4_4 \frac{\langle
\zeta_{20}\,[1+0.86 ({r}/{R})^2]\rangle_{8}}{(M/M_\odot)^2(\rho_{\rm cen}/\rho_0)}\,;
\label{t-zeta-1}
\end{align}
here we used that $M_\odot=1.99\cdot 10^{33}$~g is the solar mass, and $\eta_{20}$, $\zeta_{20}$ stand for the viscosities measured in units of $10^{20}~{\rm g\,\cdot cm^{-1}\cdot s^{-1}}$.
Finally, to evaluate $\tau$ we need to compute shear and bulk viscosities and average them appropriately over the neutron star density profile.

\section{Shear viscosity}\label{Shear viscosity}

The shear viscosity, $\eta=\eta_{e/\mu}+\eta_{n/p}$, is determined
by the term $\eta_{e/\mu}=\eta_e+\eta_\mu$, where $\eta_e$
incorporates the $ee$, $ep$, and $e\mu$ scatterings and $\eta_\mu$
includes the $\mu\mu$, $\mu e$, and $\mu p$ scatterings, and by the
nucleon term, $\eta_{n/p} =\eta_n +\eta_p$. Consider these
contributions separately.

\subsection{Lepton shear viscosity}

Leptons in neutron star matter are not superfluid. Electrons can be considered as massless because the electron Fermi momentum is much larger than the electron mass, $p_{{\rm F},e}\gg m_e$. The concentration of electrons is higher than that of muons. Thereby the lepton shear viscosity is dominated by the electron term. The lepton shear viscosity has been derived long ago by Flowers and Itoh~\cite{FlowersItoh}. Reference~\cite{Cuttler} proposed a convenient fit to their result,
\begin{align}
\eta_{e/\mu,20}^{\rm (FI)}\approx \eta_{e,20}^{\rm (FI)}= 4.2\cdot
10^{-3} \Big[\frac{\rm g}{\rm cm\cdot
s}\Big]\Big(\frac{\rho}{\rho_0}\Big)^2 T_9^{-2}\,, \label{etaE-FI}
\end{align}
$T_9=T/10^9~{\rm K}$. Equation~(\ref{etaE-FI}) was extensively exploited in the literature in various numerical simulations. The integrated viscosity entering Eq.~(\ref{t-eta-1}) evaluated with usage of the proton density profile (\ref{fx-profile}) and the Flowers-Itoh result (\ref{etaE-FI}), is then given by
\begin{align}
\langle\eta_{e/\mu,20}^{\rm (FI)} \rangle_4= 1.1\cdot 10^{-4}
\Big(\frac{\rho_{\rm cen}}{\rho_0}\Big)^2 T_9^{-2}\,.
\label{etaE-FI-aver}
\end{align}
The temperature dependence of this expression is the standard low-temperature result for the normal Fermi liquid with a short-range interaction. However, in reality, the charged lepton interaction is determined by the screened vector interaction mediated  by exchange of longitudinal and transverse plasmons, compare with Fig.~\ref{fig:NN-int}(a). By taking these effects into account, Ref.~\cite{Shternin:2008es} derived new expressions for the electron and muon shear viscosities. These expressions can be written as
\begin{align}
&\eta_l=\frac{1}{5 }n_l p_{\rmF,l}\,\tau_l =1.70\cdot 10^{34}
\Big[\frac{\rm g}{\rm cm\cdot
s}\Big]\Big(\frac{n_l}{n_0}\Big)^{4/3} \frac{\tau_l}{\rm s}\,,
\label{etal}
\end{align}
where the collision time $\tau_l$ can be expressed through the collision rate $\nu_l=\tau_l^{-1}$ determined by the lepton-lepton ($ll$)  and lepton-proton ($lp$) collisions; $p_{\rmF,l}$ is the
lepton Fermi momentum, $l=e,\mu$. The contributions of $ll$, and $lp$ collisions are dominant at small $T$, provided protons are not paired. The collision frequencies are equal to
\begin{align}
\nu_e &=A \Big(\frac{n_0}{n_p}\Big)^{\frac29}\Big(\frac{n_p}{n_e}\Big)^{\frac23}
T_9^{\frac53} (1+r)^{\frac23}\,,
\nonumber\\
\nu_\mu &=\nu_e \Big(\frac{n_e}{n_\mu}\Big)^{1/3}\,,
\label{nu-lead}
\end{align}
where $A=9.36\cdot 10^{14}~{\rm Hz}$ and $r=(p_{\rmF,e}^2+p_{\rmF,\mu}^2)/p_{\rmF,p}^2$, and $p_{\rmF,p}$ is the proton Fermi momentum. Full expressions for the lepton collision frequencies are presented in Appendix~\ref{app}.

\begin{figure}
\parbox{6cm}{\includegraphics[width=6cm]{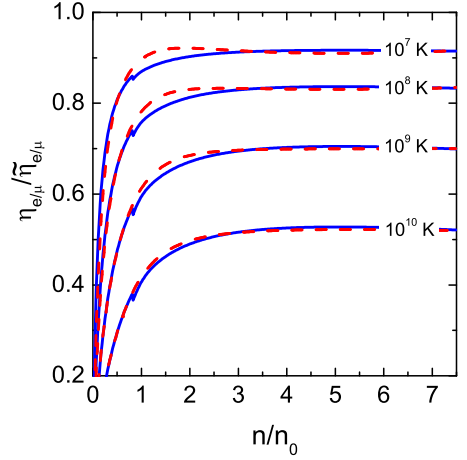}}
\caption{(Color online) Ratios of the full result for the lepton shear viscosity
$\eta_{e/\mu}$ given by Eqs.~(\ref{etal}) and (\ref{tauE-Mu}) and
the viscosity
$\widetilde{\eta}_{e/\mu}=\widetilde{\eta}_e+\widetilde{\eta}_\mu$
calculated  with Eqs.~(\ref{etaE}) and (\ref{etaMu}) as a function of
density for various temperatures (solid lines). Calculations
performed with the help of the interpolating formula
(\ref{etaE-inter}) are shown by dashed lines. }
\label{fig:ShearE-comp}
\end{figure}

Substituting Eq.~(\ref{nu-lead}) in Eq.~(\ref{etal}) we estimate the
leading contribution to the lepton shear viscosity as
\begin{align}
\widetilde{\eta}_e &=1.82\cdot10^{19}\Big[\frac{\rm g}{\rm cm\cdot s}\Big]\Big(\frac{n_p}{n_0}\Big)^{\frac{14}{9}} \Big(\frac{n_e}{n_p}\Big)^{2}
\frac{T_9^{-\frac53}}{ (1+r)^{\frac23} }\,,
\label{etaE}\\
\widetilde{\eta}_\mu &=\Big(\frac{n_\mu}{n_e}\Big)^{\frac53}\widetilde{\eta}_e\,.
\label{etaMu}
\end{align}

Note that expression (\ref{etaE}) has been  used in the literature as an approximate expression for the lepton term, e.g. in Refs.~\cite{AlfordMS12,Vidana:2012ex}; under the simplifying assumption that there are no muons in the medium, this means that one puts $n_e=n_p$ and $r=1$ in Eq.~(\ref{etaE}). For our EoS (which is close to the APR EoS for $n\lsim 4n_0$) the muon fraction is not negligible.

The ratios of the full result for $\eta_{e/\mu}$ given by Eq.~(\ref{etal}) with the collision times from Eq.~(\ref{tauE-Mu}) to the leading term $\tilde{\eta}_{e/\mu} = \tilde{\eta}_e + \tilde{\eta}_\mu$ determined by Eqs. (\ref{etaE}) and (\ref{etaMu}) are plotted in Fig.~\ref{fig:ShearE-comp} (solid lines)  as a function of the density for various temperatures. We see that for all temperatures and densities of our interest the term $\tilde{\eta}_{e/\mu}$
significantly overestimates the full result $\eta_{e/\mu}$.

The full result (shown by solid lines in Fig.~\ref{fig:ShearE-comp}) can be approximated by the following expression:
\begin{align}
&\frac{\eta_{e/\mu}}{\widetilde{\eta}_{e/\mu}}\approx
f_n(u)\,f_T(u,T)\,,\quad u=n/n_0, \label{etaE-inter}\\ &f_n(u)
=1-\frac{0.997}{(u^3 + 1.9\,u + 1)^{3/2}}\,, \nonumber\\ &f_T(u,T)
=\frac{\alpha(u)}{1 + \beta(u)\,T_9^{1/3}}\,, \nonumber\\
&\alpha(u)=0.9760+\frac{0.4925}{(u+0.5)^{2.266}}+1.055\cdot
10^{-5} u^{3.714}\,, \nonumber\\ &\beta(u) = 0.396 +
\frac{0.707}{(u+ 0.5)^{2.258}} + 2.161\cdot 10^{-5} u^{3.549}\,.
\nonumber
\end{align}
Deviation of interpolating formula (\ref{etaE-inter}) (see dashed lines in Fig.~\ref{fig:ShearE-comp}) from the full result (solid lines) ranges within 5\%  for densities $0.5\lsim n/n_0 \lsim 7.5$ and temperatures $10^{-2}\lsim T_9\lsim 10$.

The integrated viscosity entering Eq.~(\ref{t-eta-1}) evaluated with the proton density profile (\ref{fx-profile}) can be parametrized by a very simple formula
\begin{eqnarray}
\langle \eta_{e/\mu,20} \rangle_4=8.40\cdot 10^{-4}\,
\Big(\frac{n_{p,{\rm cen}}}{n_0}\Big)^{1.63}
T_9^{-1.737}\,,
\label{etaEMu-int}
\end{eqnarray}
where $n_{p,{\rm cen}}$ is the central proton  density.  This relation deviates from the full result only  within  2\%  for star masses $1\lsim M/M_\odot\lsim 2$ and temperatures $10^{-2}\lsim T_9\lsim 10$\,.

\begin{figure}
\parbox{6cm}{\includegraphics[width=6cm]{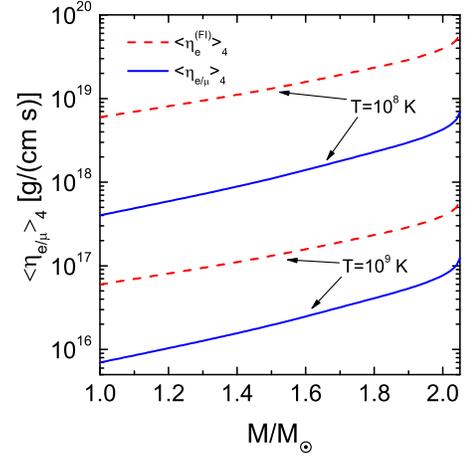}}
\caption{ \label{fig:eta1}
(Color online) Lepton  contributions to the shear viscosity integrated over the stellar density profile  $\langle \eta_{e/\mu} \rangle_4$ as a function of the neutron star mass for $T=10^9$~K and $10^8$K in the absence of the proton  pairing (dashed lines). Solid lines demonstrate calculations performed with Eqs.~(\ref{etal}) and (\ref{tauE-Mu}) and the dashed lines demonstrate calculations performed  with the Flowers-Itoh parametrization~(\ref{etaE-FI-aver}).
 }
\end{figure}

The dependence of the integrated lepton shear viscosity  on the neutron star mass is shown in Fig.~\ref{fig:eta1} for the full result, [see Eqs.~(\ref{etal}) and~(\ref{tauE-Mu}) (solid lines)] and for the Flowers-Itoh result [Eq.~(\ref{etaE-FI-aver}) (dashed  lines)] for $T=10^9$~K and $10^8$~K. We see that the proper account for the modification of the electromagnetic interaction  in the medium (Landau damping) leads to the reduction of the lepton shear viscosity by an order of magnitude. Thereby, numerous demonstrations of the $r$-mode instability performed in the literature with the Flowers-Itoh expression should be reconsidered.

\begin{figure}
\parbox{6cm}{\includegraphics[width=6cm]{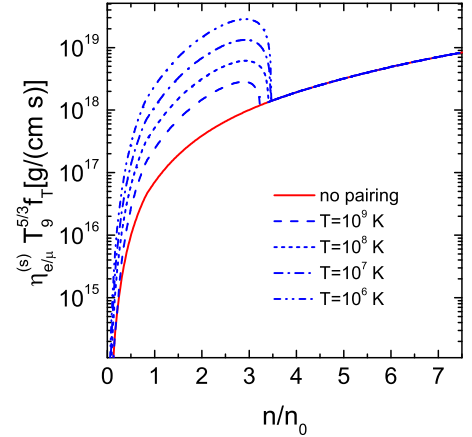}}
\caption{ \label{fig:etaSC} (Color online) Lepton contribution to the shear viscosity in the presence of the singlet proton pairing~(\ref{Eta-pairing}) scaled by the factor $T_9^{5/3}\,f_T(n,T)$ as a function of the nucleon density for various temperatures. The pairing critical temperature is the same as in Fig.~\ref{fig:gap-profiles}. The solid line shows the scaled viscosity in the absence of pairing. }
\end{figure}

A modification of the lepton shear viscosity in the presence of the singlet proton pairing can be taken into account by division of the final expression  by the overall $R$-factor,
\begin{eqnarray}
\eta_{e/\mu}\to \eta_{e/\mu}^{\rm (s)}=\eta_{e/\mu}/R_{e/\mu}(\Delta(T)/T,r) \,, \label{Eta-pairing}
\end{eqnarray}
where function $R_{e/\mu}$ is given by Eqs.~(80) in Ref.~\cite{Shternin:2008es}. The superscript ``(s)'' indicates quantities computed in presence of the pairing.

The pairing effect is illustrated in Fig.~\ref{fig:etaSC}, where the temperature-scaled shear viscosity $\eta_{e/\mu}^{\rm (s)}\,T_9^{5/3}\,f_T(n,T)$ is depicted as a function of the nucleon  density for various temperatures. Function $f_T$ is given by Eq.~(\ref{etaE-inter}). We see a strong enhancement of the lepton shear viscosity in the presence of the proton superfluidity,
growing with the temperature decrease.

\begin{figure}
\parbox{6cm}{\includegraphics[width=6cm]{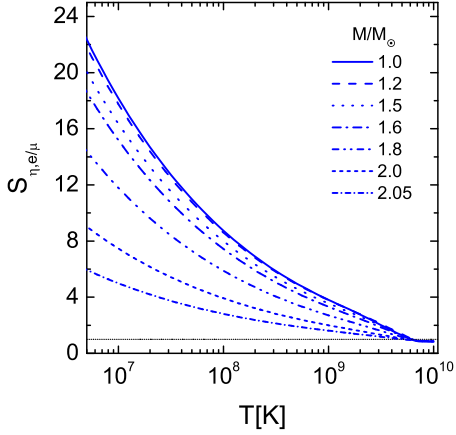}}
\caption{ \label{fig:Seta} (Color online) Integrated enhancement factor (\ref{S-eta-e}) for the lepton shear viscosity in the presence of the proton pairing  as a function of temperature for various neutron star masses. The density dependence of the critical temperature of the proton pairing is the same as in Fig.~\ref{fig:gap-profiles}. }
\end{figure}

To demonstrate the integrated effect of the superfluidity on the electron shear viscosity, in Fig.~\ref{fig:Seta} we show the ratio
\begin{eqnarray}
S_{\eta,e/\mu}=\langle \eta^{\rm (s)}_{e/\mu,20} \rangle_4 /
\langle \eta_{e/\mu,20} \rangle_4\, \label{S-eta-e}
\end{eqnarray}
as a function of the temperature for various neutron star masses. The resulting enhancement factor for $T<T_c$ depends weakly on the star mass in the range $M_\odot\lsim M \lsim 1.5\, M_\odot$ and
decreases sharply with an increase of $M$ above 1.6~$M_\odot$. The reason for this is that the region, where protons are paired, shrinks with increase of the star mass (for $M\gsim 1.5\,M_\odot$), as seen in Fig.~\ref{fig:gap-profiles}.

\subsection{Nucleon shear viscosity}

We turn now to evaluation of the nucleon contribution to the shear viscosity, first for a non-superfluid matter. The main term comes from the neutrons, because the neutron density is much higher than the proton one. As for the lepton shear viscosity, for the shear viscosity of purely neutron matter one routinely uses the results derived by Flowers and Itoh~\cite{FlowersItoh}, fitted by the analytical expression of Ref. \cite{Cuttler},
\begin{eqnarray}
\eta_n^{(\rm FI)} =1.1\cdot 10^{17}\left[\frac{\rm g}{\rm cm\cdot
s}\right](\rho/\rho_0)^{\frac94} T_9^{-2}\, \,.
\label{etaN-FI}
\end{eqnarray}

A more general result was obtained in Ref.~\cite{Shternin:2008es},
\begin{eqnarray}
\eta_n &=& \frac{n_n p_{{\rm F},n}^2}{5\,m_n^*}\tau_{n}\,,
\label{eta-n}
\end{eqnarray}
where  $p_{{\rm F},n}$ is the neutron Fermi momentum, and the relaxation time $ \tau_{n}$ is given by
\begin{eqnarray}\label{taun}
\tau_{n} =\frac{3m_N^2}{16m_n^{*\,3}T^2 S_{nn}} \,, \label{tau-n}
\end{eqnarray}
with the effective neutron-neutron scattering cross section
\begin{eqnarray}
\label{Snn} S_{nn}=\frac{m_N^2}{16\pi^2}\intop_{0}^{1} dx'
\intop_{0}^{\sqrt{1-x^{'\,2}}}dx \frac{12\,x^2\,
x^{'\,2}Q_{nn}(q,q')}{\sqrt{1-x^2-x^{'\,2}}}\,.
\end{eqnarray}
$Q_{nn}=\frac14\sum_{\rm spin}|M_{nn}|^2$ is the squared matrix element of the $nn$ scattering averaged over spins of initial and final neutrons, depending on the momenta $q=2p_{\rmF,n}\,x$ and $q'=2p_{\rmF,n}\,x'$.  To evaluate $Q_{nn}$ Refs.~\cite{Shternin:2008es,Baiko} exploited the vacuum cross sections and then evaluated  in-medium corrections owing to the Pauli blocking and the nucleon effective mass. The correction by the Pauli blocking proved to be rather small.

Combining Eqs.~(\ref{eta-n}) and (\ref{tau-n}) we obtain
\begin{eqnarray}
\eta_n &\approx& 2.05\cdot 10^{15}\Big[\frac{\rm g}{\rm cm\cdot
s}\Big]
\Big(\frac{n}{n_0}\Big)^{\frac53}\Big(\frac{m_N}{m_n^*}\Big)^4
\frac{T_9^{-2}}{[S_{nn} m_\pi^2]}\,. \label{etaN-Yak}
\end{eqnarray}
For the conversion from the convenient $m_\pi$ units ($m_{\pi}=139$~MeV)  one can use the relation $m_\pi^3\,c^3/\hbar^2=3.686\cdot 10^{11}$~g/(cm$\cdot$s).

We continue  comparison of the results for our MOPE parametrization of the $NN$ interaction and those for the FOPE parametrization. Within the FOPE model the matrix element $Q_{nn}$ in Eq.~(\ref{Snn}) is given by
\begin{align}
\label{QFOPE} Q_{nn}^{\rm FOPE}(q,q')& = f_{\pi N N}^4 \,\Big(
q^4\,[D^{(0)}_\pi(0,q)]^2 + q'^4\,[D^{(0)}_\pi(0,q')]^2
\nonumber\\ & +q^2q'^2\,D^{(0)}_\pi(0,q)D^{(0)}_\pi(0,q')\Big)
\,,
\end{align}
where $f_{\pi NN}\simeq 1/m_{\pi}$ is the pion-nucleon coupling, and $D^{(0)}_\pi(\om,q)=1/(\om^2-q^2-m_\pi^2)$ is the vacuum pion propagator; the transferred frequency is put to zero in Eq.~(\ref{QFOPE}) because $\om\sim \mu_e$ is much less than typical values $ q\sim p_{{\rm F},n}$. Then, neglecting $m_{\pi}^2$ in the denominator, which would yield correction $\propto m_{\pi}^2/p_{{\rm F},n}^2$, we get
\begin{eqnarray}
S_{nn}^{\rm FOPE}\simeq \frac{3\,m_n^2 f_{\pi NN}^4}{40\pi}\simeq \frac{1.1}{m_\pi^2}.
\label{Snn-FOPE}
\end{eqnarray}

As we have discussed,
Refs.~\cite{Voskresensky:1986af,Voskresensky:1987hm,Migdal:1990vm,Schaab:1996gd,Voskresensky:2001fd,Blaschke:2004vq,
Blaschke:2011gc,Blaschke:2013vma}  successfully exploited the in-medium modified $NN$ interaction for densities $n\gsim n_0$, which includes the attractive $NN$ interaction induced by the MOPE and the repulsive loop-corrected short-range interaction. For $n>n_{c}^{(1)}$ ($n_{c}^{(1)}< n_0$) the dominant part is given by the MOPE interaction $Q_{nn}^{\rm MOPE}$, which can be obtained from Eq.~(\ref{QFOPE}) after the replacements in the vertex $f_{\pi NN}\to f_{\pi NN}\,\Gamma$, with the vertex suppression factor $\Gamma$, and in the propagator $D_\pi^{(0)}(0,q)\to D_\pi(0,q)$, with the in-medium pion propagator given by the interpolating formula
\begin{align}
D_\pi^{-1}(0,q)\simeq -\tilde{\om}^2 - \gamma^2\,(q-k_0)^2\,, \quad
\gamma \sim 1\,,
\end{align}
valid for $|q-k_0|\ll k_0$ and for densities up to the critical density of the pion condensation $n_{c}^{\pi}$ (chosen in the above works and here to be $\sim 3n_0$). As in the model~\cite{Shternin:2008es,Baiko}, we suppress for simplicity the energy dependence of the $NN$ interaction amplitude. This dependence was taken into account in Ref.~\cite{Voskresensky:1991uv} for evaluation of the nucleon relaxation time  in the isospin-symmetric nuclear matter. The
effective gap $\widetilde{\om}^2$, which we continue to use, is shown in Fig.~\ref{omegatil}. The quantity $q=k_0\simeq p_{{\rm F},n}$ is the transferred momentum corresponding to the minimum of
the inverse pion propagator. The vertex suppression owing to the repulsive short-range $NN$ correlations depends on the Landau-Migdal parameter $g_{nn}$. As an interpolation formula, for
$\Gamma$ we use approximated Eq.~(\ref{Gameval}) with $C=1.6$.

With the MOPE interaction $Q^{\rm MOPE}_{nn}$ in Eq.~(\ref{Snn}) we obtain
\begin{align}
& S_{nn}^{\rm MOPE}=K\,S_{nn}^{\rm FOPE}\equiv K_0
\Big[1+\frac23\frac{\widetilde{\om}}{\gamma\,
p_{\rmF,n}}\Big]\,S_{nn}^{\rm FOPE}\,,
\nonumber\\ &
K_0=\frac{30\pi \Gamma^4
p_{\rmF,n}^3}{128\gamma\,\widetilde{\om}^3}\simeq 10.0
\frac{n}{n_0}\frac{\Gamma^4m_{\pi}^3}{\gamma\widetilde{\om}^3} \,.
 \label{Snn-MOPE}
\end{align}
For $n=n_0$, taking $\widetilde{\om}^2\simeq 0.98 m_{\pi}^2$, and estimating $\gamma \simeq 1$,  we find $K\simeq 0.3$. Such a degree of suppression of the MOPE $NN$ interaction amplitude
compared to FOPE one for $n\sim n_0$ is established by analysis of experiments with  atomic nuclei~\cite{Migdal1978,Migdal:1990vm,Voskresensky:1993ud} and by microscopic calculations~\cite{Blaschke:1995va,Bacca:2008yr}. With increase of the density the $NN$  interaction amplitude is suppressed by the vertex corrections induced by the repulsive short-range interaction and simultaneously enhanced by the decrease of the pion gap $\tilde{\om}$. With the growing density the pion softening becomes dominant and the factor $K$ increases, reaching 1 at $n\simeq 2.6\,n_0$. In the range of densities $2.6\, n_0\lsim n \lsim 3\, n_0$, $K$ rises  rapidly with growth of the density to the value $K\simeq 2$ for $n\simeq 3n_0$ [for $\widetilde{\om}^2(3n_0)\simeq 0.3\, m_{\pi}^2$] and remains roughly constant at higher densities
provided $\widetilde{\om}^2$ follows curve 1b in Fig.~\ref{omegatil}. If $\widetilde{\om}^2$ follows curve 1c the factor $K$ will increase further. If at $n\to n_c^{\pi}=3\, n_0$ a pion condensation occurs, then for $n>n_c^{\pi}$ in the presence of the condensation, $\widetilde{\om}^2$ follows curve 2 and the factor $K$ decreases with the density increase.

\begin{figure}
\parbox{6cm}{\includegraphics[width=6cm]{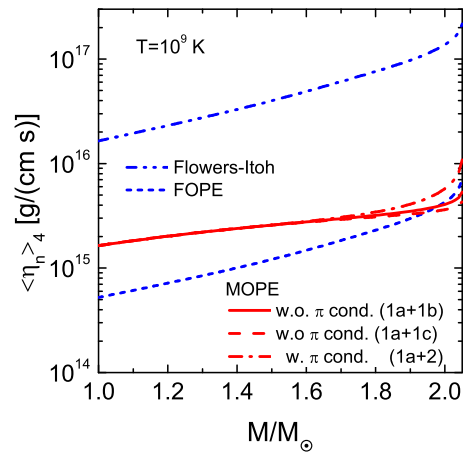}}
\caption{ \label{fig:eta} (Color online) Neutron contribution to the shear viscosity  integrated over the stellar density profile as a function of the neutron star mass at  temperature  $T=10^9$~K in the absence of   pairing.  The  MOPE calculation performed with Eq.~(\ref{Snn-MOPE}) for the pion gap $\widetilde{\omega}$ following curve 1a+1b, cf. Fig.~\ref{omegatil}, is shown by
the solid line, following curve 1a+1c, by the dashed line, and for $\widetilde{\omega}$ following curve 1a+2 by the dash-dotted line. The short-dashed line shows the calculation with Eq.~(\ref{etaN-Yak}) and the FOPE interaction given by Eq.~(\ref{Snn-FOPE}). The dash-dot-dotted line shows the calculation with the Flowers-Itoh parametrization~(\ref{etaN-FI}). }
\end{figure}

Figure~\ref{fig:eta} shows the neutron share viscosity averaged over the neutron star density profile. The dash-dot-dotted line is calculated with the Flowers-Itoh parametrization~(\ref{etaN-FI}). The dashed line depicts the result of Eq.~(\ref{etaN-Yak})
obtained with $S_{nn}^{\rm (FOPE)}$ given by Eq.~(\ref{Snn-FOPE}). The averaged viscosity calculated according to Eq.~(\ref{etaN-Yak}) with $S_{nn}^{\rm (MOPE)}$ from Eq.~(\ref{Snn-MOPE}) and the pion gap $\widetilde{\om}^2$ taken along the curve 1a+1b in Fig~\ref{omegatil} in the absence of the pion condensation is shown by the solid line. The calculations done for $\widetilde{\om}^2$ taken along curve 1a+2 in the presence of a pion condensate are shown by dash-dotted line. We see that the Flowers-Itoh result overestimates the result given by
Eq.~(\ref{etaN-Yak}) with $S_{nn}^{\rm (FOPE)}$ from Eq.~(\ref{Snn-FOPE}) by more than an order of magnitude. Compared to the FOPE result, the in-medium effects increase the averaged nucleon shear viscosity for stars with the mass  $\lsim 1.95~M_\odot$ and decrease it for heavier stars, provided there is no pion condensate in the star core. The resulting medium modified neutron shear viscosity, $\langle \eta_{n,20} \rangle_4\sim (2\cdot 10^{15}/T_9^2)\,{\rm g\,cm^{-1}\,s^{-1}}$,
is almost independent of the star mass, provided there is pion softening but there is no pion condensation. Also, we note that at all temperatures and densities of our interest the MOPE neutron shear viscosity in the absence of the nucleon superfluidity proves to be substantially smaller than the electron one, compare quantities in Figs.~\ref{fig:eta1} and~\ref{fig:eta}.

The nucleon shear viscosity in the presence of the neutron superfluidity could be considered, in principle, in full analogy with  calculations of the neutron thermal conductivity in a superfluid system performed in Ref.~\cite{Baiko}. Rough estimates show that in the presence of superfluidity the nucleon shear viscosity remains much smaller than the electron contribution. Therefore, further in this paper we neglect the nucleon contribution to the shear viscosity in our numerical simulations.

\subsection{Phonon shear viscosity in neutron superfluid}

A shear viscosity in a nuclear superfluid matter induced by phonon-phonon interactions was discussed in Ref.~\cite{Manuel:2011ed}.  Authors argued that these interactions may give a contribution  to the resulting shear viscosity for $T_{cn}>T\gsim 10^9$\,K, i.e., in the presence of the $nn$ pairing. The phonon-electron term can be neglected. However, we note that at such low temperatures the bulk viscosity term proves to be the dominant one. Therefore, owing to uncertainties in estimations of the phonon-phonon interaction we further disregard the phonon-phonon contribution to the shear viscosity in our numerical calculations.

We note that in the neutral superfluid there exist the Anderson-Bogoliubov modes, which interact with neutral particles, like superfluid neutrons. Because the number of excited neutrons is suppressed in the superfluid matter at $T\ll T_{c}$, the typical phonon-neutron collision time might be very large and the corresponding contribution to the shear viscosity might be sizable. Below we evaluate this contribution.

To find the spectrum of the phonon excitations  we use expressions for the anomalous vertices found in Ref.~\cite{KV08} as solutions of the Larkin-Migdal equations~\cite{LM63}. These equations generalize the equation for the coupling of external perturbation to the nucleon particle-hole [see diagram (b) in Fig.~\ref{fig:pion-dress}] for systems with the pairing correlations. The Anderson-Bogoliubov  modes exist as a response to the perturbation given by the temporal component of the vector current, see also Ref.~\cite{LV82}. From Eqs. (52) of the second reference in Ref.~\cite{KV08} for the anomalous scalar vertex and Eqs.~(57) and (58) there, for the vector vertex in the limit $\omega, v_{{\rm F},n}q \ll 2\Delta_{nn}$ we get
\begin{align}
&\tilde{\tau}_{V,0} =\frac{-2\, \Delta\,\om} {\om^2-\frac13
v_{{\rm F},n}^2q^2 - i\om\,\gamma_{\rm ph}(\om,v_{{\rm F},n} q)}
\,\tau^\om_{V,0}\,,
\nonumber\\
& \vec{q}\,\vec{\widetilde{\tau}}_{V,1} =
\frac{-2\, \Delta \frac13 v_{{\rm F},n}^2 q^2}
{\om^2-\frac13 v_{{\rm F},n}^2q^2 - i\om\,\gamma_{\rm
ph}(\om,v_{{\rm F},n} q)}\, \tau^\om_{V,0}\,,
\label{vertexsol-2}
\end{align}
where  $\tau^\om_{V,0}=1/a$ is the bare perturbation vertex with $a\simeq 1$ standing for the residue of the in-medium nucleon Green's function; $v_{{\rm F},n}=p_{{\rm F},n}/m^*_n$ is the neutron Fermi velocity. From these relations the phonon propagator can be defined as
\begin{eqnarray}
&&D_{\rm ph}(\om,k)=\frac{2\, \om} {\om^2-s^2\,v_{{\rm
F},n}^2q^2-i\om\,\gamma_{\rm ph}(\om,v_{{\rm F},n} q)}\,,
\label{ABmode-prop}
\end{eqnarray}
with $s=\frac{1}{\sqrt{3}}$ and the imaginary part
\begin{align}
\gamma_{\rm ph}(q) & = \frac{\pi}{6}\frac{\Delta}{T}  v_{{\rm F},n} q
\intop_{\sqrt{\frac32}}^{\infty}\frac{\rmd y}{(y^2-1)^2}
\frac{1}{(e^{\frac{\Delta}{T}y}+1)(1+e^{-\frac{\Delta}{T}y})}
\nonumber\\
& \approx \frac{2\pi}{3} v_{{\rm F},n} q e^{-\sqrt{\frac32}\frac{\Delta}{T}}\,.
\label{AB-imag}
\end{align}
The lifetime of the phonon owing to the interaction with superfluid neutrons is
\begin{eqnarray}
\tau_{\rm ph}=1/\gamma_{\rm ph} \approx\frac{3}{2\pi\, v_{{\rm F},n} q}
e^{\sqrt{\frac32}\frac{\Delta}{T}}.
\label{tau-ph}
\end{eqnarray}

\begin{figure}
\parbox{6cm}{\includegraphics[width=6cm]{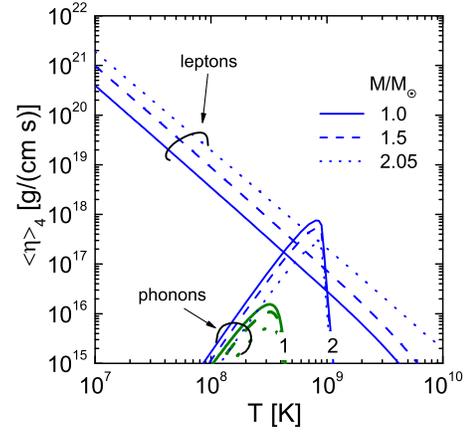}}
\caption{(Color online) The integrated shear viscosity, induced by the phonon interaction with superfluid  neutrons for the $nn$ singlet pairing as a function of the temperature for several neutron star masses in comparison with the lepton shear viscosity (proton pairing is incorporated where necessary). The curves for the phonon contribution labeled by 1 demonstrate calculations with the neutron pairing gaps shown in Fig.~\ref{fig:gap-profiles}. Curves labeled by 2 show calculations with the neutron pairing gaps taken from Fig.~1 of Ref.~\cite{Hebeler:2006kz}.} \label{fig:eta-phon}
\end{figure}

The phonon viscosity term  \cite{Khalatnikov} appearing in our case because of the interaction of superfluid neutrons with phonons is given by
\begin{align}
\eta_{\rm ph} &=
\intop\frac{\rmd^3 q}{(2\pi)^3} \frac{\tau_{\rm ph}}{15 T}
 \frac{ (s v_{{\rm F},n} )^2\,q^2}{(e^{sv_{{\rm
F},n}q/T}-1)(1-e^{-sv_{{\rm F},n}q/T})} \nonumber\\ &=
\frac{9\sqrt{3}\zeta(3)}{10\pi^3}\frac{T^3}{v_{{\rm F},n}^3}
e^{\sqrt{\frac32}\frac{\Delta}{T}}\,.
\nonumber
\end{align}
The last expression can be rewritten through the averaged lifetime (\ref{tau-ph})
\begin{align}
\bar\tau_{\rm ph}
&\approx\frac{3\,s}{2\pi\, T}
e^{+\sqrt{\frac32}\frac{\Delta}{T}} \frac{\intop_0^\infty \rmd x\,
x^3/(e^x-1)(1-e^{-x})}{\intop_0^\infty \rmd x\,
x^4/(e^x-1)(1-e^{-x})}
\nonumber\\ &=
\frac{45\sqrt{3}\zeta(3)}{4\pi^5}\frac{e^{\sqrt{\frac32}\frac{\Delta}{T}}}{T}
\label{aver-tau-ph}
\end{align}
as follows
\begin{align}
\eta_{\rm ph} =\frac{2\pi^2}{25}\frac{T^4}{v_{{\rm F},n}
^3}\bar{\tau}_{\rm ph}\,.
\end{align}
The averaged lifetime,
\begin{align}
\bar{\tau}_{\rm ph}\simeq 5.9\cdot 10^{-22}
\frac{e^{\sqrt{\frac32}\frac{\Delta}{T}}}{T_9}~{\rm s},
\end{align}
 cannot, however, exceed the ballistic time
\begin{align}
\tau_{\rm bal}\sim \frac{1~{\rm km}}{s\,v_\rmF}\simeq 1.6\cdot
10^{-5}~{\rm s}\, \Big(\frac{n_0}{n}\Big)^{\frac13}
\frac{m_n^*}{m_N}\,,
\end{align}
determined by the size of the region of the 1S$_0$  neutron pairing (put here about 1~km; cf. Fig.~\ref{fig:gap-profiles}), because phonons cannot cross the border of the superfluid region.
Taking into account this constraint we, finally, evaluate
\begin{align}
\eta_{\rm ph} &\simeq 2.1\cdot 10^{23} \Big[\frac{\rm g}{\rm
cm\cdot s}\Big] \Big(\frac{n_0}{n}\Big)
\Big(\frac{m_n^*}{m_N}\Big)^3 \nonumber\\ &\times  T_9^4\,
\frac{\min\{\bar{\tau}_{\rm ph},\tau_{\rm bal}\}}{\rm s}\,.
\end{align}
The integrated phonon viscosity is shown in Fig.~\ref{fig:eta-phon} for three values of masses of the neutron stars and for two choices for the $nn$ 1S$_0$ pairing gaps. The left slope of the curve corresponds to the ballistic regime (for the assumed 1-km size of the region with 1S$_0$ $nn$ pairing). The steep right slope corresponds to the collision  regime. For comparison, the integrated lepton term is also shown (superfluid neutron viscosity term is small and is dropped as well as the phonon-phonon interaction one). With the critical temperatures from Fig.~\ref{fig:gap-profiles} the effect of the phonon--superfluid neutron interaction can be safely neglected (see lower curves in Fig.~\ref{fig:eta-phon}). If in-medium modifications of the gaps are suppressed, with the gaps taken from Fig.~1 of Ref.~\cite{Hebeler:2006kz}, the phonon--superfluid neutron term might become dominant in a narrow  temperature region near
$10^9$\,K. This is demonstrated by the set of upper curves in Fig.~\ref{fig:eta-phon}.

\subsection{Neutrino shear viscosity}\label{subsec:nu-shear}

With the temperature increase the neutrino mean free path decreases and for sufficiently high temperatures neutrinos become trapped inside the neutron star interior. Under such conditions,
neutrinos still having a rather long mean free path may substantially contribute to the shear viscosity. The shear viscosity of the trapped neutrinos can be calculated as
\begin{align}
\eta_{\nu} &= 2\intop\frac{2\rmd^3 q}{(2\pi)^3} \frac{\tau_\nu}{15
T}
 \frac{ v_\nu^2\,q^2}{(e^{v_\nu q/T} + 1)(1 + e^{-v_\nu q/T})}
\nonumber\\ &=\frac{7\,\pi^2}{225\,v_\nu^3} T^4 \bar{\tau}_\nu
\approx 3.54 \cdot 10^{21}\Big[\frac{\rm g}{\rm cm\cdot s}\Big]
T_9^4\frac{\bar{\tau}_\nu}{\rm s} \,. \label{eta-nu-1}
\end{align}
Here in contrast to the phonon shear viscosity we have to  take into account that neutrinos are fermions with the spin $1/2$ and their speed is  $v_\nu \approx 1$. Extra factor 2 takes into account antineutrinos.  Neutrino collision times $\tau_\nu$ are different for the electron, muon and tau neutrinos: Electron-neutrinos have the shortest collision time, whereas tau-neutrinos have the longest. This difference reflects the difference in concentrations of the corresponding leptons in the neutron star matter. For simplicity we consider only the electron neutrino contribution to the  share viscosity.

In the absence of the DU reactions the neutrino collision time is mainly  determined by the inverse MU (or MMU) processes and by the inverse PU reaction, if the pion condensate is formed for $n>n_c^\pi$. Thus, in our scenario we have
\begin{align}\label{tausum}
1/\tau_\nu(q)=1/{\tau_\nu^{\rm
(MMU)}(q)+\theta(n-n_c^\pi)/\tau_\nu^{\rm (PU)}(q)}\,,
\end{align}
where $\theta (x)$ is the step function, $\theta (x)=1$ for $x\geq 0$ and $0$ for $x<0$.

Using the result for MU mean free path from Ref.~\cite{Friman:1978zq} [see Eq.~(95) there] and incorporating the in-medium modification factor (\ref{Ffun}) for the MMU processes (cf. \cite{Migdal:1990vm}) we get
\begin{align}
\tau^{\rm (MMU)}_\nu(q) &=(2.7\cdot 10^4~{\rm s})\,
\Big(\frac{n_0}{n_e}\Big)^{1/3}\frac{m_N^4}{m_n^{*\,3}m_p^*}\!\!\!
\nonumber\\
&\times\frac{\,T_9^{-4}\,F_{\rm MMU}^{-1}(n)}
{[(q/T)^2+\pi^2]\, [(q/T)^2+9\pi^2]}\,.
\label{tau-nu-MMU}
\end{align}
The PU collision time can be estimated as
\begin{align}
\tau^{\rm (PU)}_\nu(q) &=\frac{(2.9\cdot 10^{-5}~{\rm
s})\,T_9^{-2}}{(q/T)^2+\pi^2} \frac{F_{\rm DU}}{F_{\rm PU}}
\Big(\frac{n_0}{n_e}\Big)^{\frac13} \frac{m_N^2}{m_n^*\,m_p^*}\,.
\end{align}
The averaging over the neutrino momentum in Eq.~(\ref{eta-nu-1}) corresponds to the effective substitution  $q\to \bar{q}_\nu \simeq 4.4 T$ in the MMU and PU collision times~\cite{Friman:1978zq,Voskresensky:1986af}. With this replacement in Eq.~(\ref{tausum}), using Eq.~(\ref{FDUPU}) we evaluate the averaged neutrino collision time,
\begin{align}
&\bar{\tau}_{\nu} =\frac{8.7~{\rm s}}{T_9^{4}\,F_{\rm MMU}(n)}
\frac{m_N^4}{m_n^{*\,3}m_p^*}
\frac{(n_0/{n_e})^{\frac13}}{1+\chi_{\rm PU}(n,T)}\,,
\label{tau-nu-aver}\\ &\chi_{\rm PU}(n,T)=\frac{1.4\cdot
10^6\theta(n-n_c^\pi)}{T_9^2\,F_{\rm MMU}(n)}
\frac{m_N^2}{m_n^{*2}} \frac{|a_\pi|^2}{m_\pi^2}
\frac{\Gamma^2(n)}{\Gamma^2(n_0)}
\Big(\frac{n_n}{n_e}\Big)^{\frac13}\,. \nonumber
\end{align}
Substituting Eq.~(\ref{tau-nu-aver}) in Eq.~(\ref{eta-nu-1}) we finally obtain
\begin{align}
\eta_{\nu}=\frac{3.08 \cdot 10^{22}}{1+\chi_{\rm
PU}(n,T)}\Big[\frac{\rm g}{\rm cm\cdot s}\Big]
\Big(\frac{n_0}{n_p}\Big)^{\frac13}\frac{m_N^4}{m_n^{*\,3}m_p^*}F_{\rm
MMU}^{-1}(n)\,. \label{eta-nu}
\end{align}
Note that the partial MMU contribution to the neutrino shear viscosity does not explicitly depend on the temperature but only on the local density. The temperature dependence enters $\eta_\nu$ through the PU contribution ($\chi_{\rm PU}\propto 1/T^2$).

\begin{figure}
\parbox{0.44\textwidth}{\includegraphics[width=0.44\textwidth]{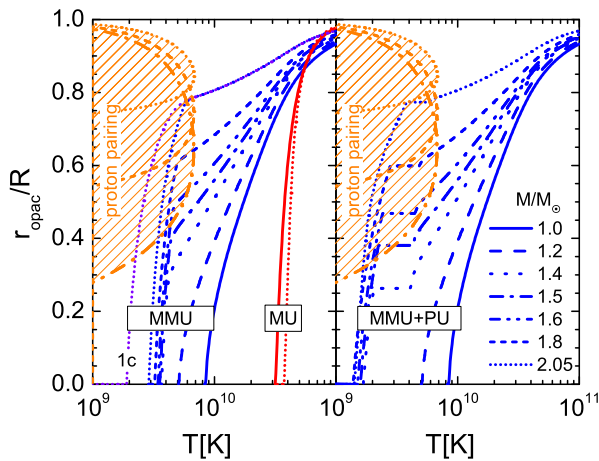}}
\caption{(Color online) The neutrino-opacity radius as a function of temperature for different neutron star masses calculated according to Eq.~(\ref{ropac}). The neutrino collision time is determined by the inverse MU processes (curves labeled as MU), by inverse MMU processes (curves labeled as MMU) [the latter processes are encoded in Eq.~(\ref{tau-nu-MMU}) by the factor $F_{\rm MMU}(n)$], and by the inverse PU processes in the presence of the MMU reactions (curves labeled as MMU+PU). The density dependence of the pion gap in the MMU enhancement factor~(\ref{Ffun}) is taken along curve 1a+1b in Fig~\ref{omegatil}, and along curve~1a+2 in presence of  pion condensate computed following curve~3. The curve labeled 1c is calculated with the pion gap taken
along curve 1a+1c. Hatched areas show regions of the star, where protons are paired, in dependence on the temperature. The regions are shown for stars with masses 1.5\,$M_\odot$, 1.8\,$M_\odot$ and 2.05\,$M_\odot$. } \label{fig:xopac}
\end{figure}

The temperature determines, though implicitly, the size of the region of the star, where neutrinos are trapped. Indeed, neutrinos can be trapped in a denser interior, whereas the outer region is already transparent for them. Thus, integrating the neutrino viscosity over the density profile we have to restrict the integration to the volume, where neutrinos are trapped,
\begin{align}
&\eta^{\rm (opac)}_\nu(r,T)=\eta_{\nu}(n(r))\,\theta(r_{\rm opac} - r)\,. \label{eta-nu-weight}
\end{align}
The radius of the region of the neutrino opacity, $r_{\rm opac}$, is determined by the relation
\begin{align}
&v_\nu\,\bar{\tau}_\nu(n(r_{\rm opac}),T)=R-r_{\rm opac}\,.
\end{align}
Using Eq.~(\ref{tau-nu-aver}) and putting $m_n^*\simeq m_p^*=m_N^*$ we can rewrite this equation in the  form
\begin{align}
\frac{r_{\rm opac}}{R}&=1 - \big(1+\chi_{\rm PU}(n(r_{\rm opac}),T)\big) \frac{m_N^4}{m_N^{*\,4}(n(r_{\rm opac}))}
\nonumber\\
&\times\frac{2.6\cdot
10^5\,T_9^{-4}}{R_6\, F_{\rm MMU}(n(r_{\rm opac}))}
\Big(\frac{n_0}{n_e(r_{\rm opac})}\Big)^{\frac13}\,.
\label{ropac}
\end{align}
The opacity radius, $r_{\rm opac}$, as given by Eq.~(\ref{ropac}), is depicted in Fig.~\ref{fig:xopac} as a function of the temperature for different values of the neutron star masses. For the MU processes (curves labeled MU) the opacity radius stays zero
independently of the neutron star mass up to the temperature $T_9\simeq 30$ and then it rises steeply to the star radius $R$ at higher temperatures. The dependence of the opacity radius $r_{\rm opac}$ on the neutron star mass  is very weak. The pattern changes drastically for the MMU reactions (curves labeled MMU). These MMU curves in Fig.~\ref{fig:xopac}  are calculated for the pion gap changing with the nucleon density along the curve 1a+1b in Fig.~\ref{omegatil}, i.e. in  absence of the pion condensation. The threshold temperature, at which $r_{\rm opac}$ starts growing, is much smaller, being sensitive to the value of the neutron star mass. For the heaviest neutron star ($2.05~M_\odot$) the opacity region appears already for $T_9>4$. We also plot $r_{\rm opac}$ for the heaviest star computed with the choice for the pion gap along the curve 1a+1c, see the curve labeled by 1c, again in the absence of the pion condensation but with a stronger pion
softening. In this case the opacity region appears at still smaller temperature, $T_9 > 2$.

Note that usually,  see
Refs.~\cite{Sawyer:1978qe,Friman:1978zq,Voskresensky:1986af,Migdal:1990vm,Voskresensky:2001fd},
the opacity temperature is determined as the temperature, at which the neutrino mean free path at some averaged density, $\bar{n}$, becomes equal to the neutron star radius, $v_\nu\,\bar{\tau}_\nu(\bar{n},T_{\rm opac})=R$. With the help of the opacity radius introduced above, we can define the opacity temperature more carefully, as the temperature, at which $r_{\rm
opac}=0$. The latter quantity is significantly smaller than the opacity temperature introduced in previous works, because the neutrino mean free path strongly depends on the density in the MMU
processes.

\begin{figure}
\parbox{6cm}{\includegraphics[width=6cm]{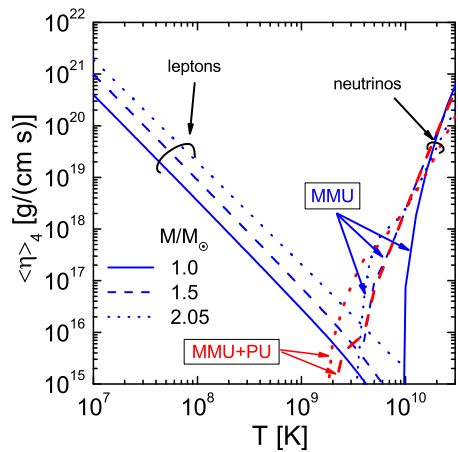}}
\caption{(Color online) The neutrino shear viscosity (\ref{eta-nu-weight}) averaged over the neutron star profile  for three values of the neutron star mass in comparison with the lepton shear viscosity (\ref{Eta-pairing}) calculated with account for the proton pairing. Curves labeled as MMU are calculated for the case where neutrinos are trapped by the inverse MMU processes only [the MMU enhancement factor (\ref{Ffun}) is taken along curve 1a+1b  in Fig.~\ref{omegatil}]. Calculations of the curves labeled as MMU+PU are performed following curve 1a+2 for MMU and curve~3 for PU. }\label{fig:eta-nu}
\end{figure}

In Eqs.~(\ref{tau-nu-aver}) and~(\ref{ropac}) we  do not include the nucleon pairing. It is permissible because the region of neutrino trapping, calculated without account for the pairing, practically, proves not to overlap with the pairing region in the star. Indeed, the neutron paring occurs at small densities only in outer parts of the star for $r\gsim 0.85 R$, which is larger than $r_{\rm opac}$ for the temperatures when the pairing exists. Proton pairing, although it reaches deeper inside the star, does not interfere with the neutrino trapping.  It is seen in Fig.~\ref{fig:xopac}, where we show proton pairing regions (hatched areas) as functions of the temperature for neutron stars with masses 1.5\,$M_\odot$, 1.8\,$M_\odot$ and 2.05\,$M_\odot$. As we see,  $r_{\rm opac}$ calculated with the MMU processes lies always outside  the pairing region for the corresponding star mass. The opacity radius calculated with inclusion of inverse PU  processes (lines labeled MMU+PU) is only minimally influenced by the proton pairing.

The neutrino viscosity calculated with the help of Eqs.~(\ref{eta-nu}), (\ref{eta-nu-weight})  and averaged over the star density profile is shown in Fig.~\ref{fig:eta-nu}. Curves labeled as MMU are calculated under the assumptions that neutrinos are trapped by the inverse MMU processes only, whereas for the curves labeled as MMU+PU, the neutrino scattering on the pion condensate is included. We see that the neutrino contribution to the shear viscosity  exceeds the lepton shear viscosity at $T_9\gsim 4$ for middle-heavy and heavy neutron stars. For the heaviest neutron star the neutrino shear viscosity starts to dominate over the lepton contribution at $T_9\gsim 3$, if the PU processes are taken into account. For the light neutron stars with masses $M\sim M_{\odot}$ the neutrino shear viscosity term  dominates over the lepton term only for $T_9 > 10$.

\section{Bulk viscosity}\label{Bulk viscosity}

We consider three sources of the bulk viscosity: the contribution owing to the $NN$ collisions, $\zeta_{\rm col}$, the soft-mode contribution, $\zeta_{\rm s.m.}$, and the radiative
viscosity term, $\zeta_{\rm rad}$. The total viscosity is the direct superposition of these contributions $\zeta=\zeta_{\rm col}+\zeta_{\rm s.m.}+\zeta_{\rm rad}$. In subsequent three
subsections we consider the bulk viscosity in the normal matter. Pairing effects will be
evaluated in Sec.~\ref{bulkpairing}.

\subsection{Collisional term}

\begin{figure}
\parbox{6cm}{\includegraphics[width=6cm]{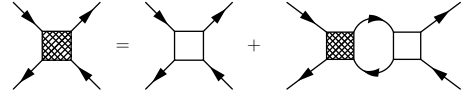}}
\caption{Diagrammatic equation for the particle-hole interaction.}
\label{fig:loceq}
\end{figure}

The contribution to the bulk viscosity induced by $NN$ collisions, calculated
at low temperatures in the framework of the Fermi-liquid theory~\cite{SB70,GP-FL}, is equal to
\begin{align}
&\zeta_{\rm col}=\frac{\pi^2n p^2_{{\rm F},N}}{18m_N^*}\tau_N
\left(\frac{T}{E_{\rm F}}\right)^4
\label{FLbulk}
\\
&\times \left[ \frac{p^2_{{\rm F},N}}{2m_N^*}\frac{\partial^2}{\partial
E_p^2}\left(\frac{p^2}{3m_N^*}+n\int\frac{d\Omega_{\vec
p}}{4\pi}T_{{\vec p},{\vec p}'}\right)\right]^2_{p=p'=p_{{\rm
F},N}}\,,
\nonumber
\end{align}
where $E_p=p^2/2m_N^*$, $E_{\rm F}= p^2_{{\rm F},N}/2m_N^*$, and $\tau_N$ is the nucleon relaxation time, which can be estimated as \cite{Bacca:2008yr} $\tau_N\sim 0.1 m^2_\pi
m_N^2/(m_N^{*\,3}\, T^2)$; cf. Eqs. (\ref{taun}), (\ref{Snn-FOPE}), and
(\ref{Snn-MOPE}). The quantity $ T_{{\vec p},{\vec p}'}$ is the symmetric part of
the dimensionless quasiparticle interaction amplitude determined
by the symbolic equation shown in Fig.~\ref{fig:loceq} with the empty
block parameterized as in Eq.~(\ref{localint}).

Retaining only zeroth harmonic of the scalar interaction we have
\begin{eqnarray}
T_{{\vec p},{\vec p}'} =\frac{C_0\,f_0}{1+C_0\,f_0 L(\omega,\vec{q},n)},
\end{eqnarray}
where $L$ is the loop function, and $f_0 $ is the zeroth harmonic
of the dimensionless scalar  Landau-Migdal parameter. For $nn$
collisions the relevant parameter is estimated as $|f^{nn}_0|\lsim
1$; e.g., see Ref.~\cite{Wambach:1992ik}. Then Eq.~(\ref{FLbulk}) can be
re-written as
\begin{eqnarray}
\zeta_{\rm col}=\frac{2}{81\pi^2} \frac{m_N^{*5}}{m_N^2 n_0}
\tau_N \, \left(\frac{n_0}{n }\right)^{1/3} T^4  f_0^2\,,
\end{eqnarray}
The $nn$ collision contribution to the viscosity proves to be
numerically small at the neutron star temperatures,
\begin{eqnarray}
\zeta_{\rm col}\sim 10^4\Big[\frac{\rm g}{\rm cm\cdot s}\Big]\,
T_9^2\, \Big(\frac{n_0}{n}\Big)^{1/3}\,\Big(\frac{m_N^*}{m_N}\Big)^2\,
f_0^2\,,
\end{eqnarray}
and can be neglected.

In the star regions where neutrinos are trapped [$r<r_{\rm opac}(T)$], neutrino collisions can contribute to the bulk viscosity, but we expect this contribution to be smaller than the neutrino shear viscosity term calculated in Sec.~\ref{subsec:nu-shear}.

\subsection{Soft-mode contribution}

The soft-mode contribution to the bulk viscosity, $\zeta_{\rm
s.m.}$, arises in the case, where the pressure of the system
depends on a parameter, which time of the relaxation to the equilibrium
value is substantially longer than the period of the density
variation~\cite{ML37,LL06}. In our case the soft-mode contribution
to the bulk viscosity arises, becasue r-modes induce a variation in
the pressure and density that drives the star matter away from the
$\beta$-equilibrium~\cite{Sawyer:1989dp}. Slow weak
interactions try to re-establish the equilibrium, i.e., to restore the equilibrium value of
electron and muon concentrations $X_{e}=n_{e}/n$ and $X_{\mu}=n_{\mu}/n$, via
charged-current processes, which contribute additively, $\zeta_{\rm s.m.}=\zeta_{{\rm
s.m.},e}+\zeta_{{\rm s.m.},\mu}$. Here we consider the neutron star at least several seconds after its formation, when the the neutrino chemical potential $\mu_\nu $ has already vanished and, therefore, there is no soft-mode contribution to the bulk viscosity from neutrinos.

Following Refs.~\cite{Haensel:2000vz,Haensel:2001mw} we define
the bulk  viscosity averaged over the perturbation period as
\begin{eqnarray}
\zeta_{{\rm s.m.},l}=n\frac{ \langle P(n+\delta n(t),X_l+\delta
X_l(t))\,\delta \dot{n}(t)\,\rangle_{\mathcal{P}} } {\langle
\big(\delta \dot{n}(t)\big)^2\rangle_{\mathcal{P}}}\,.
\label{zeta-diss}
\end{eqnarray}
Here  $P$ is the pressure, $\langle \dots \rangle_{\mathcal{P}}$
stands for the averaging over the period of a perturbation, and
$l=e,\mu$, as above.

The chemical relaxation rate takes a time of the order or longer
than $1/\om \sim 10^{-4}$--$10^{-3}$\,s,  whereas the thermal
relaxation time is much shorter. Therefore, the neutron star can
be considered in the thermal quasi-equilibrium, but not in the
chemical one. In the state out of chemical equilibrium we have to
take into account dependence of the reaction rates of MU, MMU and
DU processes on the difference of chemical potentials, $\delta
\mu_l=\mu_n-\mu_p-\mu_l$, where $\mu_n$ and $\mu_p$ are chemical
potentials of the nucleons. The bremsstrahlung reactions obviously
do not contribute to $\zeta_{\rm s.m.}$, becasue they do not change
the number of electrons and muons and cannot restore the chemical
equilibrium. According to Refs.~\cite{Haensel:1992zz,Haensel:2000vz} the dependence of the
reaction rate on the nucleon density and $\delta \mu_l$ can be
written in a separable form
\begin{align}\label{Im}
\Gamma^{\rm (r)}_l(T,n,\delta \mu_l) &=\Gamma^{\rm (r)}_{0,l}(T,n)
I^{\rm (r)}_2 (\delta\mu_l/T)\,, \nonumber\\ I^{\rm (r)}_m (\xi)
&=\intop_0^\infty\rmd x x^m J^{\rm (r)}(x-\xi)\,.
\end{align}
Here $J^{\rm (r)}$ represents the phase space of the reaction with $m^{\rm (r)}$ incoming and outgoing fermions excluding neutrinos involved in the process
\begin{eqnarray}
J^{(\rm r)}(\xi)=
\intop_{-\infty}^{+\infty}\left[\prod_{i=1}^{m^{\rm (r)}}
\frac{\rmd x_i}{e^{x_i}+1}\right] \delta\Big(\sum_{i=1}^{m^{\rm
(r)}} x_i-\xi\Big)\,.
\label{Jn}
\end{eqnarray}
Analytical expression for this integral can be found in
Ref.~\cite{GP-FL}, part 1, Appendix A.  The DU processes involve
$m^{\rm (r)} =3$ fermions, whereas the MU processes involve $m^{\rm
(r)}=5$ fermions. Then we have for example
\begin{align}
I_1^{\rm (DU)}(0)=\frac{17 \pi^4}{240}\,,\quad
I_1^{\rm (MU)}(0)=\frac{367 \pi^6}{6048} \,.
\label{Ir}
\end{align}

\begin{figure}
\centerline{
\parbox{6cm}{\includegraphics[width=6cm]{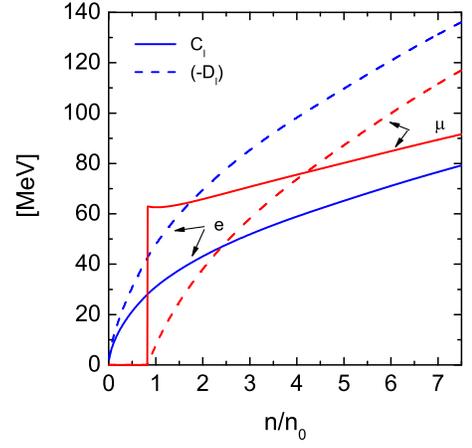}}
} \caption{\label{fig:CD} (Color online) Coefficients $C_l$ and $(-D_l)$ from
Eq.~(\ref{CD}) for electrons and muons as functions of nucleon
density for the HDD EoS.}
\end{figure}

If the rate $\Gamma^{\rm (r)}(T,n,\delta \mu_l)$ corresponds to a
reaction, in which neutrons are converted to protons, the rate for
the inverse reaction (protons to neutrons) is given by
$\Gamma^{\rm (r)}(T,n,-\delta \mu_l)$. Then the equation
determining the evolution of the slow parameter $X_l$ can be
written in the linear approximation as
\begin{align}\label{Xdot}
&n\,\delta\dot{X}_l(t) = \sum_{{\rm r}}\Big[ \Gamma^{\rm
(r)}_l(T,n+\delta n(t),\delta \mu_l(t)) \\ &- \Gamma^{\rm
(r)}_l(T,n+\delta n(t),-\delta \mu_l(t))\Big] \approx - \sum_{{\rm
r}} \lambda^{\rm (r)}_l(T,n)\,\delta\mu_l(t)\,, \nonumber
\end{align}
where
\begin{align}
\lambda^{\rm (r)}_l(T,n)&\approx  -2\frac{\partial}{\partial
\delta\mu} \Gamma^{\rm (r)}_l(T,n,0)
=-\frac{4}{T}\Gamma_{0,l}^{\rm (r)}(T,n)\, I_1^{\rm (r)}(0)\,.
\label{lam}
\end{align}
The difference of chemical potentials is  expressed through the
parameters characterizing the state of the system as
\begin{eqnarray}
\delta\mu_l(t)=\frac{\partial \delta\mu_l}{\partial n}\delta n(t)+
\frac{\partial \delta\mu_l}{\partial X_l} \delta X_l\,,
\label{dmu}
\end{eqnarray}
where all partial derivatives are taken at equilibrium. We neglect
mixing among different lepton species, see
Ref.~\cite{Haensel:2000vz} for complete consideration. From
Eqs.~(\ref{Xdot}) and  (\ref{dmu}) we find
\begin{align}
&\delta X_l(t)=-\sum_{{\rm r}}\frac{\lambda^{\rm (r)}_l(T,n)}{n^2}\,C_l\intop^t\delta n(t')
e^{(t'-t)/\tau_{X,l}}\rmd t'  \,,
\label{Xsol}\\
&\frac{1}{\tau_{X,l}}=\sum_{{\rm r}}\frac{\lambda^{\rm (r)}_l (T,n)}{n\,X_l} D_l\,,\,\,
C_l = n \frac{\partial \delta\mu_l}{\partial n}\,, \,\,
D_l = X_l \frac{\partial \delta\mu_l}{\partial X_l}\,.
\label{CtauX}
\end{align}
Here $\tau_{X,l}$ is the relaxation time of the parameter $X_l$. Partial derivatives in Eq.~(\ref{CtauX}) can be expressed through the nuclear symmetry energy, $E_{\rm sym}$, as~\cite{Haensel:2000vz}
\begin{align}
C_l &=  4(1-2\,X_p)\, n \,\frac{\partial E_{\rm sym}}{\partial n}
- \frac{p_{\rmF,l}^2}{3\mu_l}\,, \nonumber\\ D_l &= -8 X_l\,E_{\rm
sym}(n) - {p_{\rmF,l}^2}/{3\mu_l}\,. \label{CD}
\end{align}
We note that $D_l < 0$ and therefore $\tau_{X,l}$ is always
positive, becasue $\lambda^{\rm (r)}_l <0$.  Coefficients $C_l$  and
$D_l$ are shown in Fig.~\ref{fig:CD}.

Knowing the evolution of the  parameter $X_l$ given by
Eq.~(\ref{Xsol}) we can calculate the soft-mode bulk viscosity.
Expanding pressure in Eq.~(\ref{zeta-diss}) up to linear order in
perturbations we obtain
\begin{align}
\zeta_{{\rm s.m.}, l} \approx n\frac{\partial P}{\partial
n}\frac{\langle\delta n(t) \delta
\dot{n}(t)\rangle_{\mathcal{P}}}{\langle \big(\delta
\dot{n}(t)\big)^2\rangle_{\mathcal{P}}} +n\frac{\partial
P}{\partial X_l}\frac{\langle\delta X_l(t) \delta
\dot{n}(t)\rangle_{\mathcal{P}}}{\langle \big(\delta
\dot{n}(t)\big)^2\rangle_{\mathcal{P}}}\,.
\end{align}
For a harmonic perturbation with the frequency $\om$, that we
exploit in this work, the first term vanishes and the second
one yields
\begin{align}
\zeta_{{\rm s.m.},l} &\approx -\frac{\partial P}{\partial X_l}\frac{\rmd X_l}{\rmd n}
\frac{n\tau_{X,l}}{1+\om^2\,\tau_{X,l}^2} \big[1-g(\om\tau_{X,l})\big],
\label{zeta-diss-1}\\
g(x)&=\frac{x\,(1-e^{-2\pi/x})}{\pi(1+x^2)},
\label{gfun}
\end{align}
where we used the relation $\frac{\partial \delta\mu_l}{\partial
n}/\big[\frac{\partial \delta \mu_l}{\partial X_l}\big]=-\frac{\rmd X_l}{\rmd n}$, which follows from Eq.~(\ref{dmu}) that in the chemical equilibrium. Equation~(\ref{zeta-diss-1})
(without the $g$ term in squared brackets) is the result of Ref.~\cite{Lindblom:2001hd}. In the limit $\tau_{X,l}\om\gg 1$ we have $g(x\gg 1)=O(1/x^2)$ and recover the result of
Ref.~\cite{Haensel:2000vz},
\begin{align}
&\zeta_{{\rm s.m.},l}\approx \overline{\zeta}_{{\rm s.m.},l}\equiv \sum_{\rm r} \zeta_{{\rm s.m.},l}^{\rm
(r)} \,,\label{zeta-inf}\\
& \zeta_{{\rm s.m.},l}^{\rm (r)} =  -\frac{\lambda^{\rm (r)}_l (T,n)}{\om^2}\,C^2_l\,.
\nonumber
\end{align}
In this limit the viscosity becomes a superposition of partial
viscosities determined by various processes.  The limit,
$\tau_{X,l}\om\gg 1$, was usually exploited in previous works.
Below we compare the results obtained using the simplified
expression with those following from the general expression
(\ref{zeta-diss-1}).

With the help of the relation $\frac{\partial P}{\partial X_l}=-n
C_l$, Eq.~(\ref{zeta-diss-1}) is rewritten as
\begin{align}
&\zeta_{{\rm s.m.},l} \approx -\frac{C_l^2}{D_l}\frac{n_l\,\tau_{X,l}}{1+\om^2\,\tau_{X,l}^2}
\big[1-g(\om\tau_{X,l})\big]
\label{zeta-diss-2}\\
&\quad \approx 2.56\cdot 10^{34} \Big[{\rm\frac{g}{cm\cdot s} }\Big]
\frac{C_l^2\big[1-g(\om\tau_{X,l})\big]}{100~{\rm MeV}\,|D_l|}
\frac{n_l}{n_0} \frac{\tau_{X,l}/{\rm s}}{1+\om^2\,\tau_{X,l}^2}\,.
\nonumber
\end{align}

Reactions bringing the matter toward the chemical equilibrium,
which we now consider, are DU, MU, and MMU (neutron and proton
branches), so ${\rm r=DU}$, MU(n,p), and MMU(n,p). Using the
results of Refs.~\cite{Haensel:2000vz,Haensel:2001mw} we can write
\begin{align}
\Gamma_{0,l}^{\rm (DU)}(T,n) &=
\frac{G_{\rm w}^2}{4\,\pi^5}(1+3\,g_A^2)\, m_n^*\, m_p^*\, \mu_{l}\, T^5 \Theta_{npl}\,,
\label{G0DU}\\
\Gamma_{0,l}^{\rm (MU)n}(T,n) &=
\frac{G_{\rm w}^2 p_{\rmF,l}}{\pi^9 p_{\rmF,e}}\,f^4_{\pi NN}g_A^2\, m_n^{*3}\, m_p^*\, p_{{\rm
F},p}\, T^7,
\label{G0MUn}\\
\Gamma_{0,l}^{\rm (MU)p}(T,n) &= \Gamma_{0,l}^{\rm (MU)n}(T,n)
\nonumber\\
&\times \frac{m_p^{*2}}{m_n^{*2}} \frac{(3p_{{\rm F},p}+p_{{\rm
 F},l}-p_{{\rm F},n})^2}{8p_{{\rm F},p}p_{{\rm F},l}}\Theta_{pl}\,.
\label{G0MUp}
\end{align}
Here $G_{\rm w}=G_{\rm F}\cos\theta_{\rm C}$ with $G_{\rm F}\simeq
1.436 \cdot 10^{-49}$ erg$\cdot$cm$^3$,  being the Fermi weak
coupling constant, and $\theta_{\rm C}$ stands for  the Cabbibo
angle, $\sin\theta_{\rm C}=0.225$. The quantity $g_A=1.26$ is the
axial-vector coupling constant, $\mu_\mu=\mu_e\simeq p_{\rmF,e}$.
The factor $\Theta_{npl}$ in Eq. (\ref{G0DU}) is equal to $1$, if
$p_{\rmF,n}< p_{\rmF,p} + p_{\rmF,l}$, and is $0$ otherwise. It
allows the DU reaction only when the proton concentration is
higher than a critical one. The critical density, $n_c^{({\rm
DU}), l}$, for the DU process with participation of the lepton
$l=e\,,\,\mu^-$ for the HDD EoS is $n_c^{{\rm (DU)}, e}=5.05\,n_0$
and  $n_c^{{\rm (DU)}, \mu}=6.07\,n_0$; cf. Fig.~\ref{fig:RM-HDD}.
A similar factor $\Theta_{pl}$ in Eq.~(\ref{G0MUp}) allows the
proton branch of the MU process only for densities when $p_{{\rm
F},n}<3p_{{\rm F},p}+p_{{\rm F},l}$. The corresponding critical
densities are $n_c^{{\rm (MU)p}, e}=0.22\,n_0$ and $n_c^{{\rm
(MU)p}, \mu}=0.73\,n_0$, see Fig.~\ref{fig:RM-HDD}.

\begin{figure*}[ht]
\parbox{14cm}{\includegraphics[width=14cm]{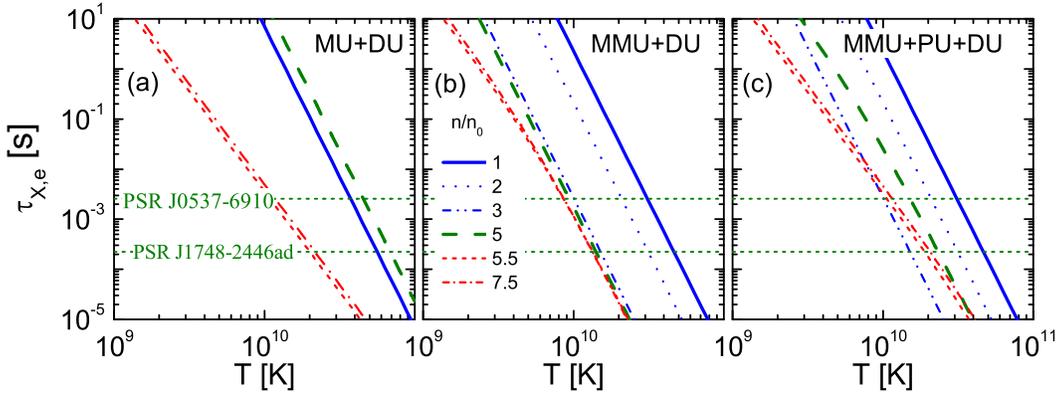}}
\caption{(Color online) The relaxation time $\tau_{X,e}$ of the electron
concentration as a function of temperature for various nucleon
densities (shown by line labels). In panel (a) only the MU and
DU (above $n_c^{{\rm (DU)},e}$) processes are taken into account.
In panel (b) only the MMU and DU
(above $n_c^{{\rm (DU)},e}$) processes are included; see
Eq.~(\ref{Gam0-MMU}). The pion gap in the MMU enhancement factor
(\ref{Ffun}) is taken along curve~1a+1b in Fig.~\ref{omegatil}. In panel (c) the PU process
(above $n_{c}^{\pi}$) is additionally incorporated following curve~3 in Fig.~\ref{omegatil}. The pion gap in the MMU enhancement factor is taken along curve~1a+2. The horizontal dashed lines show the inverse angular velocities of the fastest old-binary pulsar PSR J1748-2446ad ($\nu=716$~Hz) and the young one PSR J0537-6910 ($\nu=62$~Hz). 
} \label{fig:tx}
\end{figure*}

As discussed in Sec.~\ref{sec:NN-int}, the in-medium
nucleon-nucleoon interaction  may significantly differ from a
naive FOPE exchange. Within the MOPE model the MU partial
viscosity should be replaced with the MMU one. The ratio of the
rates of the MU and MMU processes is determined by the $F_{\rm
MMU}$ factor given by Eq.~(\ref{Ffun}):
\begin{align}
\Gamma_{0,l}^{\rm (MMU)i}(T,n) =\Gamma_{0,l}^{\rm (MU)i} \, F_{\rm MMU}(n)\,,\,\,\,\, i=n,p\,.
\label{Gam0-MMU}
\end{align}
Note that, as we have mentioned, we use for simplicity the vacuum
weak coupling vertices, because for $q_0 \gsim q$ in neutrino
vertices the $NN$ correlation corrections are rather suppressed,
(see Ref.~\cite{Migdal:1990vm,Voskresensky:2001fd}), but we correct the
strong interaction vertices. Therefore, the DU processes we consider
without in-medium modifications.

In sufficiently dense matter there might appear pion condensates.
We focus on effect of the $\pi^-$ condensation. With the
parametrization of $\widetilde{\omega}$ used above it appears for
densities $n>n_{c}^\pi=3n_0$. The rate of the PU processes, $n
+\pi^{-}_{\rm cond}\to n+l+\bar{\nu}_l$, is determined by
\begin{align}
\Gamma_{0,l}^{\rm (PU)}=\frac{G^2}{4\pi^5}\,(1+3\,g_A^2)\,f_{\pi
NN}^2\,m_p^*\,m_n^* \Gamma^2\,k_0\,|a_\pi|^2\,T^5\,,
\label{Gam-PU}
\end{align}
We use that the  momentum of the inhomogeneous pion
condensate is $k_0\simeq p_{\rmF,n}$ and  estimate the condensate
amplitude as $|a_\pi|^2\simeq |\widetilde{\om}^2|/m_{\pi}^2$, with
$\widetilde{\omega}$ given by curve~3 in Fig.~\ref{omegatil}.

The relaxation time $\tau_{X,l}$ entering Eq.~(\ref{zeta-diss-2})
is expressed through the partial relaxation times as
\begin{align}
&\frac{1}{\tau_{X,l}} = \sum_{\rm r}\frac{1}{\tau^{\rm
(r)}_{X,l}}\,, \label{tautot}\\ &\tau^{\rm (DU)}_{X,l} =
30.7\,{\rm s}\, \frac{m_N^2}{m_n^{*}m_p^*}
\Big(\frac{n_l^3}{n_e\,n_0^2}\Big)^{\frac13} T_9^{-4}
\frac{100~{\rm MeV}}{|D_l|}\,\Theta_{npl}^{-1} \,, \label{tauDU}\\
&\tau^{\rm (MU)n}_{X,l} = 1.84\cdot 10^7\,{\rm s}\,
\frac{m_N^4}{m_n^{*\,3}m_p^*} \Big(\frac{n_l^2 n_e}{n_p
n_0^2}\Big)^{\frac13} T_9^{-6} \frac{100~{\rm MeV}}{|D_l|}\,.
\label{tauMU}
\end{align}
Similarly, the relaxation time for the MMU reactions is
\begin{align}
\tau^{\rm (MMU)n}_{X,l}=\tau^{\rm (MU)n}_{X,l}/F_{\rm MMU}(n)\,,
\label{tauMMU}
\end{align}
and for the PU processes on the $\pi^-$ condensate we have
\begin{eqnarray}
\label{tauPU} \tau^{\rm (PU)}_{X,l} \simeq 6.76
\frac{m_\pi^2}{|a_\pi|^2}  \frac{\Gamma^2(n_0)}{\Gamma^2(n)}
\Big(\frac{n_e}{n_n}\Big)^{\frac13}\, \tau^{\rm
(DU)}_{X,l}\theta^{-1}(n-n_c^\pi) \,.
\end{eqnarray}
Obviously, in this expression the $\Theta_{npl}$ function in
$\tau^{\rm (DU)}_{X,l}$ should be dropped. The electron
relaxation time $\tau_{X,e}$, entering the denominator of
Eq.~(\ref{zeta-diss-1})  and  the $g$ function, is shown in
Fig.~\ref{fig:tx} as a function of the temperature for various
values of nucleon densities. The relaxation time determined by the
MU  reaction, Fig.~\ref{fig:tx}(a) shows very weak dependence on
the density for  $n<n_c^{{\rm (DU),}e}$ and drops substantially
once the DU reactions kick in. With  account for in-medium
modifications of the $NN$ interaction and the pion
condensation the lepton relaxation time gradually decreases with
the density increase, as seen from panels (b) and (c) in
Fig.~\ref{fig:tx}.  In panel (b), in the absence of the PU
process, it smoothly approaches the DU relaxation time at
$n_c^{{\rm (DU),}e}$. The main term here for $n<n_c^{{\rm
(DU),}e}$ is determined by the MMU reactions. In panel (c), in the
MMU+PU+DU case, situation is more complicated because in this case
for $n>3 n_0$ the PU process becomes efficient, whereas the MMU
rate (calculated with $\widetilde{\omega}$ given by curves
1a+2 in Fig.~\ref{omegatil}) is less than that in panel (b)
(calculated with $\widetilde{\omega}$ given by curves 1a+1b). The
dependence on the nucleon density and temperature of the muon
relaxation time is similar to that for the electron one. Thus, in
all cases for $\omega \sim 10^{3}$--$10^4$ Hz of our interest and
for $T<10^{10}$~K one can exploit the limit $\tau_{X,l}\omega \gg
1$ and use Eq.~(\ref{zeta-inf}).

The main contribution to the bulk viscosity in the $npe\mu$ matter
comes  from the DU neutrino processes, provided $n>n_c^{{\rm
(DU)},l}$. In the limit $\tau_{X,l}\om\gg 1$ it reads, cf.
Refs.~\cite{Haensel:1992zz,Haensel:2000vz},
\begin{align}
\zeta^{\rm (DU)}_{{\rm s.m.},l}
&= \frac{17 \pi^4\,C_l^2}{60\,T\om^2}\,\Gamma_{0,l}^{\rm (DU)}(T,n)
\simeq 8.35 \cdot 10^{24} \Big[{\rm\frac{g}{cm\cdot s} }\Big]
\nonumber\\
&\times \frac{m_n^*m_p^*}{m_N^2} \Big(\frac{n_e}{n_0}\Big)^{\frac13} \frac{T_9^4}{\om_4^2}
\Big(\frac{C_l}{100\,{\rm MeV}}\Big)^2 \Theta_{npl}\,,
\label{zetaDU}
\end{align}
where  $\om_4=\om/{\rm 10^4 s}$.

In absence of the DU processes (i.e. for $n<n_c^{{\rm (DU)},e}$)
the main contribution to the bulk viscosity  of non-superfluid
matter comes from the MU (or MMU) reactions. For the $n+n\to
n+n+l+\bar{\nu}$ process, employing FOPE matrix element in the
limit $\tau_{X,l}\om\gg 1$ one gets~\cite{Haensel:2001mw}
\begin{align}
\zeta^{\rm (MU)n}_{{\rm s.m.},l}
&= \frac{367 \pi ^6\,C_l^2}{1512\,T\,\om^2} \Gamma_0^{\rm (MU)n}(T,n)
\simeq 1.39\cdot 10^{19}\Big[{\rm\frac{g}{cm\cdot s} }\Big]
\nonumber\\
&\times \frac{m_n^{*\,3}m_p^*}{m_N^4} \Big( \frac{n_l\,n_p}{n_e\,n_0}\Big)^{\frac13} \Big(\frac{C_l}{100\,{\rm MeV}}\Big)^2 \frac{T_9^6}{\om_4^2}\,
\label{zetaMU}
\end{align}
for the neutron branch and
\begin{eqnarray}
\zeta^{\rm (MU)p}_{{\rm s.m.},l} =\zeta^{\rm (MU)n}_{{\rm s.m.},l}
\Big(\frac{m_p^*}{m_n^*}\Big)^2
\frac{(3p_{{\rm F},p}+p_{{\rm F},l}-p_{{\rm F},n})^2}{8p_{{\rm F},p}p_{{\rm F},l}}\Theta_{pl}
\end{eqnarray}
for the proton branch.  Thus, typically $\zeta^{\rm (MU)}/\zeta^{\rm (DU)}\sim 10^{-6}$  in the region, where DU processes are allowed. The bulk viscosity owing to the MMU processes is given by
\begin{eqnarray}
\zeta^{\rm (MMU)i}_{{\rm s.m.},l} \simeq \zeta^{\rm (MU)i}_{{\rm
s.m. },l}\, F_{\rm MMU}(n)\,,\quad i=n,p\,. \label{zetaMMU}
\end{eqnarray}
The contribution of the PU processes to the viscosity for $n>n_{c}^\pi$ is
\begin{eqnarray}
\label{zetaPU} \zeta^{\rm (PU)}_{{\rm s.m.},l} \simeq 0.15
\frac{|a_\pi|^2}{m_\pi^2}\,
\frac{\Gamma^2(n)}{\Gamma^2(n_0)}\Big(\frac{n_n}{n_e}\Big)^{\frac13}\,
\zeta^{\rm (DU)}_{{\rm s.m.},l} \,,
\end{eqnarray}
where we have used that $I_m^{\rm (PU)}=I_m^{\rm (DU)}$. With the help of Eq.~(\ref{pi-cond}) the condensate amplitude can be expressed through the pion gap taken along curve 3 in Fig.~\ref{omegatil}. Similar rates are expected for the kaon condensate Urca (KU) processes, for $n>n_{c}^{K}$, where $n_{c}^{K}$ is estimated as $\sim 3-5 ~n_0$; see Ref.~\cite{KV2003}.

\begin{figure}
\parbox{6cm}{\includegraphics[width=6cm]{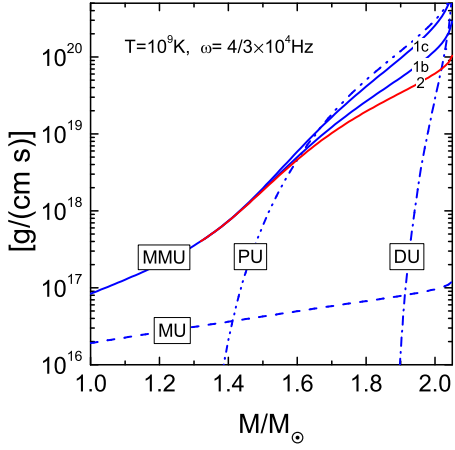}}
\caption{(Color online) Partial contributions to the profile averaged bulk
viscosity $\langle \sum_l\zeta^{\rm (r)}_{\rm
s.m.,l}(1+0.86r^2/R^2)\rangle_8$ from DU, MU, MMU, and PU
processes as functions of the neutron star mass. Three MMU lines
(labeled 1b,1c and 2) correspond to different choices of the
density dependence of the pion gap for $n>3n_0$, as shown in
Fig.~\ref{omegatil}; the PU process is calculated using the curve 3 in
Fig.~\ref{omegatil}, $T_9 =1$, $\om =\frac{4}{3}\cdot 10^4$ Hz.
\label{fig:zeta} }
\end{figure}

\begin{figure*}
\parbox{14cm}{\includegraphics[width=14cm]{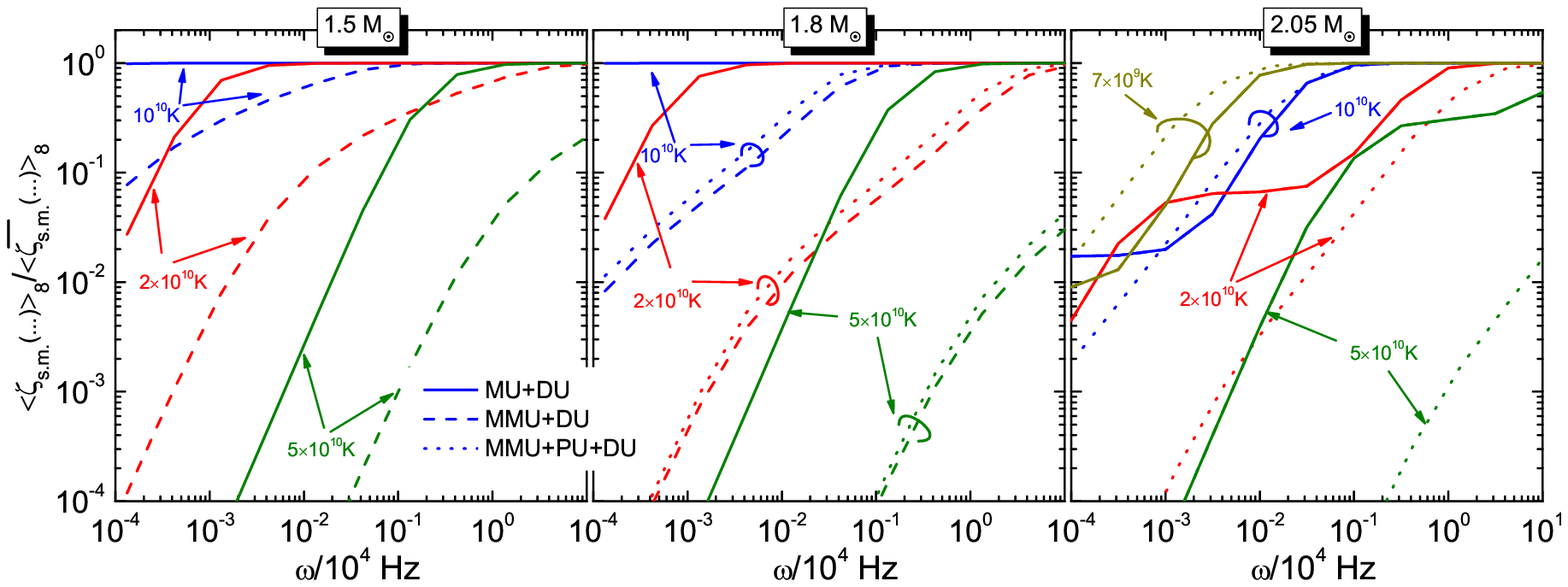}}
\caption{(Color online) Ratio of the total profile averaged bulk viscosity,
 $\langle \zeta_{\rm s.m.}(1+0.86 (r/R)^2)\rangle_8$, with the lepton contributions given by
Eq.~(\ref{zeta-diss-2}), to the same value calculated according to
Eq.~(\ref{zeta-inf}), i.e.  in the limit $\om\tau_{X,l}\gg 1$, for
the three values of neutron star masses. Notations of processes
are the same, as in Fig.~\ref{fig:tx}.} \label{fig:zeta-om-rat}
\end{figure*}

Partial contributions to the profile averaged bulk viscosity
$\langle\zeta^{\rm (r)}_{\rm s.m.,l} \rangle$  are presented in
Fig.~\ref{fig:zeta} for  DU [Eq.~(\ref{zetaDU})], MU
[Eq.~(\ref{zetaMU})], MMU [Eq.~(\ref{zetaMMU})], and PU,
[Eq.~(\ref{zetaPU})] processes as functions of the neutron star
mass for $T_9 =1$. We see that for $M_{\odot}\lsim
M<1.6M_{\odot}$ the main contribution to $\langle \zeta_{{\rm
s.m.},l}\rangle$ comes from the MMU processes. Three MMU lines
(labeled 1b, 1c, and 2) correspond to different choices of the
density dependence of the pion gap for $n>3n_0$ as shown in
Fig.~\ref{omegatil}. For $M>1.6 M_{\odot}$ the main contribution
to $\langle \zeta_{{\rm s.m.},l}\rangle$ comes from the PU
process, $\langle\zeta^{\rm (DU)}_{\rm s.m.,l}\rangle <
\langle\zeta^{\rm (PU)}_{\rm s.m.,l}\rangle$ even for the heaviest
neutron star. For $T_9=1$ the $\om$- dependence of $\langle
\zeta_{{\rm s.m.},l}\rangle$ is unimportant in the interval
$10^3\lsim \om <10^4$ Hz of our interest. Calculations presented
in Fig.~\ref{fig:zeta} are performed for $\omega_4 =4/3$ (recall
$\om=4\Omega/3$). Each partial contribution to the bulk viscosity
depends on the frequency as $\om^{-2}$. With an increase of the
temperature the $\langle\zeta^{\rm (MU/MMU)}_{\rm s.m.,l} \rangle$
increases $\propto T^6$, whereas $\langle\zeta^{\rm (PU/DU)}_{\rm
s.m.,l} \rangle$ increases $\propto T^4$.

The ratio of the total bulk viscosity averaged  over the star density profile,
$\langle \zeta_{\rm s.m.}(1+0.86 (r/R)^2)\rangle_8$, with the lepton contributions given by
Eq.~(\ref{zeta-diss-2}) to the same average of the bulk viscosity calculated according to
Eq.~(\ref{zeta-inf}) in the limit $\om\tau_{X,l}\gg 1$ is demonstrated in
Fig.~\ref{fig:zeta-om-rat}. It begins to deviate from unity  with
decrease of the frequency and/or with increase of the temperature.
The deviation from unity is more pronounced for the heavier
neutron stars.  Most of the observed rapidly rotating young pulsars
have rotation frequencies $\Omega\sim 10^2$ Hz. At such a
frequency ($\omega =\frac{4}{3}\Omega$) for the heaviest neutron
star the deviation of the ratio from unity starts for $T_9\geq 7$.
Thus, in this case the bulk viscosity should be calculated
following Eq.~(\ref{zeta-diss-2}) rather than using
Eq.~(\ref{zeta-inf}).

\subsection{Radiative contribution}

The soft-mode contribution to the bulk viscosity is determined by
dissipation of internal energy of an elementary volume, as a
result of work done against an external pressure. The latter work
is spent when the variation of the pressure in the volume is
lagged behind the variation of the density. This occurs when the
pressure depends on  parameters with  long relaxation times. In
our case these are the lepton concentrations, which are balanced
by slowly processing  weak interactions.

Another source of the energy dissipation induced by a density
perturbation  is an increase of the direct neutrino emission with
respect to  emission without the perturbation. The corresponding
contribution to the bulk viscosity, called a radiative viscosity,
 was considered in Ref.~\cite{Sa'd:2009vx} for the first time,
\begin{eqnarray}
\zeta_{\rm rad}=-n^2\frac{\langle \epsilon_{\nu}(n+\delta
n(t),\delta\mu(t))-\epsilon_{\nu}(n,0) \rangle_{\mathcal{P}}
}{\langle \big(\delta \dot{n}(t)\big)^2\rangle_{\mathcal{P}}}\,,
\label{zeta-rad}
\end{eqnarray}
where $\epsilon_{\nu}$ is the total neutrino emissivity, which is
a sum of contributions of the DU, MU/MMU and PU reactions,
$\epsilon_{\nu}=\sum_{{\rm r},l} \epsilon^{\rm (r)}_l$\,. Note
that the radiative viscosity contributes only in the regions
transparent for neutrinos, i.e., for $r>r_{\rm opac}$.

For the system out of chemical equilibrium the neutrino emissivity can be written as
\begin{align}
\epsilon_l^{\rm (r)}(n,\delta\mu_l) &=\Gamma^{\rm
(r)}_{0,l}(T,n)\, T\, \big[ I^{\rm (r)}_3(\delta\mu_l/T)+I^{\rm
(r)}_3(-\delta\mu_l/T)\big]\,.
\end{align}
The integral $I^{\rm (r)}_3$ is determined in Eq. (\ref{Im}).

Expanding Eq.~(\ref{zeta-rad}) up to second order in perturbations
we obtain $\zeta_{\rm rad}=\sum_{l,\rm r}\zeta_{{\rm rad},l}^{\rm (r)}$, where
\begin{widetext}
\begin{align}
\zeta_{{\rm rad},l}^{\rm (r)}&\approx
\Big\{\frac{n^2}{2} \frac{\partial^2 \epsilon_l^{\rm
(r)}}{\partial n^2} +\frac{\partial^2 \epsilon_l^{\rm
(r)}}{\partial \delta\mu_l^2}\, \frac{C_l^2}{2} \Big\}
\frac{\langle(\delta n(t))^2\rangle_{\mathcal{P}}}{\langle
\big(\delta \dot{n}(t)\big)^2\rangle_{\mathcal{P}}}
+\frac{\partial^2 \epsilon_l^{\rm (r)}}{\partial \delta\mu_l^2}
\frac{n}{X_l}\, C_l D_l  
\frac{\langle\delta n(t) \delta X_l(t)\rangle_{\mathcal{P}}}{\langle \big(\delta \dot{n}(t)\big)^2\rangle_{\mathcal{P}}}
+\frac{\partial^2 \epsilon_l^{\rm (r)}}{\partial \delta\mu_l^2}\,
\frac{n^2}{2X_l}\,D_l^2
\frac{\langle(\delta X_l(t))^2\rangle_{\mathcal{P}}}{\langle
\big(\delta \dot{n}(t)\big)^2\rangle_{\mathcal{P}}} \,.
\label{zeta-rad-exp}
\end{align}
\end{widetext}
Using Eq.~(\ref{Xsol}) for calculation of the averages over the perturbation period,  Eqs.~(\ref{CtauX}) and (\ref{zeta-inf}), and that ${ \partial^2 \epsilon^{\rm (r)}_l }/{\partial
(\delta\mu_l)^2}=-3\,\lambda^{\rm (r)}_l$, we obtain from Eq.~(\ref{zeta-rad-exp})
\begin{align}
\zeta^{\rm (r)}_{{\rm rad},l} &= \frac32 \zeta^{\rm (r)}_{{\rm
s.m.},l} \Big[1 -\frac{\pi (1+x^2) g^2(x)}{1-g(x)} \Big]
+\frac{n^2}{2\om^2} \frac{\partial^2 \epsilon^{\rm
(r)}_l}{\partial n^2}\,,
\label{zeta-rad-2}
\end{align}
where $x=\om\tau_X$, and $g(x)$ is defined in Eq.~(\ref{gfun}).
Taken in the limit $x\gg 1$, the first term reproduces the result
of Ref.~\cite{Sa'd:2009vx}. The second term and  the square
bracket  factor yielding the $x$ dependence of the first term are new,
derived in this work. The second term can contribute greatly,
only provided that the neutrino emissivity strongly depends on the
nucleon density.

Guided by Eq.~(\ref{zeta-rad-2}) we write
\begin{align}
\zeta^{\rm (r)}_{{\rm rad},l}=\mathcal{R}^{\rm (r)}\zeta^{\rm (r)}_{{\rm
s.m.},l}\,.
\label{R-def}
\end{align}
For r$=$MU, DU and PU processes the density dependence of the
neutrino emissivity is weak, except a narrow vicinity (of the
width $\sim \delta n$ in the density) of the thresholds for DU and
PU processes. In the limit $\om\tau_X\gg 1$, dropping the square
bracket factor and neglecting a small  second term  in
Eq.~(\ref{zeta-rad-2}), one gets
\begin{align}
\mathcal{R}^{\rm (r)}\approx \frac32\,,\quad {\rm r}=\mbox{MU, DU,
PU}\,.
\end{align}

\begin{figure}
\includegraphics[width=6cm]{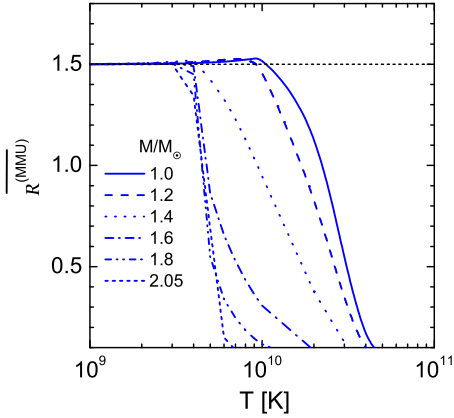}
\caption{(Color online) Factor $\overline{\mathcal{R}^{\rm (MMU)}}$, the ratio of
the averaged radiative bulk viscosity of the MMU processes given
by Eq.~(\ref{zeta-rad-2}) to the averaged soft-mode viscosity term
(\ref{zetaMMU}), both calculated in the limit $\omega\tau_{X,l}\gg
1$, as a function of temperature for various neutron star masses.
} \label{fig:Rmmu}
\end{figure}

\begin{figure*}
\parbox{14cm}{\includegraphics[width=14cm]{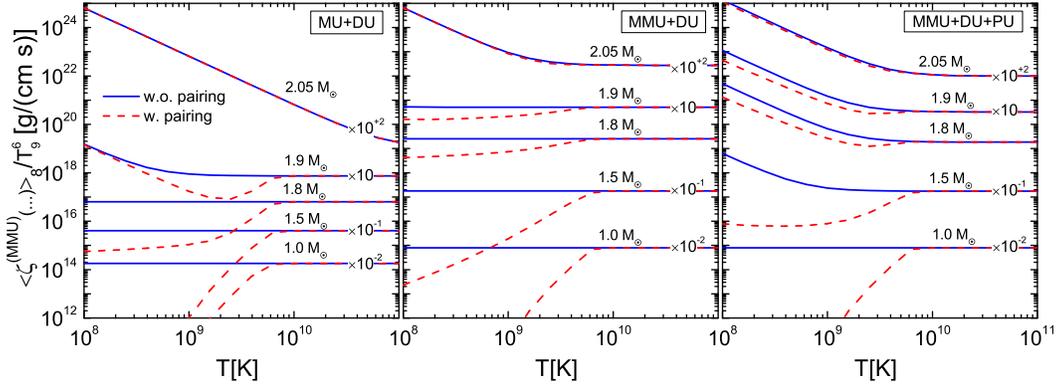}}
\caption{(Color online) The profile averaged total bulk viscosity, $\langle
\sum_l\overline{\zeta}_{\rm s.m.,l}(1+0.86 (r/R)^2)\rangle_8$,
calculated with (dashed lines) and without (solid lines) nucleon
pairing effects for several neutron star masses. The curves for
stars with different masses are scaled by the factors indicated by
the curve labels. The pairing is taken into account according
Eq.~(\ref{zeta-neq-SC}). Notations of processes are the same, as
in Fig.~\ref{fig:tx}.
 }
\label{fig:zeta-SC-rat}
\end{figure*}

The strongest density dependence is inherent in the MMU processes, because of the scaling factor $F_{\rm MMU}(n)$; see Eq.~(\ref{Ffun}). Thus, the last term in Eq.~(\ref{zeta-rad-2}) must be taken into account:
\begin{align}
\frac{\partial^2 \epsilon_l^{\rm (MMU)}}{\partial n^2} &=
\frac{ \partial^2 \Gamma^{\rm (MMU)}_{0,l} }
{ \partial n^2}\, 2T\, I_3^{\rm (MU)}(0)
\\
&=-\lambda^{\rm (MMU)}_l  \frac{I_3^{\rm (MU)}(0)}{I_1^{\rm (MU)}(0)}
\frac{T^2}{2\,} \frac{F''_{\rm MMU}(n) }{F_{\rm MMU}(n)}\,,
\nonumber
\end{align}
where $F''_{\rm MMU}(n)$ stands for the second derivative of
$F_{\rm MMU}$ with respect to the nucleon density and ${I_3^{\rm
(MU)}(0)}/{I_1^{\rm (MU)}(0)}={11513\, \pi ^2}/{7340}$. For
$\om\tau_X\gg 1$ we find
\begin{align}
&\mathcal{R}^{\rm (MMU)} =\frac32 + \frac{11513\pi^2}{29360}
\frac{T^2}{C_l^2}n^2\frac{F''_{\rm MMU}(n) }{F_{\rm MMU}(n)}
\label{R-MMU}\\
&\approx \frac32 + 2.87\times 10^{-6}\, T_9^2\, \Big(\frac{100~{\rm MeV}}{C_l}\Big)^2
n^2 \frac{F''_{\rm MMU}(n) }{F_{\rm MMU}(n)}\,.\nonumber
\end{align}
In the profile averaged viscosity this factor enters as averaged
over the star profile, cf. Eq.~(\ref{R-def}):
\begin{align}
\overline{\mathcal{R}^{\rm (MMU)}}&=\frac{\langle\sum_l\zeta^{\rm
(MMU)}_{{\rm rad},l}\, (1+0.86 r^2/R^2)\theta(r-r_{\rm
opac})\rangle_8}{
 \langle\sum_l\zeta^{\rm (MMU)}_{{\rm s.m.},l}\,(1+0.86 r^2/R^2)
 \rangle_8}.
 \label{R-MMU-aver}
\end{align}
Here the $\theta$-function takes into account that the radiative
viscosity  contributes only, if the medium is transparent for
neutrinos, i.e. only for regions with $r$ larger than the opacity
radius. The coefficient $\overline{\mathcal{R}^{\rm (MMU)}}$ is
shown in Fig.~\ref{fig:Rmmu}. We see that for light stars (with
$M\sim M_{\odot}$) the factor $\overline{\mathcal{R}^{\rm
(MMU)}}\simeq 3/2$ for $T_9<10$, and for heavy stars $\overline{\mathcal{R}^{\rm
(MMU)}}\simeq 3/2$ for $T_9<4$.
Thus, for temperatures $T_9<4$  for all masses the second term in
Eq.~(\ref{R-MMU}) proves to be small.  With increase of the
temperature the second term  in Eq.~(\ref{R-MMU}) starts to
contribute but this growth is cut by the $\theta$-function leading
to a strong suppression of the factor $\overline{\mathcal{R}^{\rm
(MMU)}}$. The values $\overline{\mathcal{R}^{\rm (PU)}}$ and
$\overline{\mathcal{R}^{\rm (DU)}}$ deviate only little from $3/2$
for temperatures $T_9<5$ and are cut at higher temperatures
similarly to the factor $\overline{\mathcal{R}^{\rm (MMU)}}$.

In contrast to the soft-mode contributions to the bulk viscosity,
which are linked with the reactions restoring the lepton
concentrations in the star, the radiative viscosity arises also
from processes not involving charged leptons, like neutrino
bremsstrahlung reactions $n+N\to n+ N+\nu+\bar{\nu}$, where
$N=n,p$. With account for the medium effects the neutrino
emissivity of these medium bremsstrahlung (MB) reactions is
\begin{align}
&\epsilon^{{\rm (MB)}n}_{\nu} \simeq \frac{41}{14175} \frac{G_{\rm
w}^2}{2\pi} g_A^2 m_n^{*4} f_{\pi NN}^4 p_{\rmF,n}\,T^8 F_{{\rm
MB}}(n)\,, \label{eps-BSn}\\ &\epsilon^{{\rm (MB)}p}_{\nu}
=\epsilon^{{\rm (MB)}n}_{\nu} \frac{m_p^{*2}}{m_n^{*2}}
\frac{p_{\rmF,p}} {p_{\rmF,n}}\,. \label{eps-BSp}
\end{align}
The factor $F_{{\rm MB}}(n)$  given by Eq.~(\ref{F-BS}) takes into
account the MOPE interaction.  The contribution to the radiative
viscosity from the (MB)n and (MB)p processes is given by
\begin{align}
&\zeta_{\rm rad}^{{\rm (MB)}}\approx 6.08\cdot10^{11}
\Big[\frac{\rm g}{\rm cm\cdot s}\Big]\,\frac{ T_9^8}{\om_4^2} \,
\\ &\quad\times n^2\frac{\rmd^2}{\rmd n^2}\left\{
\Big(\frac{m_N^{*}}{m_N}\Big)^4
\Big(\frac{n_p^{1/3}+n_n^{1/3}}{n_0^{1/3}}\Big) F_{{\rm MB}}(n) \right\}
\,.\nonumber
\end{align}
Estimations show that the averaged viscosity  $\langle\zeta_{\rm rad}^{{\rm (MB)}}\,(1+0.86 r^2/R^2)\rangle_8$ is systematically much smaller than the averaged soft-mode viscosity from the MMU reactions shown in Fig.~\ref{fig:zeta}.

\subsection{Pairing effects on the bulk viscosity}\label{bulkpairing}

In superfluid medium the neutrino production rates become
suppressed, roughly  by  exponential factors
$\xi_i=\exp(-\Delta_i/T)$ for $T\ll \Delta_i$ (in the absence of the
pairing $\xi_i$ should be put unity, $i=n,p$). Thus, we may
roughly estimate effect of the pairing on values of the DU, PU, and
MU/MMU and Bn/MBn, Bp/MBp partial contributions to the bulk
viscosity as
\begin{align}
\zeta_{{\rm s.m.},l}^{\rm (DU/PU)(s)}  &=\zeta_{{\rm s.m.},l}^{\rm
(DU/PU)}\min[\xi_n,\xi_p]\,, \nonumber\\ \zeta_{{\rm s.m.},l}^{\rm
(MU/MMU)i(s)}&=\zeta_{{\rm s.m.},l}^{\rm
(MU/MMU)i}\,\xi_p\,\xi_i\,. \label{zeta-neq-SC}
\end{align}
Suppression factors for  Bn/MBn, Bp/MBp quantities should be
similar to those for the MU/MMU reactions. As above, superscript
``(s)'' indicates quantities computed in the presence of pairing.
More generally,  the pairing effects are usually included in terms
of the so called $R$-factors~\cite{YLS99} taking into account the
Pauli blocking effects. The complete consideration should involve
the normal and anomalous Green's functions. It is not yet carried
out for the two-nucleon processes. Simplifying, we further use the
exponential suppression factors.

For the radiative contribution to the bulk viscosity we exploit a simple relation
\begin{align}
\zeta^{\rm (r,s)}_{{\rm rad},l}=\mathcal{R}^{\rm (r)}\zeta^{\rm
(r,s)}_{{\rm s.m.},l}\,, \label{R-def1}
\end{align}
which follows from Eq.~(\ref{R-def}) and (\ref{zeta-neq-SC}) if one
neglects density dependence of the gaps.

The effect of the proton superconductivity on the bulk viscosity
is illustrated  in Fig.~\ref{fig:zeta-SC-rat}. As we see, this
effect is more pronounced  for the light and middle-heavy neutron
stars, with $M<1.8\,M_\odot$. In the heavy neutron stars, for
$M>1.9\,M_\odot$, the viscosity is mainly determined by the DU
reactions, which occur in the central region of a star, where the
proton pairing is absent (cf. Fig.~\ref{fig:gap-profiles}).

In the presence of the pairing there appear extra efficient
processes of the one-nucleon origin, the so-called
pair-breaking-formation (PBF) processes, suggested in
Refs.~\cite{Flowers:1976ux,Voskresensky:1987hm,Senatorov:1987aa}. In the
case where nucleons are paired in the $^1{\rm S}_0$ state the
neutrino emissivity is as follows~\cite{KV08}
\begin{align}
\epsilon^{\rm(PBF)i}_{\nu}&=\frac{8 G_{\rm w}^2\,g_A^2}{35\pi^5}\,
\frac{p_{\rmF,i}^3}{m_i^*}\, \Delta_i^7 \,
I\Big(\frac{\Delta_i}{T}\Big)\,,i=n,p \,, \nonumber\\
I(z)&=\intop_1^\infty\frac{\rmd y }{\sqrt{y^2-1}}
\frac{y^5}{(1+e^{z\,y})^2}\,.
\end{align}
In the limit $T\ll \Delta_i$ the corresponding contribution to the radiative viscosity is given by
\begin{align}
\zeta_{\rm rad}^{\rm (PBF)i} &\approx 4.74\cdot
10^{18}\Big[\frac{\rm g}{\rm cm\cdot s}\Big]
\frac{T_9^{1/2}}{\om_4^2} \nonumber\\ &\times n^2\frac{\rmd}{\rmd
n^2} \Big\{
\frac{n_i}{n_0}\frac{m_N}{m_N^*}\Big(\frac{\Delta_i}{\rm
MeV}\Big)^{\frac{13}{2}}\,\xi_i^2 \Big\}\,.
\end{align}
This contribution proves to be much smaller than those for the MMU
and PU processes, and therefore can be dropped.

\begin{figure}
\centerline{\includegraphics[width=6cm]{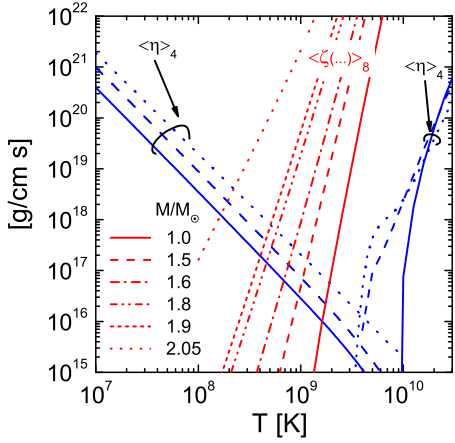}}
\caption{(Color online) The profile averaged shear, $\langle\eta\rangle_4$, and
bulk, $\langle \zeta(1+0.86 (r/R)^2)\rangle_8$,  viscosities
calculated accounting for the nucleon pairing, cf.
Eqs.~(\ref{etaEMu-int})--(\ref{S-eta-e}), (\ref{zeta-neq-SC}), and
(\ref{R-def1}) as functions of the temperature for various
neutron star masses. Calculations of the bulk viscosity  are done
for $\omega=\frac{4}{3}\cdot 10^4$~Hz and MMU+DU reactions are
included. The pion gap  in the MMU contribution is taken along
curves 1a+1b in Fig.~\ref{omegatil}.
} \label{fig:zeta-eta-com}
\end{figure}

In Fig.~\ref{fig:zeta-eta-com} we collect our results for the
density profile averaged shear and bulk viscosities. The
viscosities are plotted as functions of the temperature for
various star masses. The bulk viscosity $\langle \zeta(1+0.86
(r/R)^2)\rangle_8$ is computed accounting for the nucleon
pairing; see Eqs.~(\ref{zeta-neq-SC}) and (\ref{R-def1}). The main
contributions come from the soft mode term (MMU and DU processes
are included) and the radiative one (the latter term being $\simeq
1.5$ times higher than the former). The  main contributions to the
shear viscosity are for $T\lsim 3\cdot 10^9$~K the lepton term
$\langle\eta^{(\rm s)}_{e/\mu}\rangle_4$ [see~(\ref{etaEMu-int}) and
(\ref{S-eta-e})], and  for higher temperatures, the neutrino
radiative term $\langle\eta_{\nu}\rangle_4$ [see~(\ref{eta-nu}) and
(\ref{eta-nu-weight})]. The bulk viscosity is a very rapidly rising
function of the temperature and neutron star mass. At temperatures
above $\sim 10^9$ for all masses the bulk viscosity exceeds the
shear viscosity term. For heavy stars with masses $>1.9\,M_\odot$,
where the bulk viscosity is dominated by the DU reactions, it
exceeds the shear viscosity term already at $T>(2-5)\cdot 10^8$~K.
For lower temperatures the dominant contribution to the
dissipation comes from the lepton shear viscosity.

\section{Critical spin frequencies of  neutron stars}\label{spin}

\begin{figure}[t]
\centerline{\includegraphics[width=6cm]{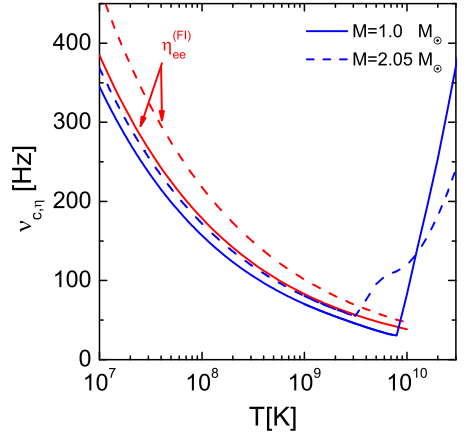}}
\caption{(Color online) The  critical spin frequency of the r-mode
instability, $\nu_{c,\eta}$, calculated  using
Eq.~(\ref{nuB-eta}), i.e., accounting for the shear viscosity
only, as a function of the temperature for two values of neutron
star masses. Calculations of the lepton term are performed with
the help of Eqs.~(\ref{etaEMu-int}) and~(\ref{S-eta-e}) including
the proton pairing effects. The neutrino shear viscosity is taken
according to Eqs.~(\ref{eta-nu}) and~(\ref{eta-nu-weight}). Curves
labeled $\eta_{ee}^{\rm (FI)}$ are computed with the electron
shear viscosity taken in the Flowers-Itoh
parametrization~(\ref{etaE-FI-aver}).  } \label{fig:nucrit-eta-nu}
\end{figure}

The critical frequency for the $r$-mode instability follows from
the solution of the equation
\begin{eqnarray}
\tau^{-1}_G(\nu_c)=\tau_\eta^{-1}(\nu_c)+\tau_\zeta^{-1}(\nu_c)\,,
\label{nuc-eq}
\end{eqnarray}
see Eq. (\ref{t-mode}), $\nu =\Omega/2\pi$. Using Eqs.~(\ref{t-G-1}) and (\ref{t-zeta-1}) we are able to express $\nu_c$ through auxiliary quantities $\nu_{c,\eta}$ and $\nu_{c,\zeta}$,
\begin{align}
\nu_c^6=\nu_{c,\eta}^6+\nu_c^2\,\nu_{c,\zeta}^4\,,
\label{nuc-full}
\end{align}
where $\nu_{c,\eta}$ follows from the equation
\begin{eqnarray}
\tau^{-1}_G(\nu_{c,\eta})= \tau_\eta^{-1}(\nu_{c,\eta})\,,
\label{nuB-eta}
\end{eqnarray}
which includes only the shear viscosity, and
$\nu_{c,\zeta}$ from the equation
\begin{eqnarray}\label{zetac}
\tau^{-1}_G(\nu_{c,\zeta})=\tau_\zeta^{-1}(\nu_{c,\zeta})\,,
\end{eqnarray}
including only the bulk viscosity. The full solution of
Eq.~(\ref{nuc-full}), $\nu_c$, is always larger than
$\max\{\nu_{c,\eta},\nu_{c,\zeta}\}$ but the difference is small,
reaching a maximum, when $\nu_{c,\eta}=\nu_{c,\zeta}=\nu_c^{\rm
cros.}$ with $\nu_c\approx 1.15 \nu_{c}^{\rm cros.}$.

As is seen from Fig.~\ref{fig:zeta-eta-com},
for $T_9< 0.2-2$, depending on the star mass,  the bulk viscosity term
is smaller than the shear term. Thus, for sufficiently low
temperatures solution $\nu_{c,\eta}$ of  Eq.~(\ref{nuB-eta})
approximates well the full solution of Eq.~(\ref{nuc-eq}). At such
temperatures the main term in the shear viscosity is the lepton
term, $\eta_{e/\mu}^{(\rm s)}$, which includes Landau damping
effects following Eq.~(\ref{etaEMu-int}) and the proton
superfluidity factor $S_{\eta,e/\mu}$ from Eq.~(\ref{S-eta-e}) in
the region of the proton pairing.  Using the interpolating formula~(\ref{etaEMu-int})
we get a simple expression convenient for
estimations
\begin{align}\label{realeta}
\nu_{c,\eta}=\frac{153\,{\rm Hz}} {R_6^{3/2}} \frac{ (n_{p,{\rm
cen.}}/n_0)^{0.272}} {(\rho_{\rm cen}/\rho_0)^{\frac{1}{3}}}
\frac{S^{1/6}_{\eta,e/\mu}}{T_9^{0.290}}\,.
\end{align}
Simplifying further, one may use  that $(n_{p,{\rm
cen}}/n_0)^{0.272}/ {(\rho_{\rm
cen}/\rho_0)^{\frac{1}{3}}}\simeq X_{p,{\rm cen}}^{1/3}$.

The solution of Eq.~(\ref{nuB-eta}) (with all contributions to the
shear viscosity  that we considered in this paper) is shown in
Fig.~\ref{fig:nucrit-eta-nu} for stars with masses from 1 to
2.05~$M_\odot$. The phonon shear viscosity is included with the
neutron pairing gaps shown in Fig.~\ref{fig:gap-profiles} (see the
curves labeled by 1 in Fig.~\ref{fig:eta-phon}) and yields only a
minor contribution. The nucleon term in the shear viscosity is
small and is dropped, therefore. We see that the  critical spin
frequency $\nu_{c,\eta}$ decreases with an increase of the
temperature, varying from $\nu_{c,\eta}=350-380$~Hz at $T=10^7$~K
to the value of the maximum  observed spin frequency for young
pulsars, $62$~Hz, at $T=1.5-3\cdot 10^9$~K. The mass dependence of
$\nu_{c,\eta}$ is only moderate for temperatures $T_9\lsim 3$.
Note that values of $\nu_{c,\eta}$ obtained by us are less than
those found in Ref.~\cite{Gusakov:2013aza} by a factor varying
from 1.5 at $T=10^7$~K to 2 at $T=10^9$~K. Simplifying
consideration, Ref.~\cite{Gusakov:2013aza} used the density
profile taken {\em ad hoc}, which corresponds effectively to a
substantially higher central density of the star. Moreover, they
ignored density dependence of the proton pairing gap, which
results in an increase of the lepton shear viscosity term, since
then the proton superfluidity holds up to central densities in all
stars. In our case the lepton shear viscosity drops substantially
for $n>3.5 n_0$ due to the switching off of the proton
superfluidity, cf. Fig.~\ref{fig:etaSC}. We have checked that with the
assumptions used in Ref.~\cite{Gusakov:2013aza} we recover their
results. For comparison in Fig.~\ref{fig:nucrit-eta-nu} we also
show the results obtained with the Flowers-Itoh parametrization of
the lepton share viscosity [see Eq.~(\ref{etaE-FI-aver})], that
often was exploited in the literature. The  critical spin
frequency in this case can be estimated as
\begin{align}
\nu_{c,\eta}^{\rm (FI)}=\frac{108\,{\rm Hz}}{R_6^{3/2}\,
T_9^{1/3}}\,.
\end{align}
We see from Fig.~\ref{fig:nucrit-eta-nu} that  a more realistic
value given by Eq.~(\ref{realeta}) is essentially less than the value
$\nu_{c,\eta}^{\rm (FI)}$. At higher temperatures, $T_9> 3-10$,
the main contribution to the shear viscosity is given by the neutrino shear viscosity term determined by Eqs. (\ref{eta-nu}) and (\ref{eta-nu-weight}), which results in an increase of $\nu_{c, \eta}$ with an increase of the temperature.

For temperatures $T\gsim 10^9$~K the bulk viscosity is typically
larger than the shear viscosity (cf. Fig.~\ref{fig:zeta-eta-com})
and it is legitimate to consider another approximation of Eq.~(\ref{nuc-eq}), given by
Eq.~(\ref{zetac}). Using Eqs.~(\ref{t-G-1}) and~(\ref{t-zeta-1}) we
present the solution of Eq.~(\ref{zetac}) as
\begin{align}
\nu_{c,\zeta}&=\frac{68\,{\rm Hz}}{R_6^{3/4}}
\frac{\langle\zeta_{20}^* [1+0.86
r^2/R^2]\rangle_8^{1/4}}{(M/M_\odot)^{\frac12}\,
(\rho_c/\rho_0)^{\frac12}}\,. \label{nuB-zeta}
\end{align}
Here the bulk viscosity $\zeta_{20}^*$ is the sum of all soft-mode
and radiative contributions evaluated at the frequency $\om_4=4/3$.

\begin{figure}
\centerline{\includegraphics[width=8.5cm]{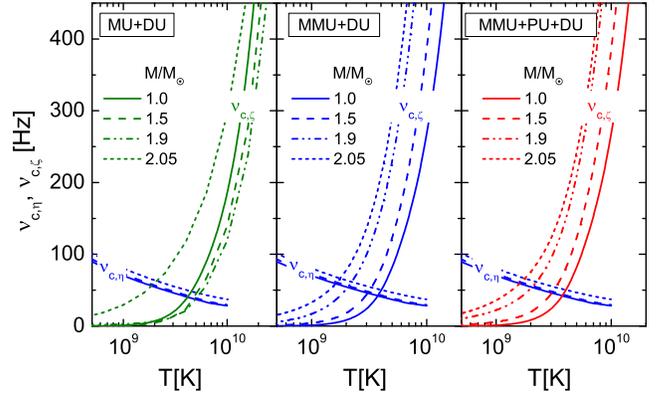}}
\caption{(Color online) The  critical spin frequencies for the
$r$-mode instability determined by the shear viscosity,
$\nu_{c,\eta}$ from Eq.~(\ref{nuB-eta}), and by the bulk
viscosity, $\nu_{c,\zeta}$ from Eq.~(\ref{nuB-zeta}), as functions
of the temperature for various star masses and various processes
included in the calculation of the bulk viscosity term. Notations
of processes are the same, as in Fig.~\ref{fig:tx}. Nucleon
pairing is included, as in Eqs.~(\ref{etaEMu-int})
and~(\ref{S-eta-e}) for the shear viscosity, and, as in
Eq.~(\ref{zeta-neq-SC}) for the bulk viscosities, with the
critical temperatures shown in Fig.~\ref{fig:gap-profiles}. }
\label{fig:nucrit-eta-zeta}
\end{figure}

The dependence of $\nu_{c,\zeta}$ on the temperature for various
neutron star masses is illustrated in
Fig.~\ref{fig:nucrit-eta-zeta}. For comparison we plot also the
critical spin frequency determined by the shear viscosity,
$\nu_{c,\eta}$.  We see that at low temperatures ($T_9<2-6$
depending on the star mass) $\nu_{c,\eta}>\nu_{c,\zeta}$, and
$r$-modes are stabilized by the shear viscosity, whereas at higher
temperatures the $r$-modes are damped  by the bulk viscosity,
$\nu_{c,\eta} < \nu_{c,\zeta}$. For the case where only MU+DU
processes are included  in calculation of the bulk viscosity (left
panel in Fig.~\ref{fig:nucrit-eta-zeta}) the value of
$\nu_{c,\zeta}$ is only weakly dependent of the star mass, if $M<
1.9\,M_\odot$, and it decreases slightly with the mass growth. The
dependence of the value $\nu_{c,\zeta}$ on the star mass becomes
sharp, when the DU processes become operative, i.e. at $M>
1.9\,M_\odot$. The pattern changes if the pion softening effect
is taken into account. For the MMU+DU processes (middle panel in
Fig.~\ref{fig:nucrit-eta-zeta}) the value $\nu_{c,\zeta}$
increases with  increase of the star mass and the effect of the
switching on of the DU reaction is not so pronounced, as  it was
in MU+DU case. Inclusion of the PU processes (right panel in
Fig.~\ref{fig:nucrit-eta-zeta}) does not change $\nu_{c,\zeta}$
much compared to the MMU+DU case. As everywhere above, to compute
MMU+DU processes we use curves 1a+1b in Fig.~\ref{omegatil} and
for MMU+PU+DU ones we exploit curves 1a+2 for MMU and 3 for PU.

\begin{figure}
\centerline{\includegraphics[width=7cm]{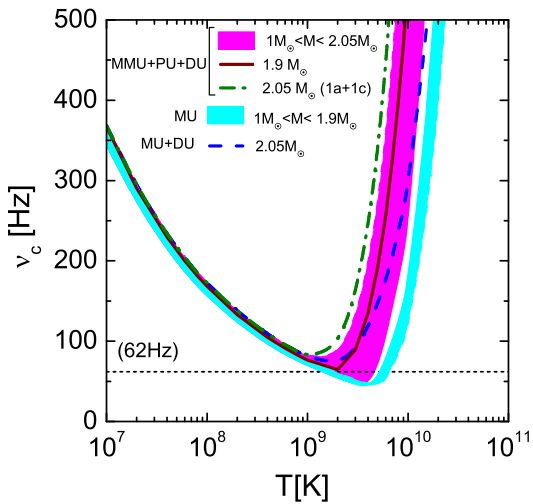}} \caption{
(Color online) The critical frequency of the $r$-mode instability calculated using
the full Eqs.~(\ref{nuc-full}) as a function of the temperature
for a band of neutron star masses. The critical values
$\nu_{c,\eta}$ and $\nu_{c,\zeta}$ given by  Eqs.~(\ref{nuB-eta})
and~(\ref{nuB-zeta}) are the same, as  in Fig.~\ref{fig:nucrit-eta-zeta}.  Notations of processes are the same, as in Fig.~\ref{fig:tx}. The dashed line shows calculation with  MU+DU
for $M=2.05~M_{\odot}$. The solid line demonstrates calculation done
for the threshold mass of the DU processes within MMU+PU+DU
scenario. The dashed-dotted line shows calculation with MMU computed
with curve 1a+1c in Fig.~\ref{omegatil} for $M=2.05~M_{\odot}$. }
\label{fig:fignuB-all}
\end{figure}

Now we  solve the full equation~(\ref{nuc-full}) with all
contributions to the shear and bulk viscosities (cf.
Fig.~\ref{fig:nucrit-eta-zeta}) taken into account. The solutions
of Eq.~(\ref{nuc-full}) are presented in Fig.~\ref{fig:fignuB-all}
as functions of the temperature for neutron star masses from
$M_{\odot}<M<2.05M_{\odot}$. Within the minimal cooling  scenario
when in-medium effects are not included, $\nu_{\rm c, min}$
exceeds $\nu_{\rm max}^{\rm young}$ only for the masses
$M>2.03M_{\odot}$ (when DU reaction is already efficient), i.e.,
very close to the maximum mass, $2.05M_{\odot}$. However, if
initially the star passes through the instability region, the
developed $r$-modes blow off some part of the star matter. So, its
final mass can hardly be very close to the maximum mass.
Alternatively, the experimental value of the frequency of the
pulsar PSR J0537-6910 could be explained within the minimal
cooling scenario, if  one exploited the EoS that allows for a
lower DU threshold density. Reference~\cite{Alford:2012yn},
involving the data on $\dot{\nu}$ and $\nu$, presented arguments,
that within their analysis, curve AB in
Fig.~\ref{fig:eta-nu}(a) is shifted to the left for the relevant
values of the amplitude $a_{\rm sat}$ (the lower is $a_{\rm sat}$
the more pronounced is this shift), which could also make it possible to
explain stability of PSR J0537-6910 within the minimal cooling
paradigm. Within the nuclear medium cooling scenario, the DU
processes are not needed to explain the stability of PSR
J0537-6910. We may explain it for $M>1.804M_{\odot}$, if
calculations of MMU processes are done using curves 1a+2, and 3
for PU, and for $M>1.776M_{\odot}$ provided we use curves 1a+1c to
calculate MMU, without PU. For the MMU case (for curves 1a+1b for
MMU without PU), PSR J0537-6910 mass should be $M>1.84 M_{\odot}$.
Any restrictions on the values of the amplitude $a$ are not
required; see our recent note~\cite{Kolomeitsev:2014epa}.

Recall that many old recycled rapidly rotating pulsars in LMXB
have much  higher spin frequencies ($\nu \sim 200-716$\,Hz). The
maximum observed frequency is 716 Hz, for PSR J1748-2446ad.
Estimated values of internal temperatures lie in the range
$10^{-2}\lsim T_9\lsim 0.3$; see Table~1
in Ref.~\cite{Gusakov:2013aza}. Some of the data enter a region of
stability $\nu<\nu_c$, whereas others are far outside the
stability range. Thus one should try to search other mechanisms
leading to a substantial increase of $\nu_c\simeq \nu_{c,\eta}$
for $T\lsim 3\cdot 10^8$~K. Below we elaborate  such a
possibility.

\section{Bose condensates with non-zero momentum and braking of the neutron star rotation}
\label{Bose condensates}

\subsection{Low-lying Bose excitations in neutron stars}

There might exist many low-lying bosonic modes of different nature
in the neutron star crust and interior. Some of these modes are
gapped, some are gappless, as, e.g., sound modes with the
frequency $\epsilon \propto k$ for $k\to 0$. The presence of the
low-lying collective bosonic excitations is particularly important
for understanding of thermal and transport properties of accreting
neutron stars with temperatures in the range $10^7$--$10^9$ K, see
Ref.~\cite{Chamel:2012ix}. The transition between the outer crust
and the inner crust corresponds to the density $n_{\rm drip}\simeq
4 \cdot 10^{11}$ g$/$cm$^{-3}$, at which neutrons start dripping
out of nuclei. At densities $n>n_{\rm drip}$ there appear to be
superfluid transverse and longitudinal lattice phonons [$\epsilon
(k)=v_{\rm ph}k$]. They prove to be very strongly mixed, and the
speed of transverse lattice modes ($v_{\rm ph}=v_{\perp}$) is
greatly reduced. Soundlike shear modes in neutron-star crusts
with velocities  $\sim 10^{-3}$--$10^{-2}$ (in $c=1$ units) have
been proposed to play a role in the interpretation of
quasiperiodic oscillations (QPO) observed in giant flares from
soft gamma repeaters (SGR)~\cite{Strohmayer:2006py}. The velocity
of sound excitations in a two-dimensional slab phase of the pasta
(at densities $n\sim 0.5 n_0$) is estimated~\cite{DiGallo:2011cr}
to be $\simeq 0.04$.

In the normal neutron Fermi liquid the velocity of the zero scalar
sound  $v_0=s v_{\rm F}$, where $s\geq 1$, depends on a value of
the zero harmonic of the scalar Landau parameter $f_0$, $s\to 1$
for $f_0\to 0$. Similarly, for the spin sound, the value $s$
depends on the zeroth harmonic $g_0$ of the spin-spin interaction.
The velocity of the first-sound mode in the neutron Fermi  liquid
is determined by the values of the zeroth and first scalar Landau
parameters. Using the values of these parameters from
Ref.~\cite{Wambach:1992ik} we get $v_1\sim (0.3\mbox{---}1) v_{\rm
F}$ depending on the density.

In the presence of an external magnetic field there exist low-lying
spin wave excitations with frequencies $\epsilon =\om_L
\left(1+\frac{(1+g_0)^2}{3g_0}\frac{(kv_{\rm
F})^2}{\om_L^2}\right)$, where $\om_L =\gamma_0H$ is the Larmor
frequency, $H$ is the magnetic field, $\gamma_0=e/2m_N^*$, and $e$
stands for the electric charge~\cite{Ketterson}.

For the superfluid neutrons paired in the 1S$_0$ state at
densities $n\lsim 0.7 n_0$ besides the first sound [related to the
Anderson-Bogoliubov mode with the energy $\epsilon =v_{1,{\rm s}}
k$ for $k\to 0$, and $\epsilon (k)\to 2\Delta_{nn}$ for large $k$]
there exists a second-sound mode, $\epsilon (k\to 0)=v_{2,{\rm s}}
k$, with a still lower velocity $v_{2,{\rm s}} \to v_{1,{\rm
s}}/\sqrt{3}$ for $T\to 0$, where $v_{1,{\rm s}} \to v_{\rm
F}/\sqrt{3}$, and $v_{2,{\rm s}}\to 0$ for $T\to T_{c,n}$. There
also exists a Schmid mode with $\epsilon (k)\simeq 2\Delta_{nn}$,
see Ref.~\cite{Kulik}.

In the 1S$_0$ proton superfluid there exists the Carlson-Goldman
mode with the frequency $\epsilon (k)\to 2\Delta_{pp}$, and the
damping rate $\gamma \to 0$ for sufficiently large $k$. This mode
starts with the value $\epsilon (k_{m})=0 $ at a finite value of
the momentum $10\sqrt{T\Delta_{pp}}/v_{\rm F}\lsim k_m\lsim p_{\rm
F}$. Recently it has been argued \cite{Baldo-Ducoin-11} that  a
pseudo-Goldstone mode associated with the superconducting protons
in neutron star matter may exist owing to the screening of the
Coulomb field  by the electrons. The proton Fermi velocity is
substantially lower than the neutron one. The typical value of the
slop of the branch for $k\to 0$ is $\sim \sqrt{3}v_{{\rm F},p}$.

For densities $n\gsim n_{0}$, neutrons are paired in  the 3P$_2$ state at
$T<T_{c,n}({\rm 3P_2})$. In this case there may exist a low-lying
Bose excitation mode with frequency $\epsilon
(k=0)=\Delta_{nn}({\rm 3P_2})/\sqrt{5}$,~\cite{Leinson:2012pn}.
Estimations of the value of the pairing gap are very uncertain. As
we have mentioned, there are arguments~\cite{Schwenk} that the
value of the 3P$_2$ pairing gap is tiny, $\Delta_{nn}({\rm 3P_2})
< 10^{-2}$ MeV. Then this spectrum branch lies very low.

Note that with participation of the bosonic excitations  may occur neutrino (resonance) reactions~\cite{Voskresensky:1986af}. However their contribution to the total emissivity is not large compared to MMU, PU, and DU reactions. Therefore, we do not add their contribution to the radiative bulk viscosity.

\subsection{Condensates of bosonic excitations with non-zero momentum}

\begin{figure*}
\centerline{\includegraphics[width=14cm]{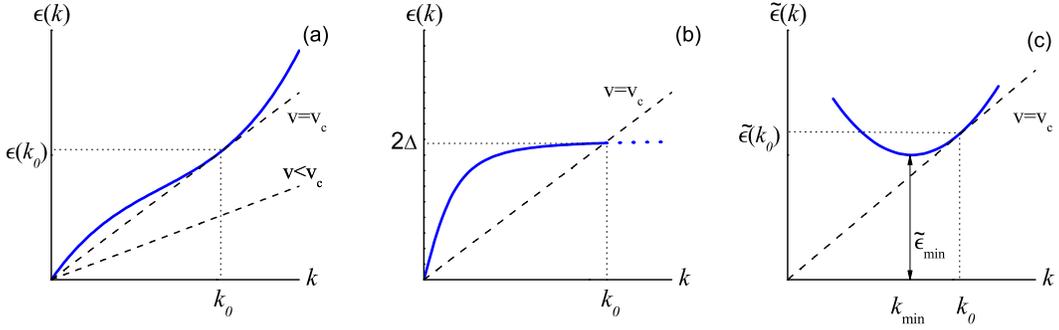}}
\caption{(Color online) Examples of the dispersion law of the low-lying bosonic excitations. Panel~(a) shows the  spectrum $\epsilon(k)$ of soundlike excitations, like assumed quasiparticle excitation (levon) spectrum in Bose gases~\cite{BP12,Voskresensky:1993uw} (cf. the zero-sound~\cite{Vexp95} and the first sound excitation spectra in normal Fermi liquids); panel~(b) demonstrates the Anderson-Bogoliubov mode in the Fermi system with pairing~\cite{Kulik}; panel (c) shows the spectrum of roton-like excitations $\tilde{\epsilon}(k)=\epsilon(k) +q \mu_e$, as in the case of the p-wave $\pi^+$, $\pi^-$,  or $K^-$ condensates, counted from the value of the chemical potential $\mu_e$,  $q=\pm 1$. }
\label{fig:spec-excit}
\end{figure*}

In our further consideration we assume that there exists a low-lying bosonic quasiparticle mode of collective excitations in some range of momenta. For instance, this can be a sound-like mode [see Fig.~\ref{fig:spec-excit}(a)], like an assumed quasiparticle excitation (levon) mode in Bose gases (cf.~\cite{BP12,Voskresensky:1993uw}), or a zero-sound mode of a Fermi liquid continued to a momentum $k_0\sim p_{\rm F}$, at which it touches the line $\epsilon =kv_{\rm F}$, or the first-sound mode; or the mode behaving similar to the Anderson-Bogoliubov mode in the Fermi system with pairing~\cite{Kulik} or Carlson-Goldman mode in superconductors [see Fig.~\ref{fig:spec-excit}(b)]; or a roton-like mode having a local minimum at $k=k_{\rm 0}$, as in $^4$He, and in the case of the p-wave $\pi^+$, $\pi^-$, or $K^-$ condensates [see Fig.~\ref{fig:spec-excit}(c)]. For the charged quasiparticles in the neutron star matter the energy is counted from the chemical potential; i.e., $\tilde{\epsilon}(k)=\epsilon(k) + q\,\mu_e$ plays a role of the excitation energy, where $\mu_e$ is the electron chemical potential and $q=+ 1$ and $-1$ for the positively and negatively charged particles, respectively. We consider now a possibility of appearance of a condensate of excitations with $k=k_0$, when the system moves with velocity above  the Landau critical value [$v_{c,{\rm L}}=\epsilon (k_0)/k_0$ or $v_{c,{\rm L}}=\widetilde{\epsilon} (k_0)/k_0$].

The key idea was formulated in Refs.~\cite{Pitaev84,Voskresensky:1993uw}. When the medium moves with a velocity $v>v_{c,{\rm L}}$ it may become energetically favorable to transfer a part of the momentum from the particles of the moving medium to a condensate of bosonic excitations with a
finite momentum provided the spectrum of excitations is such that they may possess a rather large momentum but a small energy. Transfer of the momentum to the condensate would diminish the
velocity of the system from the point of view of an external observer. For a rotating system, a part of the angular momentum can be transferred to the boson subsystem~\cite{Voskresensky:1993uw}, whereas an external observer will see a slowing down of the rotation. A common rotation may be reached only after the passing of a long time provided the damping rate is low.

For better understanding of the phenomenon, following Ref.~\cite{Voskresensky:1993uw}, we consider a fluid element of the medium with the mass density $\rho$  moving with a non-relativistic constant velocity $\vec{v}$. For the probing function of the condensate  of a complex scalar boson field, chosen in the simplest form of a running wave
\begin{align}
\label{running} \varphi =\varphi_0 e^{-i\epsilon
t+i\vec{k}\vec{r}}
\end{align}
with the amplitude $\varphi_0$, the momentum $k$, and the energy $\epsilon$,  the Lagrange density acquires the form
\cite{Migdal1978}
\begin{align}\label{exLagr}
L_{\rm b} =D^{-1}(\epsilon , k)|\varphi|^2 -\half\Lambda
|\varphi|^4\,.
\end{align}
Here $D^{-1}(\epsilon , k)$ is the inverse Green's function of the collective excitation, $\Lambda$ is the self-interaction coupling constant of excitations which may, in general, depend on $\epsilon (k)$ and $k$  and on the assumed spatial structure of the field $\varphi$. For excitations of a hadronic nature with assumed field structure (\ref{running}), one may expect that $\Lambda ={\rm const}\sim 1 $. If the system is superfluid, then it is
characterized by an order parameter. In this case we can use Eq. (\ref{exLagr}), additionally assuming that the field describing the excitation does not interact with the order parameter of the superfluid.

The quasiparticle energy $\epsilon (k)$ is determined by the dispersion equation
\begin{align}
D^{-1}(\epsilon , k)=\epsilon^2 -m_{\rm b}^2 -k^2-\Re\Sigma^R(\epsilon, k)=0\,,
\end{align}
where $m_{\rm b}$ is the boson mass, and $\Re\Sigma^R (\epsilon(k),k)$ is the real part of the retarded self-energy, and the width $\Im \Sigma^R (\epsilon(k),k)$ is assumed to be
negligibly small. To specify the problem let us consider a cold system, $T=0$, and
neutral excitations. The energy density of the boson sub-system follows then as
\begin{align}
E_{\rm b}=\epsilon \partial L_{\rm b}/\partial \epsilon -L_{\rm b}\,.
\end{align}
The momentum conservation yields
\begin{align}
\rho \vec{v} =\rho \vec{v}_{\rm fin} +\vec{k} Z^{-1}_0
|\varphi_0^2| \,, \label{momentumcons}
\end{align}
where
\begin{align}
Z^{-1}_0 (k) =\left(2\epsilon -\frac{\partial\Re\Sigma^R}
{\partial \epsilon}\right)_{\epsilon(k),k}>0\,.
\end{align}

In the absence of the condensate, the energy density of the liquid element was $E_{\rm in}=\rho v^2/2$, whereas in the presence of the condensate, which takes a part of the momentum, the energy
becomes
\begin{align}
\label{Ef}
E_{\rm fin}=\frac{\rho v_f^2}{2} +\epsilon (k_0)Z^{-1}_0 |\varphi_0|^2 +\frac{\Lambda|\varphi_0|^4}{2}.
\end{align}
The last two terms appear owing to the classical field of the condensed excitations. The energy density gain is
\begin{align}
\delta E &= E_{\rm fin} -E_{\rm in}
= -{Z_0^{-1} (k)}\left({\vec{v}\vec{k} -\epsilon (k)}\right)
|\varphi_0|^2 +\half \widetilde{\Lambda}|\varphi_0|^4\,,
\nonumber\\ \widetilde{\Lambda} &=\Lambda +Z^{-2}_0(k)  k^2
/\rho\,. \label{E-gain}
\end{align}
Note that above equations hold also for $\Lambda =0$. If the function $\epsilon(k)/k$ has a minimum at $k=k_0$,  it becomes energetically favorable to generate excitations, when the speed of the fluid exceeds the Landau critical value $v>v_{c,{\rm L}} =\epsilon (k_0)/k_0$ for $\vec{k_0}\parallel \vec{v}$, see Fig.~\ref{fig:spec-excit}. These bosonic excitations may then
develop a classical condensate field characterized by the finite momentum $k_0$. The condensate field amplitude is found from the energy minimization.

The energy density gain because of the appearance of the condensate is
\begin{align}
\label{condE} \delta E = -  {Z_0^{-1}(k_0)}\left({v\,k_0 -\epsilon
(k_0)}\right)\varphi_0^2\, \theta (v-v_{c,{\rm L}})
 + \half \widetilde{\Lambda} \varphi_0^4\,.
\end{align}

The amplitude of the condensate field is found by  minimization of Eq.~(\ref{condE}) with respect to $\varphi_0$. Finally we get
\begin{align}
\varphi_0^2 = \frac{Z_0^{-1}(k_0)\left(vk_0 -\epsilon
(k_0)\right)}{\widetilde{\Lambda}}\theta (v-v_c)\,.
\end{align}
When the condensate of excitations is formed the resulting velocity of the non-condensate matter becomes
\begin{align}
\label{finvel} v_{\rm fin} =v_{c,{\rm L}} +\frac{v-v_{c,{\rm
L}}}{1+c_4}\,,\quad c_4 = [Z^{-1}_0 (k_0)]^2
k_0^2/(\Lambda\rho)\,.
\end{align}

For the probed function of the condensate field different from the running wave we get for the energy gain $\delta \mathcal{E}=\int \delta E dV$ owing to appearance of the condensate
\begin{align}
\label{condE2} \delta \mathcal{E} =-\alpha_1 V \frac{vk_0
-\epsilon (k_0)}{Z_0 (k_0)}\varphi_0^2\theta (v-v_{c,{\rm L}})
+\alpha_2 V\frac{\widetilde{\Lambda} \varphi_0^4}{2}\,,
\end{align}
with coefficients $\alpha_{1,2}\sim 1$ depending on the assumed structure of the field. For example, for the field of the form $\varphi =\varphi_0 \cos (k_0 z)$ one has $\alpha_1 =1/2$  and
$\alpha_2 =3/4$.

In case of the charged excitations, their energy should be counted from the value of the chemical potential. The above consideration still holds, if we replace $\epsilon (k)\to \widetilde{\epsilon}(k)$ everywhere.

\subsection{Evaluation of the critical spin frequency}\label{critfreqcond}

\begin{figure*}
\includegraphics[width=15cm]{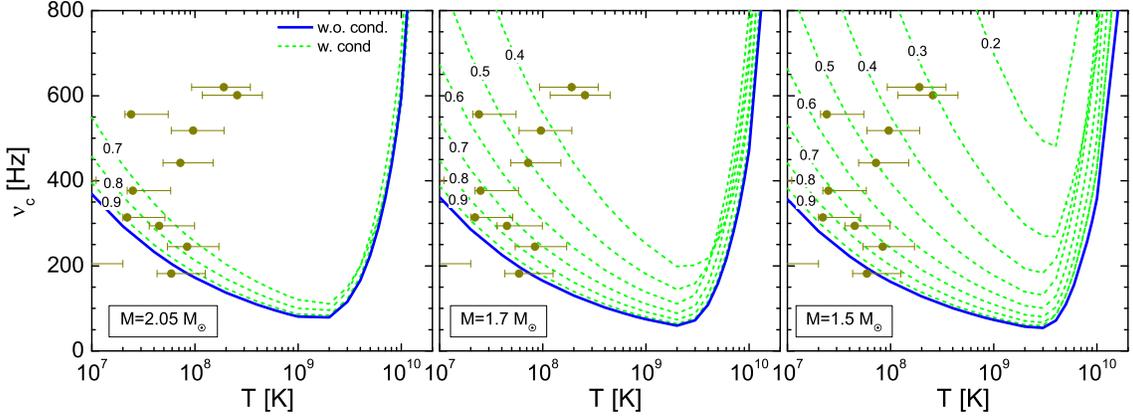}
\caption{(Color online) The  critical spin frequency of the r-mode instability as a function of the temperature, calculated according to Eq.~(\ref{Omfin-sharp}) accounting for the possibility of a condensation of excitations  for $r<r_c$ for various values of $r_c$ (dashed lines) and without condensation (solid line) for $M=2.05\,M_{\odot}$, $1.7\,M_{\odot}$ and 1.5\,$M_{\odot}$ on three panels from left to right. Numbers near the dashed curves indicate values of $r_c/R$. The minimal chosen values of $r_c/R$ correspond to the boarder of the proton superfluid region, as shown in Fig.~\ref{fig:gap-profiles}. The choice MMU+PU+DU is exploited as in calculation shown in Fig.~\ref{fig:fignuB-all}.}
\label{fig:nuBcondin}
\end{figure*}

The above expressions hold also for rotating systems~\cite{Voskresensky:1993uw}, for which one should put $\vec{v}(r) = [\vec{\Omega},\vec{r}\,]$. Choosing the appropriate probing function for the condensate field~\cite{Voskresensky:1993uw}, we may obtain the  critical angular velocity, at which the condensate appears in the star rotating as the whole, as $\Omega_{c,{\rm L}}(R)\alpha_3  R = v_{c,{\rm L}}$ with $\alpha_3\simeq 0.42$ and $R$, being the star radius.  For example, in regions of the neutron and proton superfluids, the critical velocities $v^n_{c,{\rm L}}\sim
\Delta_{nn}/p_{{\rm F},n}$ and $v^p_{c,{\rm L}}\sim  \Delta_{pp}/p_{{\rm F},p}$ and the corresponding critical angular velocities $\Omega^n_{c,{\rm L}}\sim \Delta_{nn}/(p_{{\rm F},n}R)$
and $\Omega^p_{c,{\rm L}}\sim \Delta_{pp}/(p_{{\rm F},p}R)$ might be very low.

Following Ref.~\cite{SaTome} the differential rotation is an unavoidable feature of nonlinear $r$-modes. We may apply Eq.~(\ref{finvel}) also for a sphere undergoing a differential rotation,
in which the density changes with the distance from the star center. Then the angular velocity of the medium becomes dependent of the distance from the star center,
\begin{align}
\Omega_{\rm fin}(r) =\min\Big\{\Omega,\Omega_{c,{\rm L}}(r) +\frac{\Omega
-\Omega_{c,{\rm L}}(r)}{1+\alpha_4 \, c_4}\Big\}\,,
\label{Omfin}
\end{align}
where the critical angular velocity follows from the equation $\Omega_{c,{\rm  L}}(r)=v_{c,{\rm L}}(n(r))/(r\,\alpha_3)$ and $\alpha_4 \sim 1$.

Thus, a layer of the star, in which a branch of bosonic excitations is populated by  condensate with a nonzero momentum, may rotate with the angular velocity $\Omega_{\rm fin}$ smaller than the angular velocity $\Omega$ of the outer part of the star (mantle) seen by a distant observer. For instance, for $c_4\gg 1 $, one would have $\Omega_f \simeq \Omega_{c,{\rm L}}(r)$ in the interval of $r$, where $\Omega>\Omega_{c,{\rm L}}(r)$, even if the initial angular velocity were $\Omega\gg \Omega_{c,{\rm L}}$. Such a differential rotation of the star must be taken into account in averaging (\ref{aver}), entering the gravitation (\ref{tG}) and viscous times (\ref{tau-eta}), (\ref{tau-zeta}). Recovering from Ref.~\cite{Lindblom:1998wf} the dependence of these times on the angular velocity, we present the corrected gravitation, shear, and bulk viscosity times as
\begin{align}
&\frac{1}{\widetilde{\tau}_{G,\eta,\zeta}(\Omega)}=
\frac{\chi_{G,\eta,\zeta}(\Omega)}{\tau_{G,\eta,\zeta}(\Omega)}
\,, \label{chi-rescale}\end{align} where the scaling factors are
given by
\begin{align}
&\chi_{G}(\Omega)=\frac{\langle\rho\big(\frac{\Omega_{\rm
fin}}{\Omega}\big)^4
\rangle_6^2}{\langle\rho\big(\frac{\Omega_{\rm
fin}}{\Omega}\big)^2 \rangle_6 \langle\rho \rangle_6},
\quad 
\chi_{\eta}(\Omega) = \frac{\langle\eta\big(\frac{\Omega_{\rm
fin}}{\Omega}\big)^2\rangle_{4}}
{\langle\rho\big(\frac{\Omega_{\rm
fin}}{\Omega}\big)^2\rangle_{6}}
\frac{\langle\rho\rangle_{6}}{\langle\eta\rangle_{4}}, \nonumber\\
&\chi_{\zeta}(\Omega) = \frac{\langle\zeta(1+0.86
r^2/R^2)\big(\frac{\Omega_{\rm fin}}{\Omega}\big)^4\rangle_{8}}
{\langle\zeta(1+0.86 r^2/R^2)\rangle_{8}}
\frac{\langle\rho\rangle_{6}}{\langle\rho\big(\frac{\Omega_{\rm
fin}}{\Omega}\big)^2\rangle_{6}}\,. \label{chis}
\end{align}
In expression for $\chi_{\zeta}$ we exploit that $\zeta\propto 1/\om^2\propto 1/\Omega^2$.

Replacing $\tau_{G,\eta,\zeta}$ in Eq.~(\ref{nuc-eq}) with $\tilde{\tau}_{G,\eta,\zeta}$ we obtain the modified equation for the critical spin frequency
\begin{align}
\nu_c^6=\tilde{\nu}_{c,\eta}^6+\nu_c^2\,\tilde{\nu}_{c,\zeta}^4\,,
\label{nuc-eq-cor}
\end{align}
where in contrast to Eq.~(\ref{nuc-full}) we define
\begin{align}
\tilde{\nu}_{c,\eta}=\nu_{c,\eta}
\Big(\frac{\chi_\eta(\nu_{c,\eta})}{\chi_G(\nu_{c,\eta})}\Big)^{1/6}\,,
\,\,
\tilde{\nu}_{c,\zeta}=\nu_{c,\zeta}
\Big(\frac{\chi_\zeta(\nu_{c,\zeta})}{\chi_G(\nu_{c,\zeta})}\Big)^{1/4},
\label{nuc-cut}
\end{align}
with $\nu_{c,\eta}$ and $\nu_{c,\zeta}$ as the solutions of the original Eqs.~(\ref{nuB-eta}) and~(\ref{nuB-zeta}).

If, simplifying, we assume the weakness of the self-interaction of bosonic excitations and put $\Lambda=0$, we have $\Omega_{\rm fin}(r)= \Omega_{c,{\rm L}}(r)\to 0$ in the condensate regions. Thus, for a small critical angular velocity, $\Omega_{c,{\rm L}}(r)$, we can roughly approximate in the integrals in Eq.~(\ref{chis})
\begin{equation}
\Omega_{\rm fin}(r)/\Omega\approx\theta(r_c-r)\,.
\label{Omfin-sharp}
\end{equation}
Here the critical radius $r_c$ marks the inside boundary of the region with the condensate of excitations, and  we neglected a thin mantle layer, where the condensate is absent and the rotation frequency coincides with the initial frequency. For such a distribution we find
\begin{align}
\chi_G =\frac92\xi^7-\frac72 \xi^9\,,
\chi_\eta =\frac{7}{4\xi^2}-\frac{5}{4}
+\frac{1}{9-7 \xi^2},
\label{chis-cut}
\end{align}
where we denoted $\xi=r_c/R$, and we used that~\cite{Gusakov:2013aza} in  presence of a nucleon pairing the density dependence of the shear viscosity can be roughly presented as $\eta\propto \rho^2$. Numerical calculations show that for rough estimations  the factor $\chi_\zeta$ can be written as $\chi_\zeta\simeq a(M,T)/\chi_G$ with $\chi_G$ taken from Eq.~(\ref{chis-cut}). For $r_c\gsim 0.5 R$ and for temperatures $0.1<T_9<10$, we get $a(M,T)\simeq 1$. Note that for $T_9 \ll 1$ the effect of the bulk viscosity on the values of $\nu_c$ is minor and can be neglected. Now, with the corrected times we are able to recalculate curve $\nu_c (T)$ and then compare it with the data on the rapidly rotating recycled pulsars in LMXB.

Solutions of Eqs.~(\ref{nuc-eq-cor}) and~(\ref{nuc-cut}) for the angular velocity distribution (\ref{Omfin-sharp}) are shown in Fig.~\ref{fig:nuBcondin}. To be specific we associate here the
condensate of excitations with a soft mode in the proton superfluid. The minimal values of $r_c/R$ shown in Fig.~\ref{fig:nuBcondin} correspond to the boarder of the proton superfluid region for the corresponding star mass shown in Fig.~\ref{fig:gap-profiles}, i.e. with these values we simulate a maximal possible effect. Curves corresponding to larger values of $r_c$ for the star of the given mass effectively simulate more complex effects, which we do not discuss explicitly, e.g. finite $\Lambda$ and $\Omega_{c,{\rm L}}\neq 0$. Curves $\nu_c (T)$ shift up, if one uses assumption of the differential rotation in the stars induced by a condensation of bosonic excitations on a certain time scale. The minimal values of $\nu_c$, reached in the interval $T_9\sim 2-3$, are increased. We see that at least a part of the data on the recycled pulsars in the LMXB can be now accommodated. For $M=1.5 M_{\odot}$ and for the choice $r_c/R <0.45$ all the data points get to the region stable with respect to the $r$-mode excitation.

We should stress that we used very rough approximations [e.g., we exploited the model ansatz (\ref{Omfin-sharp}) putting $\Omega_{c,{\rm L}} \to 0$] and thus demonstrated only a possibility of an increase of the value $\nu_c$ because of the condensation of excitations with the finite momentum. A full solution of the problem needs a detailed analysis that goes beyond the scope of this work.

Finally, note that in the rotating star with $\Omega >\Omega_{c,{\rm L}}$ the condensates of excitations may participate in the neutrino reactions that causes additional contributions to the neutrino emissivity and the radiative bulk viscosity; e.g., if bosonic excitations couple with nucleons, the reactions of the PU type are possible.

\subsection{$\pi^{+}_s$ condensation with non-zero momentum}

In the neutron star matter  there appears a $\pi^{+}$ quasiparticle branch with a negative energy. At densities $n>n_{c}^+$ ($n_{c}^+\lsim n_0$) the minimum  value of $\epsilon (k)$ for $k=k_0\sim p_{{\rm F},n}$ is such that $\widetilde{\epsilon} (k_0)=\epsilon (k_0)+\mu_e $ becomes negative. This may induce the reaction $p\to n+\pi^{+}_{\rm cond}$ in the neutron star matter~\cite{Migdal1978,Migdal:1990vm}, leading to formation of the so-called $\pi^{+}_s$ condensate with a non-zero momentum, corresponding to the minimum of $\epsilon(k)$.

We assume that the condensate  has appeared long ago at the stage of the hot star formation and, therefore, it most probably exists initially in the form of disoriented domains. When temperature
decreases, in a rotating star all domains become oriented in one direction parallel to the matter flow, diminishing thereby the velocity of the non-condensate matter owing to the angular momentum
conservation. The resulting system contains, thus, two (condensed and noncondensate) fluids rotating with different angular velocities.

In the medium at rest ($v=0$) the (initial) condensate  energy is given by
\begin{align}
E^{\rm cond}_{\rm in}(v=0) = -\frac{\widetilde{\epsilon}^2 (k_0)
}{2\Lambda} \frac{\theta[-\widetilde{\epsilon}
(k_0)]}{Z_0^{2}(k_0)}\,.
\end{align}
The final energy density $E_{\rm fin}$ of the medium moving initially with constant velocity $\vec{v}$ is determined from Eqs.~(\ref{momentumcons}) and (\ref{Ef}), now for $\epsilon(k_0)<0$. For the field of the form (\ref{running}) one finds the energy density gain~\cite{Voskresensky:1993uw}
\begin{align}
\delta E=[Z_0^{-1}(k_0)]^2\left[\frac{\widetilde{\epsilon}^2
(k_0)}{2\Lambda}-\frac{(k_0 v-\widetilde{\epsilon} (k_0))^2
}{2\widetilde{\Lambda}}\right]\,,
\end{align}
counted from the value $\rho v^2/2$, $\vec{k}_0\parallel \vec{v}$, $Z^{-1}(\epsilon(k_0))>0$. The critical velocity $v_c$ is found from the condition $\delta E<0$. In the limit of a large $\Lambda$ we get
\begin{align}\label{vLinf}
v_{c,{\rm L}}^{\pi}=\frac{[Z_0^{-1}(k_0)]^2|\widetilde{\epsilon}
(k_0)| k_0}{2\Lambda \rho}\,;
\end{align}
therefore, $v_{c,{\rm L}}^{\pi} \to 0$ for $\Lambda \to \infty$.

For a tiny $\Lambda$ we find
\begin{align}\label{vL0}
v_{c,{\rm L}}^{\pi} =Z_0^{-1}(k_0)\frac{|\widetilde{\epsilon}
(k_0)|}{\sqrt{\Lambda\rho}}\to \infty\,,
\end{align}
and the condensate domains remain disordered.

\subsection{$\pi^-$ and $K^{-}$ condensations with non-zero momenta}

In the neutron star matter at densities $n>n_{c}^\pi$ (in the given work we supposed $n_{c}^\pi =3n_0$) the minimal value of the $\pi^-$ energy, $\epsilon (k_0)>0$, may decrease below the electron chemical potential $\mu_e$, then the reaction $n\to \pi^-_{\rm cond} +p$ produces a $\pi^-$ condensate with the momentum $k_0\sim p_{{\rm F},n}$, for $Z^{-1}(\epsilon(k_0))>0$.

Besides the pion condensate, a $K^-$ condensate may appear in the neutron star interiors for $n> n_{c}^K\sim (3-5) n_0$. The $K^-$ condensate arises in reaction $e\to K^-_{\rm cond}$, provided the value at the minimum on the $K^-$ branch $\epsilon (k_0)>0$ decreases with a density increase below the electron chemical potential $\mu_e$. The $K^-$ condensate may appear either with the momentum $k_0=0$ or with $k_0 \neq 0$~\cite{KVK95,KV2003}. In the latter case, following estimate \cite{KV2003}, the typical value of the condensate momentum is  $k_0\sim m_{\pi}$, where $m_{\pi}$ is the pion mass. All the expressions derived above for the $\pi^+_s$ condensate continue to hold also for the $\pi^-$ and $K^-$ condensates after the replacement $\widetilde\epsilon(k_0)\to \epsilon (k_0)-\mu_e$ therein.

The next possibility is a $\pi^0$ condensation appearing for densities $n>n_c^0$ at $\epsilon (k_0) =0$. However, for $\epsilon =0$, $Z_0^{-1}=0$ and this possibility is not of interest here
because the motion of the system does not affect it.

The following remark is in order. One could think that one should put $r_c =0$ (see Sec.~\ref{critfreqcond})  considering the stars with the inhomogeneous condensates of charged pions/kaons extending up to the star's center (for $M>1.31 M_{\odot}$ in our model). However, to find the dependence of $\nu_c (T)$ in the above consideration we exploited the case $\Lambda =0$, and for the pion/kaon inhomogeneous condensates $v_{c,{\rm L}}^{\pi/K}\to \infty$ for $\Lambda \to 0$, see Eq.~(\ref{vL0}). A realistic consideration for $\Lambda \neq 0$ is, thus, more involved.

\subsection{A possible mechanism for acceleration of rotation of old accreting pulsars}

One usually assumes that  old pulsars are slowed down owing to the magnetic dipole radiation loosing, simultaneously, their magnetic fields.  At a longer times, however, old pulsars in LMXB are spin up owing to the accretion of matter from a companion star~\cite{Patruno:2010qz}. This spin up might be compensated for by the $r$-mode emission. Below we propose a supplementary mechanism of the pulsar acceleration.
\begin{figure}
\parbox{6cm}{\includegraphics[width=6cm]{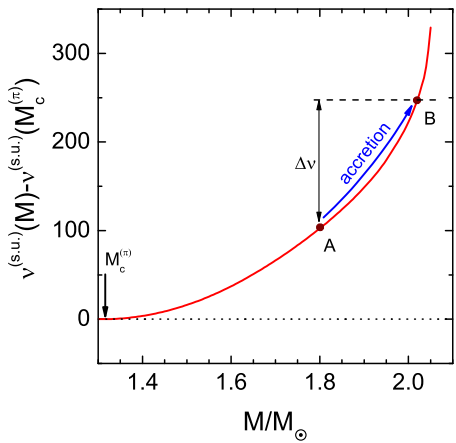}}
\caption{(Color online) Additional spin-up frequency of an accreting pulsar in the LMXB, which could be reached owing to accumulation of the pion condensation in the course of accretion. } \label{fig:nupi-spinup}
\end{figure}

Assume that after some time, when the neutron star was substantially slowed down below $\Omega_{c,{\rm L}}$ (here we  assume that $\Omega_{c,{\rm L}}$ is so low that we can neglect the effect of the resulting  rotation), it has a mass $M_A$ and already a rather low value of the magnetic field. Then during a long period of accretion the pulsar central density and the total mass increase slowly. If $M_A$ exceeds $M_c^{\pi/K}$ in the course of the accretion, a meson condensate with a finite momentum $k_0$ starts forming. It could be a p-wave charged pion condensate~\cite{Migdal:1990vm} or a p-wave $K^-$ condensate studied in Refs.~\cite{KVK95,KV2003}. To be specific, we further speak about pion condensation.

With a toroidal form of the condensate field (see Ref.~\cite{Voskresensky:1993uw}), the condensate component (for $r<r^{\pi}_c$) gets the  finite angular momentum
\begin{align}
L_{\pi,c}=\frac{\pi^2}{2}\intop_0^{r^{\pi}_c} Z_0^{-1}(k_0) k_0\,
|\varphi_0|^2\, r^3\rmd r\,, \label{Lpi-up}\end{align} where
$r^{\pi}_c$ is the radius  of the sphere occupied by the
condensate. Owing to the angular momentum conservation, the
noncondensate part of the matter spins up and after passing of
some dissipation time  the whole star becomes involved in this
rotation. The resulting angular velocity $\Omega_{\pi}^{\rm
(s.u.)}$  of the star (simplifying, we assume that
$\Omega_{\pi}^{\rm (s.u.)}=const$) is then determined by the
relation
\begin{align}
\Omega_{\pi}^{\rm
(s.u.)}\,\frac{8\pi}{3}\intop_{0}^R\,\rho\,r^4\rmd r =
 L_{\pi,c}\,. \label{Ompi-up}
\end{align}
Taking for estimation $Z_0^{-1}(k_0)\simeq 2m_\pi$, $k_0\simeq p_{\rmF}$ and $|\varphi_0|^2=|a_\pi|^2\simeq \tilde{\om}^2_\pi$ given by line 3 in Fig.~\ref{omegatil}, and substituting Eq.~(\ref{Lpi-up}) in Eq.~(\ref{Ompi-up}) we can express the spin-up frequency of the star as
\begin{align}
\nu_\pi^{\rm (s.u.)}&=\frac{4.3\cdot 10^3\,{\rm Hz}}{R_6\,(n_{\rm
cen}/n_0)^{2/3}} \frac{\intop_0^{x_c^\pi}
f^{1/3}(x) x^3 \frac{\tilde{\om}^2_\pi}{m_\pi^2}\rmd
x}{\intop_{0}^1 f(x) x^4\rmd x}\,,
\end{align}
here $x_c^\pi=r_c^\pi/R=\sqrt{1-(n_c^\pi/n_{\rm cen})^2}\theta(n_{\rm cen}-n_c^\pi)$\,.

The additional spin-up frequency, which may be reached owing to the accumulation of the pion condensation in the course of accretion is demonstrated in Fig.~\ref{fig:nupi-spinup}. Assume that the neutron star was slowed below $\nu_{c,{\rm L}}$ and its mass is $M=M_A >M_c^{(\pi)}$. Let the mass of the pulsar at the present time be $M_B$. Then, owing to the proposed mechanism the pulsar gets additional frequency $\Delta \nu$, as shown in Fig.~\ref{fig:nupi-spinup}. For $M_A =1.8 M_{\odot}$ and $M_B =2.02 M_{\odot}$ we estimate the change of the frequency as $\Delta \nu =150$\,Hz. The higher is the  mass of the pulsar $M_B$ reached in the course of the accretion at fixed $M_A$, the larger is $\Delta \nu$.


\section{Conclusion}\label{Conclusion}

In this paper we studied different mechanisms of deceleration of
rotating neutron stars and we considered stability conditions
against the excitation of $r$-modes. Computations are performed using
the HDD equation of state from Ref.~\cite{Blaschke:2013vma}, which
for densities $n\lsim 4n_0$ coincides with the HHJ fit of the APR
EoS but becomes stiffer for higher densities yielding the maximum
neutron star mass  $\simeq 2.05 M_{\odot}$, the latter value is in
agreement with observations.

For calculations of different partial contributions to viscosities
and to neutrino emissivities one often uses the free one pion
exchange model and the free nucleon-nucleon ($NN$) cross-sections
corrected by the Pauli blocking effects. We exploit more complex
$NN$ interaction based on the Migdal's Fermi-liquid  approach with
 account of {\em in-medium pion exchange}, which
incorporates an attractive pion-nucleon interaction and repulsive
$NN$ correlations. Included together, these effects result in an
increase of the $NN$ interaction amplitude with the density growth
for densities $n\gsim n_0$, where $n_0$ is the nuclear saturation
density; cf. Ref.~\cite{Migdal1978,Migdal:1990vm}. In the
literature this effect is usually called {\em the pion-softening
effect}. This pion-softening effect may result in  the pion
condensation for $n>n_c^{\pi}>n_0$.  Simplifying, we do not
distinguish effects of $\pi^+_s,\pi^-$ and $\pi^0$ condensates
studied in the literature~\cite{Migdal1978,Migdal:1990vm}, just
assuming that the pion condensation in the neutron star matter
occurs for $n>n_{c}^{\pi}=3n_0$ and  focusing on reactions
associated with the $\pi^-$ condensate. Owing to the lack of
knowledge about the $NN$ interaction for densities $n\gg n_0$, we
also consider the case where there is no pion condensation and the
pion softening either continues or saturates for $n\gsim 3n_0$.

 {\em Nucleon superfluidity} effects are incorporated, where necessary.
The contributions to the shear and bulk viscosities of different
processes are recalculated within our model and novel dissipation
mechanisms are suggested.

The most important contribution to the shear viscosity for $T\lsim
10^{9}$\,K  comes from the {\em lepton shear viscosity}. When one
incorporates the in-medium polarization effects~\cite{Shternin:2008es},
this lepton term proves to be less,
typically by an order of magnitude, compared to the quantity
estimated by Flowers and Itoh~\cite{FlowersItoh}; see
Fig.~\ref{fig:eta1}. In the proton superfluid region the lepton
term increases several times for $T=10^9$~K (see
Fig.~\ref{fig:etaSC}), compared to the same quantity computed for
non-superfluid matter. We calculated  the {\em $NN$ shear
viscosity} in nonsuperfluid matter and argued that it changes
rather moderately with a density increase.  The $NN$ term  proves
to be substantially less than the lepton term in the
temperature-density interval of our interest. Thus, we argue that
the $NN$ shear viscosity term can be dropped. Relying on these
findings we argued that calculations of the shear viscosity
performed in the number of papers, which used the Flowers--Itoh
results, yield too high critical angular velocities for the
$r$-mode stability at low temperatures $T\lsim 10^9$\,K and must be
reconsidered.

A novel term to the shear viscosity related to the interaction of {\em phonons with paired nucleons} was calculated. It proves
to be not as important as the lepton shear viscosity for the
values of the pairing gaps, which we exploit. A suppression of the
gaps by in-medium effects is included in our analysis, which causes
their model dependence.  If these suppression effects were minor,
the phonon-paired neutron term would contribute to the resulting
shear viscosity in the temperature range, $T\sim
(5\mbox{--}8)\cdot 10^8$~K; see Fig.~\ref{fig:eta-phon}.

Also, we introduced a novel {\em neutrino shear viscosity} term
contributing in a part of the neutron star interior, where
neutrinos are trapped at sufficiently high temperatures. This term
increases, if we exploit the medium modified pion exchange instead
of the free one pion exchange. Although the neutrino  term is
larger than other contributions to the shear viscosity for
temperatures $\gsim (3\mbox{--}5)\cdot 10^9$~K (see
Fig.~\ref{fig:eta-nu}), it proves to be much smaller than the bulk
viscosity computed for the same temperatures. The neutron star
cooling calculations demonstrate that temperatures $\sim
(3\mbox{--}5)\cdot 10^9$~K are reached for $t\gsim 10^{-3}$yr.
This opens an opportunity to observe, in the future, the neutrino
radiation from  newly born neutron stars during a longer time than
that follows from the conservative estimation, $\sim 10$~s.

The bulk viscosity is presented, as the sum of  three
contributions: the collisional term, the soft-mode term and the
radiation term. The {\em collisional bulk viscosity} is found to
be small and can be safely neglected. The {\em soft-mode bulk
viscosity} term is related to the weak interaction reactions
occurring in the non-equilibrium system on charged currents. With
our HDD EoS, the one-nucleon direct Urca (DU) reactions with the
neutrino production occur only for the most massive neutron stars
(for $M>1.9\, M_{\odot}$). The matrix element of the two-nucleon
modified Urca (MU) reactions are strongly increased provided we
take into account the pion softening with a nucleon density
increase. Following the notation of previously published works we call
so-calculated MU  reaction, as medium modified Urca (MMU)
reactions. With our estimates, {\em the bulk viscosity in the MMU
reactions increases up to three to four orders of magnitude for
$n\sim 3n_0$} (see Fig.~\ref{fig:F-fact}), which is in accord with
the corresponding increase of the emissivity of the MMU process
incorporated in the nuclear medium cooling scenario successfully
exploited in
Ref.~\cite{Schaab:1996gd,Blaschke:2004vq,Grigorian:2005fn,Blaschke:2011gc,Blaschke:2013vma}.
A contribution to the bulk viscosity from the pion Urca (PU)
reaction processing  on the pion condensate is included for
densities $n>n_{c}^{\pi}$. With our estimations, in presence of
the pion condensation  for the most massive neutron stars the
profile averaged bulk viscosity term related to the MMU process
becomes only by an order of magnitude less than that due to the
PU (see curves 2 and PU in Fig.~\ref{fig:zeta}), superfluidity is
suppressed, where we show partial contributions to the soft-mode
bulk viscosity from different reactions  (the effects of
superfluidity are switched off in Fig.~\ref{fig:zeta}). In the
absence of the pion condensation but with the saturated or
continued pion softening with increase of the density, the profile
averaged bulk viscosity term related to the MMU process still
increases (see curves 1b and 1c in Fig.~\ref{fig:zeta}). We also
re-analyzed the applicability of the usually made assumptions on
relation between the relaxation time of perturbed lepton
concentrations, $\tau_{X,l}$, and the $r$-mode frequency $\om$,
$\om\tau_{X,l}\ll 1$. The ratio of the total bulk viscosity
averaged  over the star density profile to the same average of the
bulk viscosity calculated in the limit case $\om\tau_{X,l}\ll 1$
begins to deviate from unity  with decrease of the frequency
and/or with increase of the temperature; see
Fig.~\ref{fig:zeta-om-rat}.  Most of observed rapidly rotating
young pulsars have rotation frequencies $\Omega\sim 10^2$ Hz. At
such a frequency, for the heaviest neutron star the deviation of
the ratio from unity starts for $T\geq 7\cdot 10^9$.

We calculated {\em the radiative bulk viscosity contribution} from
 the neutrino reactions going on the charged and neutral weak
currents. Previously only reactions on charged currents were
considered with the $NN$ interaction not including in-medium
modifications. We show that with paring gaps, which we exploit,
processes on the neutral currents, such as the nucleon-nucleon
bremsstrahlung and the pair braking-formation (PBF) processes,
contribute only a little. The resulting radiative bulk viscosity
proves to be of the same order as the soft-mode viscosity for
temperatures $T\lsim (5\cdot 10^9\mbox{--}10^{10}$)\,K (see
Fig.~\ref{fig:Rmmu}), where the ratio of the profile averaged
radiative MMU bulk viscosity term  to the MMU soft-mode term is
shown. This ratio tends to 1.5 for low temperatures.

The exponential suppression factors  for the soft-mode and the
radiative bulk viscosities are incorporated in the regions with
the $NN$ pairing. Finally, with all effects included, the
resulting profile-averaged viscosity  for the typical $r$-mode
frequencies $\sim 10^4$ Hz is determined mainly by the lepton
shear viscosity term for $T\lsim (2\cdot 10^8\mbox{--}2\cdot
10^9)$K (in dependence on the neutron star mass) and by the bulk
viscosity for higher temperatures; see
Fig.~\ref{fig:zeta-eta-com}. The bulk viscosity is mainly
determined by the MMU reactions and, supplementary, by the PU and
DU reactions, whenever these reactions  occur.

With all the processes included we calculated critical spin
frequencies, above which the star is unstable with respect to the
$r$-mode excitation. The  critical spin frequency as a function of
the temperature  $\nu_c(T)$, computed in a broad range of the
neutron star masses is presented in Fig.~\ref{fig:tx}. This is the
key figure demonstrating results of our present study.

We showed that with the bulk viscosity calculated within the
minimal cooling paradigm~\cite{Page:2006ly}, when the most
efficient neutrino processes are the MU and the PBF ones, the
value of the  spin frequency of the most rapidly rotating young
pulsar PSR~J0537-6910 ($\nu=62$~Hz) is above the minimum on
curve $\nu_c (T)$. The observed frequency of pulsar
PSR~J0537-6910 would be hardly explained, if the pulsar were in
unstable region. Only when the DU reactions become efficient (for
$M>2.03 M_{\odot}$ with our EoS, i.e. very close to the maximum
mass of 2.05\,$M_{\odot}$), the minimum of $\nu_c (T)$ overwhelms
the frequency of PSR~J0537-6910. However, because the star has
passed the $r$-mode unstable region during its early evolution, it
may hardly keep its mass very close to the maximal possible value.
Another possibility~\cite{Alford:2012yn} to explain the data on
PSR~J0537-6910 within the minimal cooling paradigm exists if the
trajectory $\nu (T)$ is appropriately shifted to the left from the
minimum of the $\nu_c (T)$. To find actual $\nu (T)$ dependence
one should solve the system of dynamical equations for the
rotation frequency $\nu (t)$, temperature $T(t)$, and $r$-mode
amplitude $a(t)$.

The low values of the observed frequencies of the young pulsars
are  naturally explained within the nuclear medium cooling
scenario of Refs.~\cite{Blaschke:2004vq,Grigorian:2005fn,Blaschke:2011gc,Blaschke:2013vma},
which we exploit in the present work. With the pion softening
effects included, the frequency 62~Hz proves to be below the
minimum of $\nu_c (T)$ provided the star mass is $\gsim
(1.77\mbox{--}1.84)\,M_{\odot}$, whereas the DU process appears
for  $M>1.9 M_{\odot}$. At the same time, we observe that the
frequencies of most rapidly rotating recycled pulsars in LMXB are
substantially above curve $\nu_c (T)$ at relevant temperatures
$T\sim 10^8$~K. This statement is insensitive to any possible
enhancement of the bulk viscosity (in the MMU, PU and DU
processes). Also, any  artificial increase of the lepton shear
viscosity in  reasonable limits does not help to rise the line
$\nu_c (T)$ above the data points.

As a possibility to explain the stability of the rotation of the
pulsars in LMXB,  we suggested a new mechanism allowing to {\em
transfer  a part of the angular momentum to inhomogeneous
condensates of bosonic excitations } and/or to the inhomogeneous
charged pion/kaon condensates. For this transfer to
occur~\cite{Pitaev84,Voskresensky:1993uw,Vexp95,BP12}, one needs
bosonic modes with low energies at sufficiently large momenta to
exist in the neutron star matter. When a part of the  angular
momentum is transferred to the inhomogeneous condensate, as the
consequence of the angular momentum conservation, the remaining
non-condensate part of the matter participating in the $r$-mode
generation will rotate with a lower angular velocity. More
specifically,  we took into account condensation of excitations
occupying the mode at a finite momentum in the proton superfluid.
Then the part of the star, where there is the proton pairing, may
rotate with a much lower speed than the other part of the star.
We modeled the effect in simplifying assumptions that the critical
velocity for the appearance of the condensate of excitations is
very low and a part of the star, where the condensate occurred, is
fully stopped. Our results presented in Fig.~\ref{fig:nuBcondin}
show that, with such effects taken into account, the critical
frequency $\nu_c$ may increase substantially and the data on the
rapidly rotating old pulsars may fall into the $r$-mode stable
region.

One usually assumes that  old pulsars  are slowed down owing to
the magnetic dipole radiation, simultaneously losing the magnetic
field.  At a longer time period old recycled pulsars in LMXB are
spin up owing to the accretion of matter from a companion star~\cite{Patruno:2010qz}.
We proposed a supplementary mechanism of
the pulsar acceleration owing to the charged pion/kaon condensation
with non-zero momentum accumulated in the course of the accretion.
Because it is energetically favorable for the condensate to move as
a whole in a one particular direction, the non-condensate matter
would be driven in the opposite direction, which results in an
acceleration of the non-condensate part of the star, as seen by a
distant observer.

Concluding, the data on young rapidly rotating pulsars can be
naturally explained provided one incorporates pion-softening
effects with increase of the density. As for the old rapidly
rotating recycled pulsars in LMXB, in spite of a lot of works
performed by many researchers including the present research,
further more detailed calculations, as well as the search for new
stability mechanisms, are needed to arrive at a convincing
explanation of their $r$-mode stability.

\acknowledgments
The study presented here was partially motivated by discussions on the workshops ``Neutron Stars: Nuclear Physics, Gravitational Waves and Astronomy'' organized by the Institute of Advanced Studies at the University of Surrey.  The work was supported by the Ministry of Education and Science of the Russian Federation (Basic part), by Grants No. VEGA~1/0457/12 and No. APVV-0050-11 and by ``NewCompStar'', COST Action No.~MP1304.

\appendix
\section{}\label{app}
The lepton collision times are given by
\begin{align}
\tau_e=\frac{\nu_\mu-\nu'_{e\mu}}{\nu_e\nu_\mu-\nu'_{e\mu}\nu'_{\mu
e}}\,, \quad \tau_\mu=\frac{\nu_e-\nu'_{\mu
e}}{\nu_e\nu_\mu-\nu'_{e\mu}\nu'_{\mu e}}\,
\label{tauE-Mu}
\end{align}
with the collision frequencies
\begin{align}
\nu_e &= A
\Big(\frac{n_0}{n_p}\Big)^{\frac29}\Big(\frac{n_p}{n_e}\Big)^{\frac23} T_9^{\frac53} (1+r)^{\frac23}
+ B \Big(\frac{m_N}{m_N^*}\Big)^{\frac12} \Big(\frac{n_0}{n_p}\Big)^{\frac16} T_9^2 \,
\nonumber\\
&\times \Big(f_{+,e} + \Big[2
+\frac{m_N^*}{m_N}\Big(\frac{n_0}{n_e}\Big)^{\frac23}\Big]\,f_{-,e}
+ f_{-,\mu} \Big)\,,
\label{nue}\end{align}
\begin{align}
\nu_\mu &= A\Big(\frac{n_0}{n_p}\Big)^{\frac29}\Big(\frac{n_p^2}{n_e n_\mu}\Big)^{\frac13}
T_9^{\frac53}(1+r)^{\frac23}
\nonumber\\
&+B \Big(\frac{m_N}{m_N^*}\Big)^{\frac12} \Big(\frac{n_0}{n_p}\Big)^{\frac16} T_9^2 \,
\Big(\frac{n_e}{n_\mu}\Big)
\nonumber\\
&\times\Big[\Big( 2+ \Big(\frac{n_\mu}{n_e}\Big)^{\frac23}+ \frac{m_N^*}{m_N}
\Big(\frac{n_0}{n_e}\Big)^{\frac23} \Big)  f_{-,\mu} +
\Big(\frac{n_\mu}{n_e}\Big)^{\frac23} f_{+,\mu} \Big]\,,
\label{numu}\end{align}
\begin{align}
\nu'_{e\mu} & 
 =B \Big(\frac{m_N}{m^*_N}\Big)^{\frac12} \Big(\frac{n_\mu}{n_e}\Big)^{\frac23}
\Big(\frac{n_0}{n_p}\Big)^{\frac16} T_9^2[f_{+,\mu}+f_{-\mu}]\,,
\label{nupemu}\\
\nu'_{\mu e}&= B \Big(\frac{m_N}{m^*_N}\Big)^{\frac12} \Big(\frac{n_\mu}{n_e}\Big)
\Big(\frac{n_0}{n_p}\Big)^{\frac16} T_9^2[f_{+,\mu}+f_{-\mu}]\,,
\label{nupmue}\\
 &\quad f_{\pm,l} =
\frac{2}{\pi}{\rm arctan} s_l \pm \frac{2}{\pi}\frac{s_l}{1+s_l^2}
\,,
\nonumber\\
&\quad A=9.36\cdot 10^{14}~{\rm Hz}\,,\quad B/A=0.118 \,,
 \nonumber\\
&\qquad s_l=152 \, \frac{m_N}{m_N^*}\, \Big(\frac{n_l \,
n_e^3}{n_0^2 n_p^2}\Big)^{1/6} \,.
\nonumber
\end{align}
In Eqs.~(\ref{nue},\ref{numu},\ref{nupemu},\ref{nupmue})
we retain  only leading order terms in the electromagnetic coupling constant $\alpha$,
with $A\propto \alpha^{5/3}$ and $B\propto \alpha^{3/2}$.


\end{document}